\documentclass[letterpaper, 10 pt, doublecolumn, conference]{IEEEtran}  % Comment this line out
\usepackage[none]{hyphenat}
\usepackage{tabularx}
\usepackage{graphicx}
\usepackage{subcaption}
\usepackage{float}
\usepackage{cuted}
\usepackage{amsmath}

\usepackage{array} % provides m{} column type
\usepackage[hyphens]{url}
\usepackage[
    breaklinks=true,
    colorlinks=true,
    allcolors=blue,
    linktoc=all
]{hyperref}
\makeatletter
\newcommand{\nodottedtocline}[5]{%
  \vskip \z@ \@plus.2\p@
  {\leftskip #2\relax \rightskip \@tocrmarg \parfillskip -\rightskip
   \parindent #2\relax \@afterindenttrue
   \interlinepenalty\@M
   \leavevmode
   \@tempdima #3\relax
   \advance\leftskip \@tempdima \null\nobreak\hskip -\leftskip
   {#4}\nobreak\hfill
   \nobreak\hb@xt@\@pnumwidth{\hss #5}\par}
}

\renewcommand{\l@section}[2]{%
  \vspace{1.7mm}% adjust to taste (negative = tighter)
  \nodottedtocline{1}{0em}{1.5em}{\bfseries #1}{#2}%
}

\makeatother

\hypersetup{pdftex,colorlinks=true,allcolors=blue}
\usepackage{hypcap}

\usepackage{graphics} % for pdf, bitmapped graphics files
\usepackage{epsfig} % for postscript graphics files
\title{\LARGE \bf FlexNGIA 2.0: Redesigning the Internet with Agentic AI\\
\Large Protocols, Services, and Traffic Engineering Designed, Deployed, and Managed by AI}

\author{Mohamed Faten Zhani$^{1}$, Younes Korbi$^{2,3}$ and Yamen Mkadem$^{2,3}$\\
         {$^{1}$Department of Computer Engineering,}
         {Center for Intelligent Secure Systems (IRC-ISS),}\\
         {King Fahd University of Petroleum \& Minerals (KFUPM), Saudi Arabia}\\
         {$^{2}$FlexNGIA, Tunisia}\\
         {$^{3}$ISITCom, University of Sousse, Tunisia} \\
         {\tt\small mohamed.zhani@kfupm.edu.sa}
         {\tt\small \{ykorbi, ymkadem\}@FlexNGIA.net}
}

\begin{document}

\onecolumn

\maketitle
\thispagestyle{empty}
\pagestyle{empty}

\providecommand{\keywords}[1]
{
  \small	
  \textbf{\textit{Keywords---}} #1
}

%The~keynote details are available at \url{https://goo.gl/ppdsVP}.}

%\twocolumn[{
%\begin{strip} %to use two columns for the abstract and table of content (\usepackage{cuted})
\vspace{1\baselineskip}

\begin{abstract}

\textbf{The escalating demands of immersive communications, alongside advances in network softwarization and AI-driven cognition and generative reasoning, create a pivotal opportunity to rethink and reshape the future Internet.
In this context, we~introduce in this paper, FlexNGIA 2.0, an Agentic AI-driven Internet architecture that leverages LLM-based AI agents to~autonomously orchestrate, configure, and evolve the network. These agents can, at runtime,  perceive, reason, coordinate among themselves to dynamically design, implement, deploy, and adapt communication protocols, Service Function Chains (SFCs), network functions, resource allocation strategies, congestion control, and traffic engineering schemes, thereby ensuring optimal performance, reliability, and~efficiency under evolving conditions.\vspace{0.5\baselineskip}}

\textbf{The paper first outlines  the overall architecture of FlexNGIA 2.0 and its constituent LLM-Based AI agents. For~each~agent, we~detail its design, implementation, inputs and outputs, prompt structures, interactions with tools and other agents, followed by~preliminary proof-of-concept experiments demonstrating its operation and potential. The results clearly highlight the~ability of~these LLM-based~AI~agents to~automate the~design, the implementation, the deployment, and the performance evaluation of~transport protocols, service function chains, network functions, congestion control schemes, and resource allocation strategies.\vspace{0.5\baselineskip}
}
\textbf{Building on these capabilities, FlexNGIA 2.0 paves the way for a new class of Agentic AI-Driven networks, where fully cognitive, self-evolving AI~agents can autonomously design, implement, adapt and optimize the network's protocols, algorithms, and behaviors to efficiently operate across complex, dynamic, and heterogeneous environments.}
\textbf{To bring this vision to reality, we also identify in~this~paper key research challenges and future directions toward achieving fully autonomous, adaptive, and agentic AI-driven networks.
}\\


% TCP \and  QUIC \and  SCTP \and  FlexNGIA \and  Holoportation \and  Telepresence \and  Virtual Reality \and  Augmented Reality}
% \PACS{PACS code1 \and PACS code2 \and more}
% \subclass{MSC code1 \and MSC code2 \and more}
\end{abstract}
%\end{strip}
\vspace{1\baselineskip}

%Table of content
%\newpage
\tableofcontents
\twocolumn % switch back to two-column mode

\vspace{1.5cm} 
%\newpage
%}]

\newpage

\addtocontents{toc}{\protect\setcounter{tocdepth}{2}} %hide subsubsection

\newpage
\newpage
%------------------------------------------------------
\section{Introduction}
%------------------------------------------------------
Immersive communications, including real-time virtual and augmented reality, holography, haptics, represent the~forthcoming pinnacle of digital demand. Yet the stringent performance requirements of these applications, such as ultra-low latency, high reliability, and massive bandwidth, remain far beyond the capabilities of today’s Internet.  Fortunately, recent advances in network softwarization and programmability, combined with breakthroughs in~AI-driven cognitive, Large Language Models (LLMs), and generative reasoning, could provide an~unprecedented opportunity to fundamentally rethink and reshape the architecture of the Internet to~make~it able to meet the ambitious performance, reliability, and responsiveness demanded by~these~next-generation applications.

Before exploring this opportunity, it is important to examine the core limitations of today’s Internet architecture. They can be summarized as~follows.

\vspace{0.5em}\noindent\textbf{$\bullet$ Best-effort Service Model:} 
Today’s Internet relies on a best-effort delivery model, where Internet Service Providers offer basic transit services without any performance guarantees. This model is inadequate for emerging applications that demand deterministic performance.

\vspace{0.5em}\noindent\textbf{$\bullet$ Rigid Communication Protocols:} The Internet’s communication stack relies on rigid, monolithic protocols. For~instance, transport protocols (e.g.,~TCP~\cite{rfc793TCP}, QUIC~\cite{QUIC-UDP}, SCTP~\cite{SCTP-rfc4960}) offer a fixed bundle of services (e.g.,~reliability, in-order delivery, congestion control) without the flexibility to customize them based on application needs. They are also based on a two-endpoint model, which clashes with modern distributed applications connecting multiple endpoints, resulting in flows being managed independently in~the~network, even when they are interdependent or~even~belong to the same application. 
Another fundamental limitation is the lack of cross-layer collaboration. For~instance, the transport and network layers operate in silos with limited visibility into each other’s state and goals.
This isolated design prevents holistic optimization across all flows and~leads to~suboptimal or conflicting behaviors.

\vspace{0.5em}\noindent\textbf{$\bullet$ Lack of Customizable Network Functions and~Services:} 
The current core network of the Internet primarily focuses on data delivery and lacks the inherent capability to~dynamically integrate new network functions and~services to~support application-specific requirements. Ideally, the~network would allow the~instantiation of customizable functions and services, such~as~real-time content analysis and~processing, content-based routing, advanced packet caching and~forwarding, and AI/ML-driven services.

\vspace{0.5em}\noindent\textbf{$\bullet$ Congestion Control - A Key Bottleneck for Performance:} 
Congestion control (CC), i.e.,~dynamically adjusting the~sending rate, plays a critical role in determining overall application performance. Numerous CC algorithms have been proposed in the literature, each tailored to~specific environments, runtime conditions and application requirements (e.g.,~wired vs.~wireless networks, WAN vs.~LAN, high-bandwidth links)~\cite{Korbi2024-CC-IEEEAccess, jacobson1988, ha2008, cardwell2017bbr, casetti2002westwood, CC-QoE-2023}. This~implies that a~CC~scheme tailored for specific network conditions may suffer significant performance degradation when those conditions change. In today’s Internet, however, CC~algorithms are typically fixed at deployment and remain static regardless of the network conditions and application requirements \cite{linux-default-cc-2025,microsoft_set_nettcpsetting_2025,microsoft_cubic, ns3tcp2022,ns3tcp2022}, and this lack of~adaptability constitutes a major performance bottleneck.

%------------------------
To address these limitations, a novel fully-\textbf{Flex}ible \textbf{N}ext-\textbf{G}eneration \textbf{I}nternet \textbf{A}rchitecture (FlexNGIA 1.0) was recently proposed~\cite{FlexNGIA2019,FlexNGIAWebsite}. FlexNGIA overcomes the~Internet’s rigid protocols, lack of guarantees, and limited in-network intelligence by~enabling per-application Service Function Chains (SFCs) with guaranteed performance. An~SFC is an ordered sequence of customized network functions (NFs) orchestrated through tailored communication protocols. The originality of FlexNGIA lies in its ability to compose SFCs with NFs that can operate across any layer of the OSI model, while being governed by fully customizable protocols adapted to the specific requirements of each application.

Most of these customizations in FlexNGIA 1.0, such~as designing SFCs, network functions, communication protocols, congestion control and traffic engineering schemes, is~performed manually by developers or network engineers. Once deployed, these components cannot be modified at~runtime, limiting the network’s ability to adapt to changing conditions. In~addition, the FlexNGIA 1.0 management framework, like its traditional counterparts~\cite{rfc3411,openstack,opendaylight,tungstenfabric,onap}, is composed of modular, self-contained functional components for resource allocation, failure management, and monitoring. These modules operate in a largely static, rule-based manner, relying on pre-programmed algorithms that can only be tuned by adjusting parameters, without the~ability to modify the underlying logic or behavior.

In this paper, we build on the FlexNGIA 1.0 foundation to propose FlexNGIA 2.0, an agentic AI architecture for~the~next-generation Internet, in~which LLM-based~AI~agents can autonomously design, implement, adapt, and manage SFCs, network functions, communication protocols, congestion control and traffic engineering schemes, and resource allocation algorithms, all at runtime. These~agents are equipped with machine learning, reasoning, and natural language understanding~\cite{derouiche2025agenticai, bornet2025agentic}. They can collaborate, learn, and adapt, continuously evolving at runtime the underlying logic, behavior, and algorithms that drive communication protocols, network functions, service management, and traffic engineering schemes.

FlexNGIA 2.0 redefines the network architecture by~embedding cognitive intelligence and reasoning, enabling AI~agents to redesign the network's protocols, logic and~algorithms on the fly. This game-changing capability delivers the flexibility, intelligence, and responsiveness needed to adapt in real time to different network environments, and~application demands. In~this~context, the~contributions of~this~paper could be summarized as~follows:
\begin{itemize}
    %\item We motivate the shift towards Agentic AI and present the anatomy of an LLM-based AI agent 
    \item We introduce the FlexNGIA 2.0 architecture and the role of its key composing LLM-based AI agents.
    
    \item We present a detailed design for each agent, including its internal architecture, implementation, inputs and outputs, prompt structures, and interactions with other agents and tools, highlighting their potential to autonomously support complex network tasks such as protocol design, service function chaining, congestion control, and~resource allocation.
    
    \item We implemented a prototype of the key agents\footnote{The code of the prototype, the developed agents, experiments is available upon request. Please use FlexNGIA Contact Page \cite{FlexNGIAWebsite}}, and we demonstrate through experiments their~efficiency. These experiments highlight the potential of~AI~Agents powered by advanced LLM models with reasoning and problem-solving capabilities (e.g.,~GPT-5 Thinking~\cite{openai_introducing_gpt5_2025}, DeepSeek-R1-Distill-Llama-70B~\cite{deepseek2025}) to autonomously design, customize, implement and manage network protocols and services.
    Specifically, we show their efficiency in \textbf{(1)}~analyzing application requirements, designing and~implementing a~complete custom end-to-end SFC, including all its composing network functions along~with a~tailored transport protocol; \textbf{(2)}~dynamically deciding among different CC~schemes, and even design and implement new schemes at runtime to better match the application’s specific requirements and~network conditions; and~\textbf{(3)}~guiding the~resource allocation strategy to~satisfy applications' performance and ensure the~operator's operational, sustainability, and~economic objectives.
    
    \item We finally highlight current limitations, identify key research challenges, and outline future directions toward evolving FlexNGIA~2.0 and building a fully autonomous, Agentic AI-driven Internet.
\end{itemize}

\textit{It is important to emphasize that the main contribution of this work is the proposed LLM-based agentic AI architecture for future autonomous networks. The presented implementation of some of the proposed agents and the experimental results should be interpreted as a proof-of-concept demonstrating feasibility and initial potential rather than a fully mature or production-ready system. A comprehensive investigation of robustness, reliability, scalability, security, and operational constraints of these AI agents and LLM-driven networking systems remains an open research direction, as further discussed in the key challenges and future directions Section of this paper.}\\

In the remainder of this paper, we begin by providing an~overview of FlexNGIA 1.0 (Section~\ref{section:FlexNGIA1Overview}). We then introduce FlexNGIA 2.0, an agentic AI-based Internet Architecture fully managed by~AI~agents (Section~\ref{sec:FlexNGIA2}). Next, we detail the design of its key internal AI agents, demonstrating their operation through a proof of concept and~experimental results (Section~\ref{sec:AgenticAIinAction}). Following this, we outline key research directions to evolve FlexNGIA 2.0’s agentic AI design (Section~\ref{sec:KeyResearchDirections}). The paper concludes with~a~summary discussion~(Section~\ref{sec:Conclusion}).

%Furthermore, traditional congestion control mechanisms in TCP~\cite{} rely primarily on indirect signals, such as timeouts or duplicate acknowledgments, which may misinterpret the actual network state (e.g., mistaking non-congestion-related losses for congestion~\cite{Korbi2024-CC-IEEEAccess}). This often results in overly conservative rate reductions, which degrades application performance. Additionally, these schemes typically ignore the state of concurrent flows competing for bandwidth, as well as the relevance and priority of the data being transmitted by each flow. Collectively, these limitations make current congestion control mechanisms a significant bottleneck in achieving optimal application performance.

%------------------------------------------------------
\section{FlexNGIA 1.0 Overview}\label{section:FlexNGIA1Overview}
%------------------------------------------------------

%FlexNGIA (Flexible Next-Generation Internet Architecture) is a novel architectural framework designed to support the stringent requirements of future network applications such as virtual reality, holoportation, augmented reality, and telepresence. These applications demand ultra-low latency, high reliability, availability, and performance, challenges that current Internet architectures are not equipped to meet.

\begin{figure}[tb]
    \centering
    \includegraphics[width=1.00\columnwidth]{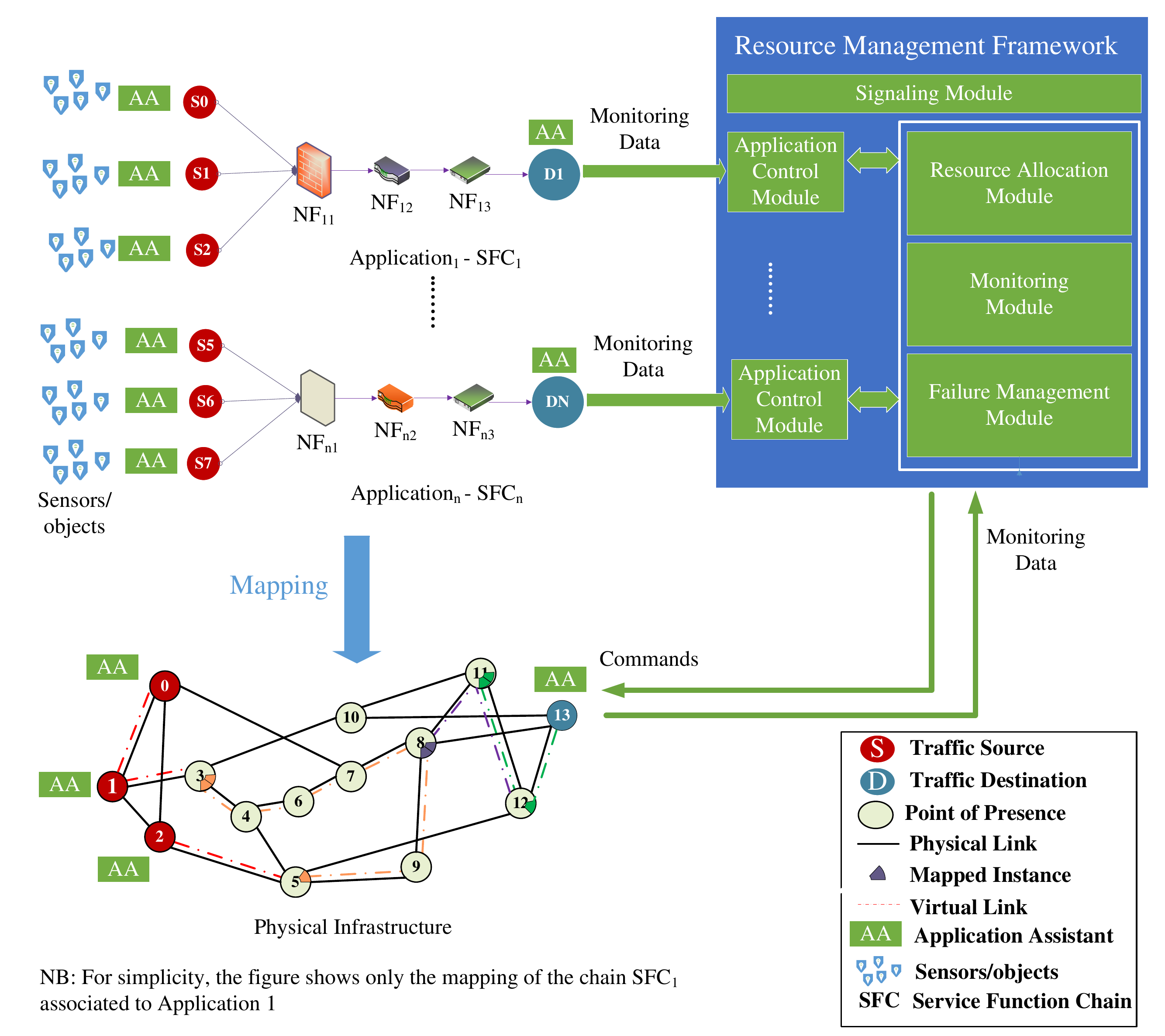}
    \caption{FlexNGIA 1.0 management framework.}
    \label{fig:FlexNGIA1.0ManagementFramework}
\end{figure}

The FlexNGIA architecture is motivated by the need to~overcome the aforementioned fundamental limitations of~today’s Internet, particularly its rigid protocol stacks, lack of service guarantees, and limited in-network intelligence. Its~key features can be summarized as follows:\\
\textbf{$\bullet$ Service Function Chain per application:} Rather than relying on traditional best-effort data delivery, the FlexNGIA architecture leverages in-network computing to deploy a~dedicated Service Function Chain for each application. This~SFC consists of a tailored set of network functions that steer traffic from sources to destinations while meeting the specific needs of the application (see, the example of SFC$_1$ associated with application$_1$ in~Fig.~\ref{fig:FlexNGIA1.0ManagementFramework}). Within this~model, applications' developers can define reliability and performance requirements (e.g.,~throughput, packet loss, end-to-end delay, and jitter) and specify customized network functions and communication protocols that best support these requirements. This SFC-based per-application approach enables fine-grained control, performance optimization, and~customized network services. \\
\textbf{$\bullet$ Advanced network functions:} In FlexNGIA, service function chaining supports the dynamic composition of~a~wide range of network functions, selected and ordered to~best serve the needs of each application. These functions can operate at any layer of the network stack and are not limited to traditional routing or forwarding tasks. For instance, it is possible to design application-aware network functions that include advanced capabilities such as context- and application-aware traffic analysis, processing and engineering, AI/ML-driven decision-making, real-time data analysis and processing, or media-specific operations like video/audio compression, cropping, and rendering. Network functions could also operate at lower layers. For~instance, transport and traffic engineering network functions can manage packet scheduling, grouping, caching, and retransmission to optimize delivery under varying network conditions. Another type of network functions could focus on monitoring and measurement to enable in-network continuous collection, filtering, and AI-based analysis of monitoring data in order to support adaptive management of the network, services and~applications. \\
\textbf{$\bullet$ Communication Protocol Customization:} Rather than relying solely on legacy protocols like TCP or~QUIC, FlexNGIA enables the design and deployment of communication protocols specifically adapted to each SFC and tailored to the unique requirements of its associated application. These protocols can operate at individual OSI layers or in cross-layer designs to enable inter-layer coordination. Additionally, they can incorporate interactions and collaboration between endpoints and the network functions of~the~SFC in order to enhance overall performance and reliability. They can also support multi-endpoint communication models, moving beyond the two-endpoint model of TCP and QUIC. For each designed protocol, a~custom packet header is defined, enabling the inclusion of~metadata and even control commands to be executed directly in the data plane and network functions.

%
%that enable coordinated communication across multiple devices
 %assumed by traditional protocols like TCP. 
%to enhance coordination, improve performance, and fully leverage in-network computing capabilities. 

By enabling rapid design and deployment of customized protocols and services, FlexNGIA moves beyond rigid layering and outdated Internet principles, and it opens the~network infrastructure to innovation and deep customization, empowering developers to optimize communication for~diverse and evolving application requirements.

\textbf{$\bullet$ FlexNGIA management framework:} 
The FlexNGIA proposes a management framework comprising several core modules~(Fig.~\ref{fig:FlexNGIA1.0ManagementFramework}). The Signaling module initiates the SFC creation upon application requests. The Application Control module manages each application’s SFC, estimating traffic and resources while adapting the chain based on continuous monitoring data. The Resource Allocation module provisions resources to meet operator goals like performance, costs, and~energy efficiency~\cite{moufakirITU2022, Racheg-ICC2017, Tashtarian2019, ZhaniGreenSLACNSM2014,AlomariIJNM2023}. The Failure Management module detects and predicts failures to ensure reliability and availability~\cite{Aidi-CNSM2018,AlomariSurvJNSM2023}. Finally, the Monitoring module gathers  data on physical and virtual components.

FlexNGIA 1.0 addresses fundamental Internet issues but depends on static modules and manual intervention for key tasks, limiting its flexibility, intelligence, and responsiveness to complex, real-time challenges. To overcome these shortcomings, we introduce, in the following, FlexNGIA~2.0, an agentic~AI framework that enables a fully intelligent, autonomous, and adaptive network management.

\section{FlexNGIA 2.0 Architecture}\label{sec:FlexNGIA2}

This section presents FlexNGIA~2.0, first discussing the~shift from static components to agentic AI, then describing the~building blocks of~LLM-based agents, and finally introducing the~complete FlexNGIA~2.0 architecture.

\subsection{FlexNGIA 1.0 to 2.0: A Shift Toward Agentic AI}

%----------------------------------------------------------------------------------
\begin{table*}[htb]
\scriptsize
\centering
\caption{Comparison of Existing Literature, FlexNGIA 1.0, and FlexNGIA 2.0}
\begin{tabularx}{\textwidth}{|X|X|X|X|}
\hline
\textbf{Aspect} & \textbf{Existing Literature} & \textbf{FlexNGIA 1.0} & \textbf{FlexNGIA 2.0} \\
\hline
\textbf{Architecture Type} & Static, protocol-centric & Modular, rule-based & Agentic AI, cognitive and adaptive \\
\hline
\textbf{Functional Design and Control} & Predefined, hard-coded & Modular, developer-defined & Runtime-adapted by autonomous agents \\
\hline
\textbf{Service Function Chaining (SFC)} & limited chain (packet forwarding, IDS, Firewall) & Designed and configured by network engineers  & Dynamically designed and adapted at runtime by AI agents \\
\hline
\textbf{Network Function Design} & Rigid & Designed and configured by network engineers & AI agents generate and evolve functions at runtime \\
\hline
\textbf{Communication Protocols Design} & Legacy Protocols (e.g., TCP) & Developer-designed & AI agents generate and evolve protocols at runtime \\
\hline
\textbf{Runtime Adaptation for the SFC, Network Functions \& Communication Protocols} & Limited or none & Not supported & Fully supported via agent-based decision-making augmented by LLMs\\
\hline
\textbf{Customization} & Manual, time-consuming & developer-led & Fully automated, AI-driven \\
\hline
\textbf{Context/Semantic Awareness} & Absent or minimal & Minimal & Agents interpret semantic, contextual, and runtime information \\
\hline
\textbf{Learning \& Intelligence} & Not integrated & Absent & Built-in learning, reasoning, and adaptation \\
\hline
\textbf{Scalability \& Flexibility} & Poor in dynamic environments & Limited by static design & High - agents scale and adapt the chain and the protocols \\
\hline
\textbf{Responsiveness to Change} & Manual reconfiguration needed & Requires redeployment & Real-time response to dynamic conditions \\
\hline
\textbf{Human Involvement} & High (design, tuning, updates) & High (especially for protocols/Network Functions/SFC) & Minimal - system adapts autonomously \\
\hline
\end{tabularx}
\label{tab:flexngia_comparison}
\end{table*}
%----------------------------------------------------------------------------------

FlexNGIA 1.0 is composed of modular, self-contained functional components that perform predefined tasks using static algorithms. Moreover, many critical responsibilities in~FlexNGIA~1.0, such~as designing service functions, developing communication protocols, implementing CC~schemes, and building network functions, are  handled manually by developers and~network engineer. This places the burden of customization and adaptability on manual engineering rather than on the system itself. Once these components are designed and deployed, they are typically fixed and~cannot be modified at runtime, significantly limiting the~system’s ability to respond to different network conditions and application requirements.

In contrast, FlexNGIA 2.0 adopts an agentic AI architecture, where autonomous AI agents are equipped with techniques like machine learning, reasoning, and natural language understanding, and are able to reason, set goals, choose strategies, plan to achieve them, learn, collaborate and adapt in real time to dynamic network conditions and evolving application requirements. These agents can also access and deploy specialized tools, such as measurement utilities, analysis modules, or protocol components, to perceive and act upon their environment. They maintain internal memory to store relevant context, including historical observations, prior decisions, and learned behaviors, enabling long-term reasoning and informed decision-making. Through machine learning, they can recognize patterns and optimize strategies over time. Natural language understanding allows them to interpret human-provided guidance or documentation when available, while reasoning capabilities support goal-driven planning and coordination with other agents. This combination of tools, memory, learning, and interaction enables these agents to operate in a decentralized yet cooperative manner, continuously improving their ability to manage complex, dynamic networking scenarios.

In short, FlexNGIA 2.0 AI agents can design and manage at runtime service function chains, network functions, communication protocols, congestion control schemes, and~resource allocation algorithms. By interpreting semantic, contextual, and real-time information, they dynamically adapt application protocols, network functions, and service chains, enabling more flexible, intelligent, and scalable networks.

This evolution represents a fundamental shift from rigid, deterministic modules, protocols, and algorithms toward a~cognitive system designed to autonomously evolve and meet the evolving network conditions and application demands.

Table~\ref{tab:flexngia_comparison} provides a comparative overview highlighting key differences between existing literature, FlexNGIA~1.0, and~the~advanced capabilities introduced in~FlexNGIA~2.0.

%---------------------------------------------------------------------------------
\subsection{LLM-based AI Agent: Building Blocks \& Design Principles} %Anatomy of an 
%---------------------------------------------------------------------------------
LLMs have become powerful systems for generating and~processing human-like text, driving advances in natural language generation, code synthesis, and~beyond~\cite{vaswani2017attention,devlin2019bert}. However, LLMs lack the capacity to interact with the external world, perform actions, or~autonomously select and use tools to acquire new information.  Bridging this gap requires agentic behavior~\cite{wang2023survey}, which integrates environmental perception, reasoning, memory, and tool-driven action to~enable adaptive, goal-directed operation. Building on~this~concept, \textit{LLM-based agents} are designed to extend LLMs with these agentic capabilities, thereby enabling autonomous, context-aware decision-making and~interaction with the environment.

\begin{figure}[hb]
    \centering
    \includegraphics[width=1.0\columnwidth]{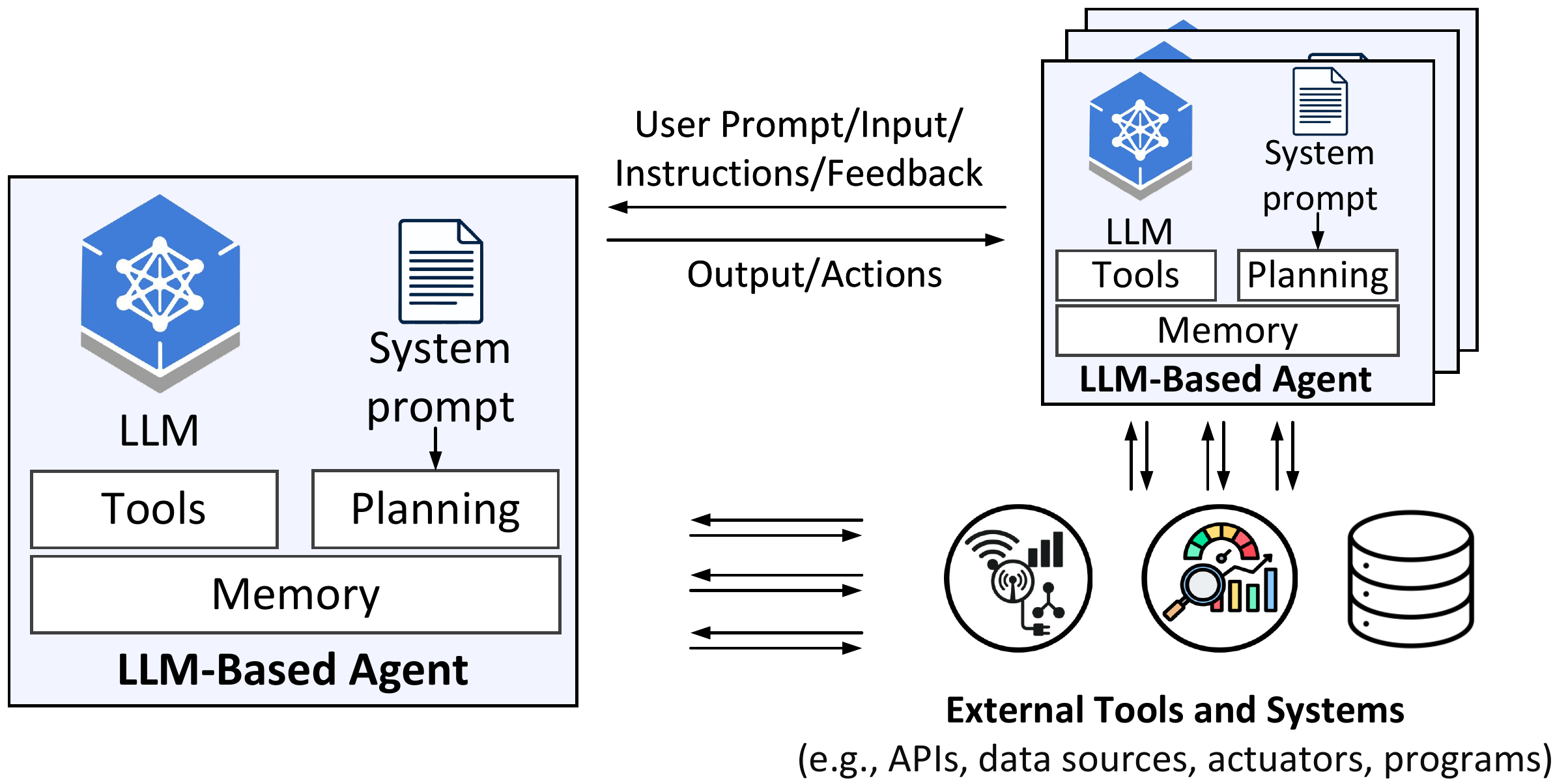}
    \caption{Building Blocks of an LLM-Based Agent and~Its~Interaction with External Tools.}
    \label{fig:Agent-Building_Blocks}
\end{figure}

As shown in~Fig.~\ref{fig:Agent-Building_Blocks}, at the heart of an LLM-based agent lie the following fundamental building blocks~\cite{superannotateLLMagents2025}: %that together empower its capabilities:
\begin{itemize}
    \item \textbf{Brain:}  This component consists of an LLM that serves as the central "brain" responsible for interpreting user requests thanks to its ability to process and understand natural language input, reasoning about tasks, and~generating appropriate responses.
    
    \item \textbf{Memory:} A persistent store of past actions, interactions, and context to inform future reasoning, planning, and~decision-making.
    
    \item \textbf{Planning:} This component is responsible for breaking down complex tasks into manageable subtasks, determining a logical step-by-step execution plan, evaluating options, and performing reflective reasoning. The planning module generates these detailed plans by prompting an LLM, either the same one used by the brain or a separate LLM~instance. 
    %Therefore, it does not necessarily operate as a separate, standalone model.
    %The planning module typically leverages an LLM, either the same one used by the brain or a separate instance, by~prompting~it to~generate detailed plans. Therefore, it does not necessarily operate as a separate, standalone LLM.
    %(i.e., , sometimes with specialized prompts or orchestrated logic, rather than being a fully separate model. 
    %Effective planning often relies on prompt engineering to guide the LLM toward optimal results. Notable approaches include Chain of Thought (CoT), which encourages step-by-step reasoning \cite{wei2022chain}. Tree of Thought (ToT), which is an extension of Chain of Thought prompting that explores coherent intermediate “thought” steps to solve problems~\cite{yao2023tree}, and hierarchical or decision-tree planning, which evaluates possible options before selecting a final plan,
    
    \item \textbf{Tools:} This component serves as~an~interface that allows the agent to interact with external tools and~APIs, such as databases, network measurement utilities, resource management systems, enabling it to gather information, execute actions, and complete tasks \cite{schick2023toolformer}.
\end{itemize}

Among the above components, the planning module is crucial as it establishes the strategy and the steps that direct the agent toward achieving its intended objective. A key part of this planning involves prompt engineering: designing and refining the instructions, known as~\textit{prompts}, provided to~the~agent. In the following, we present a more detailed on~this~critical aspect of~LLM-based AI model design.

\vspace{0.5em}\noindent\textbf{Prompt engineering:} \textit{A prompt} is a textual instruction or input provided to the agent, while \textit{prompt engineering} refers to the process of designing, structuring, and refining that prompt to optimize the agent’s reasoning and output. Well-crafted prompts are critical, as they directly influence the agent’s performance, the quality and relevance of its results, and the accuracy with which it achieves the intended objectives.

The prompt given to an LLM-based agent may include, but is not limited to: the agent’s role (to provide context), a~set of~goals, a planning formulation technique, the~desired output format, and the specific task or~instructions. \textit{A~planning formulation technique} refers to the style or~structure of prompts used to guide an LLM in breaking down, organizing, and creating a coherent plan to achieve a~goal. Notable techniques include Chain of Thought, which guides the agent to reason through problems in logical, sequential steps~\cite{wei2022chain}; Tree of Thought (ToT), an extension of Chain of Thought prompting that explores multiple coherent intermediate “thought” steps to solve problems~\cite{yao2023tree}; Hierarchical planning focuses on decomposing tasks into subtasks arranged in~a~layered hierarchy, enabling abstraction and refinement~\cite{kojima2023LLMszeroshot}, while decision-tree planning models decisions and possible outcomes as a tree of decision points, evaluating different paths to select the optimal plan~\cite{yao2023tree}.

%--------------------------------------------------------------------------------
\begin{figure*} [!t]
	\centering
		\includegraphics[width=1.0\textwidth]{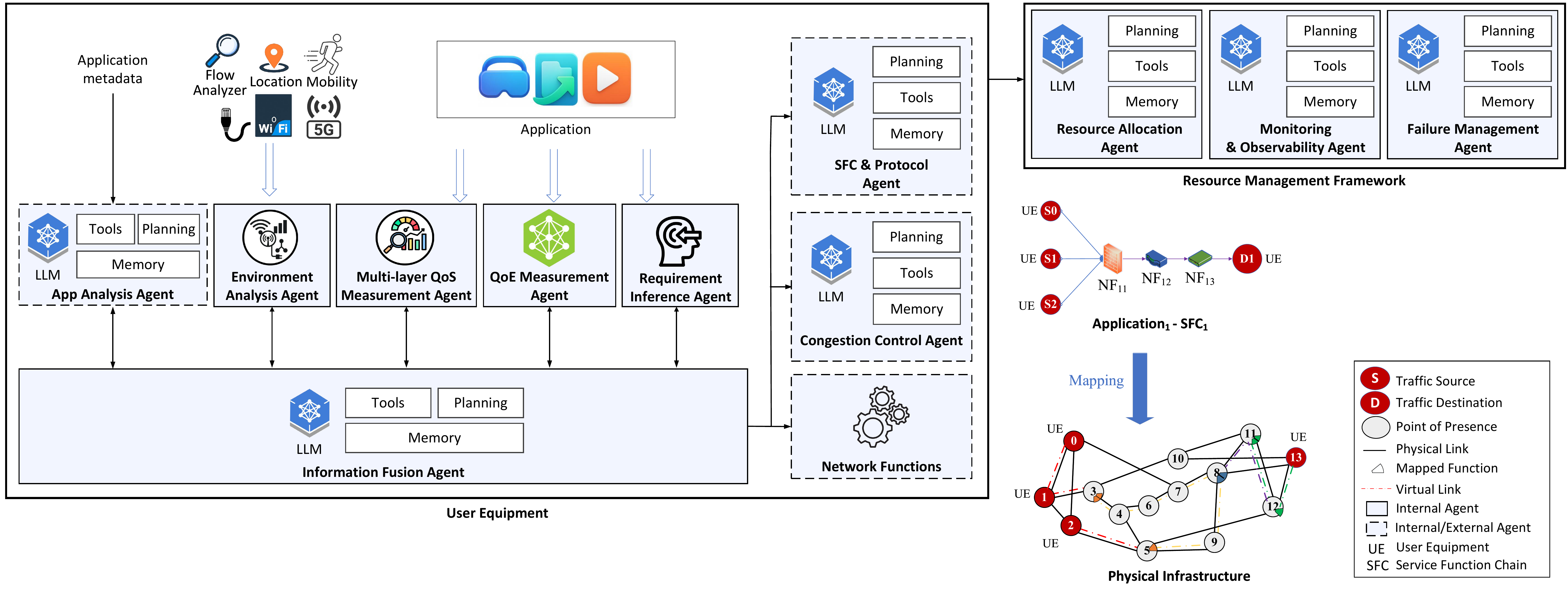}
	\caption{FlexNGIA 2.0 - Agentic AI-based Management Framework.}
	\label{fig:AI-CC-Framework}
\end{figure*}
%--------------------------------------------------------------------------------

AI agent prompts can be categorized into two primary types:

\begin{itemize}
    \item \textit{System Prompt:}  It establishes the agent’s fundamental identity, role, behavior and operational constraints.  It~typically remains consistent across interactions and ensures that the AI maintains a coherent persona, tone, and style. It is defined by the agent’s developers as~part of~its~initial design and configuration. It can also include instructions for using tools, storing memory, and applying a planning formulation technique to enforce step-by-step reasoning toward accomplishing a goal or task.
    
    \item \textit{User Prompt/input:} Contains task-specific instructions, queries, or inputs provided by the end users (if the agent is interacting with humans), other agents or external systems. It varies dynamically depending on the current task or interaction and directs the agent to perform specific actions or generate particular outputs.
\end{itemize}

For each request or interaction, the LLM receives the system prompt, user prompt/input, and pertinent information retrieved from the memory module (e.g.,~past interactions, decisions), all of which are combined and processed jointly to produce the model’s response. As~such, the model always recalls previous interactions and stored overarching context (via~the~memory), its role and constraints (via the system prompt) while addressing the current input (via~the~user prompt/input). In~this~paper, the word “prompt” will refer to the system prompt, which includes placeholders or~references to incorporate inputs. Fig.~\ref{fig:CCA-prompt} is~an~example of a well-structured system prompt that clearly states the goal, outlines the step-by-step reasoning process to accomplish it (using chain-of-thought reasoning), and contains placeholders (indicated by red text within curly brackets) to retrieve inputs.

It is also worth noting that agents can communicate and coordinate with each other using protocols like~Agent2Agent~(A2A)~\cite{google2025a2a} and~Agent Communication Protocol~(ACP)~\cite{ibm2024acp}. Model~Context~Protocol~(MCP)~\cite{anthropic2024mcp} could be used to~provide agents with standardized access to~external tools and data sources.

%As shown in Figure \ref{fig:Agent-Building_Blocks}, these components work together within the LLM-based agent architecture: the LLM (brain) operates alongside planning, tools, and memory, interacting with the environment and external systems through inputs, outputs, and feedback loops.

%\noindent \textbf{Environment (Interacting Entities):} Encompasses all external components that the agent perceives, communicates with, and affects. This includes: \textbf{(i) Other LLM-Based Agents} that may collaborate or compete via structured message exchanges; \textbf{(ii) a Knowledge-Driven Analysis System} responsible for analyzing incoming data streams, identifying patterns, and generating context-aware insights—that can be fed back into the agent’s reasoning loop; and \textbf{(iii) External Systems} such as APIs, data sources, and actuators that interface with the physical or digital world. The Knowledge-Driven Analysis System not only supplies the agent with processed knowledge but can also refine its own analytical models based on agent outputs, creating a continuous feedback cycle. Together, these entities form a dynamic interaction substrate—sending observations, feedback, or tasks to the agent, and receiving actions or updates in return.

%---------------------------------------------------------------------------------
\subsection{FlexNGIA 2.0 - Agentic AI Architecture for the Future Internet}
%---------------------------------------------------------------------------------

FlexNGIA 2.0 introduces an~Agentic~AI-Driven  network architecture that embeds autonomous, intelligent LLM-based agents capable of reasoning, planning, and adapting in~real time. The framework leverages LLM-based agents to~design, manage, and optimize network functions, protocols, and~services dynamically, addressing changing network conditions and application requirements. 
The main components of the proposed architecture are illustrated in Fig.~\ref{fig:AI-CC-Framework} and described as follows:

\textbf{$\bullet$ App Analysis Agent:} Operating primarily in offline mode, this agent analyzes the~application and~determines their functional requirements, user satisfaction criteria, and~performance expectations, based on~the~data available about the application (e.g.,~documentation, user feedback). 

Based on this analysis, the agent has two key~tasks. The~first is to~identifies the multi-layer Quality of Service (QoS) metrics that are most relevant to the specific application. These metrics span several layers of the protocol stack. At the network layer, this includes parameters such as latency tolerance, jitter, bandwidth, and~packet loss. At~the~transport layer, relevant indicators include round-trip time, retransmission rates, congestion window behavior, and timeouts. At the application layer, metrics may cover response time, throughput, error rates, and user engagement measures. The agent also identifies the relevant Quality of Experience (QoE) metrics that should be monitored like~responsiveness, video/audio quality, or user engagement indicators. 
The~second task of~this~agent is to~build a~Target QoS Profile that~outlines the target values of~the~key performance metrics like the maximum allowable latency, jitter, required bandwidth, and tolerable packet loss.
    %, which guide the operation of other agents and the overall system’s operation.

Once the App Analysis Agent identifies the key relevant metrics and the Target QoS Profile, they are shared with other agents to guide their operation.
%. This information is then proactively shared with other agents to guide their operations. For instance, the Multi-layer QoS Measurement Agent and the QoE Measurement Agent use it to determine which metrics to monitor at runtime. Meanwhile, the SFC \& Protocol Agent and~the~CC Agent rely on the Target QoS Profile to inform their decisions, aiming to maintain system behavior within the specified target values.
    
%---------------------------------------------
\textbf{$\bullet$ Environment Analysis Agent:} 
    %This agent is responsible for monitoring and analyzing the software stack (e.g., OS version, running processes, applications), hardware components and technologies (e.g., CPU load, memory usage, battery level), and environmental context (e.g., location, network conditions, signal strength and quality, location, mobility patterns) of the User Equipment (UE). It collects real-time telemetry data to support adaptive decision-making, resource optimization, and context-aware service delivery.
This agent is tasked with monitoring and analyzing the software stack, hardware components, runtime state, and contextual environmental parameters of the User Equipment (UE).
The software stack includes information about the deployed OS, protocols, CC~scheme, and running applications or services. The~hardware components refer to UE's physical resources like the CPU, GPU, memory, battery, and~onboard sensors. The runtime state reflects the UE’s current operational and networking conditions, including resource utilization, power mode, number of active processes, and  network and flow~statistics. %such as the number of concurrent traffic flows, protocol usage, port activity, and real-time bandwidth consumption. 
Finally, the contextual environmental parameters encompass external factors such as network signal strength, user mobility, GPS location, and the presence of nearby wireless devices or interference. Together, these dimensions provide a~comprehensive, real-time view of~the~UE’s status, serving as a critical input for other agents to~design protocols and network functions that are dynamically adapted to the current UE's environment. 
    %In addition, this information can be relayed to network entities such as the Radio Access Network (RAN) and User Plane Function (UPF) to optimize resource scheduling, traffic steering, and slice selection based on the real-time behavior and needs of the UE. By enabling continuous feedback and closed-loop adaptation, this agent plays a foundational role in achieving context-aware, resilient, and user-centric networking in next-generation mobile systems.
    
    %    This includes technologies (e.g., 4G/5G, wireless/wired), network conditions (signal quality), mobility, as well other factors related to the end-user device like the total resource capacity and usage (e.g., cpu,memory, bandiwdth), component characteristics (processor and others), the number of flows competing for bandwidth (including TCP or UDP), applications priorities compared to each other.

   % \item \textbf{Media Adapter:} Extracts visual content from the running application and preprocesses video frames for perceptual and semantic quality assessment.

\textbf{$\bullet$ Multi-layer QoS Measurement Agent:} This Agent is responsible for continuously monitoring the performance of the application by collecting and analyzing QoS metrics across multiple protocol layers.
Rather than operating with a fixed monitoring strategy, the agent dynamically adapts its monitoring scope and methods based on the Target QoS Profile received from the App Analysis Agent. This allows the QoS Measurement Agent to focus on the most impactful indicators for each flow, reducing unnecessary overhead.
As an autonomous agent, this agent is capable of deploying and~activating the appropriate measurement tools at runtime. %This includes a diverse set of~measurement tools operating across multiple layers. 

\textbf{$\bullet$ QoE Measurement Agent:} The QoE Measurement Agent evaluates QoE by~measuring user-perceived quality, which often reflects factors beyond traditional network metrics. It collects the specific QoE metrics defined and provided by the App Analysis Agent, enabling targeted and relevant monitoring tailored to each application. Leveraging advanced AI-based tools, such as Vision-Language Model (VLM), Image Quality Assessment (IQA) and Video Quality Assessment (VQA) models, the agent predicts perceived quality automatically and in real time, rather than relying solely on delayed or sparse human feedback. %This AI-driven approach allows the system to assess not only the visual and semantic quality of media content but also to adapt quickly to changing conditions. Additionally, the agent integrates direct user feedback when available, combining objective AI predictions with subjective input to provide a comprehensive understanding of QoE in dynamic environments.
    
%-------------------------------------------------------
\textbf{$\bullet$ Requirement Inference Agent:} This module plays a~crucial role in FlexNGIA 2.0 architecture by continuously analyzing at runtime the application, user interactions, and~contextual data to dynamically infer and update the precise QoS and performance requirements for each active flow. By~leveraging real-time monitoring and intelligent reasoning, this agent translates raw measurements and semantic context into actionable requirement profiles that reflect the current needs of the application and its users. These inferred requirements are then shared with other agents, enabling them to adapt and optimize their operation to~meet the~evolving application requirements.

%-------------------------------------------------------
\textbf{$\bullet$ Information Fusion Agent:} The Information Fusion Agent (IFA) plays a central role in aggregating and synthesizing the data collected by all the aforementioned agents to create a~comprehensive report, called an~\textit{IFA~report}~(~Fig.~\ref{fig:AI-CC-Framework}). It~contains a coherent, comprehensive view of system performance, sate, and user experience. It~includes key measurements, along with contextual correlations and analysis, producing a semantically meaningful dataset that other agents can use to guide their operations. 
%It collects metrics from various measurement agents, like the QoS and QoE measurement agents, along with real-time network feedback and application-level information, including session state and user interaction data. 

%-------------------------------------------------------
\textbf{$\bullet$ SFC \& Protocol Agent} Given a particular application and its requirements (as described in the IFA Report), this agent is responsible for designing and developing the~SFC, its composing network functions and its associated communication protocols. More precisely, it can build the~chain leveraging existing network functions or~by~creating and implementing new ones tailored to the specific requirements of the application. Additionally, it~could select one of~the~legacy protocols (e.g.,~TCP, QUIC) for the SFC, or~decide to design and implement a~new~communication protocols adapted to~the~application, specifying how packets are processed and forwarded as they traverse the sequence of network functions, and considering different factors like the~SFC's network functions and~the~type of the application (e.g.,~point-to-point vs.~multipoint). This ensures that the~SFC is perfectly tailored to~each application.

%-------------------------------------------------------
\textbf{$\bullet$ Congestion Control Agent:} The CC Agent dynamically adjusts the congestion control behavior to align with the Target QoS Profile provided by the App Analysis Agent. This~agent uses the information provided in~the~IFA report like the current environment state, QoS and QoE metrics, specific application requirements, and historical decision data. Based on these data, it can decide to tune the parameters of the current CC scheme, switch to an alternative CC~scheme, or~even design and deploy a new CC scheme tailored to~the~current network conditions and~application requirements.

%-------------------------------------------------------
\textbf{$\bullet$ Resource Allocation Agent:} The Resource Allocation (RA) Agent is responsible for dynamically managing and optimizing the allocation strategy of computational, bandwidth, storage, and energy resources across the physical infrastructure. 
Its primary goal is to ensure that the resource allocation algorithm satisfies SFC-specific requirements and QoS constraints while also meeting operational, sustainability, and economic objectives. These objectives may include improving energy efficiency, maximizing the utilization of green and renewable resources, increasing the number of successfully mapped SFCs, minimizing operational costs, and maximizing revenues. 
Furthermore, the Resource Allocation Agent tightly coordinates with the~SFC~\&~Protocol~Agent, ensuring that protocol selection and service chain deployment remain feasible given the~current resource constraints. 
Thus, resource allocation decisions are not made in isolation but are tightly coupled with protocol selection and SFC composition.

\textbf{$\bullet$ Monitoring \& Observability Agent:} The~Monitoring \&~Observability Agent is responsible for continuously collecting, analyzing, and interpreting telemetry and performance data across the network and service function chains. Its role is to maintain real-time awareness of network health, resource utilization, and service quality, enabling proactive detection of anomalies, congestion, or~potential failures. It~provides critical information to other agents to~ensure their smooth operation and~to~guide their decisions.
%The agent leverages advanced analytics, AI/ML techniques, and correlation mechanisms to transform raw data into actionable insights, supporting both reactive and predictive management. 
%feedback to~the~SFC~\&~Protocol Agent and Resource Allocation Agent, ensuring that service chains, protocols, and resource allocations can be dynamically adapted based on accurate and up-to-date information. %By~maintaining a~comprehensive view of the system's state, the~Monitoring \&~Observability Agent underpins resilience, reliability, and~intelligent self-optimization across~FlexNGIA~2.0.

\textbf{$\bullet$ Failure Management Agent:} The Failure Management Agent is responsible for detecting, diagnosing, and~mitigating failures across the service function chains, network function and~the~physical infrastructure. 
%It continuously monitors system components, identifies anomalies, and correlates events to determine the root causes of disruptions. 
%Leveraging predictive analytics, AI/ML~reasoning, and historical failure patterns, the agent can anticipate potential issues before they impact service quality. Upon detection of a fault, it can trigger automated recovery actions such as rerouting traffic, reallocating resources, or dynamically reconfiguring SFCs and protocols to maintain service continuity. Crucially, the Failure Management Agent coordinates closely with the Monitoring \&~Observability Agent, Resource Allocation Agent, and~SFC~\&~Protocol Agent, ensuring that detection, resource adjustment, and recovery actions are aligned across the system. This collaborative operation minimizes downtime, optimizes system resilience, and upholds end-to-end performance objectives. By integrating real-time detection, predictive analysis, and cross-agent coordination for autonomous remediation, the Failure Management Agent is a cornerstone of FlexNGIA 2.0’s self-healing and highly reliable network architecture.

%particularly in scenarios where resource prices vary across regions of the infrastructure.
% Operational objectives → maximize the number of mapped SFCs, minimize costs, optimize utilization.
% Performance objectives → guarantee QoS/QoE, minimize latency, maximize throughput.
% Sustainability objectives → energy efficiency, maximize use of green/renewable resources.é

\begin{figure}[!b]
    \centering
    \includegraphics[width=1.0\columnwidth]{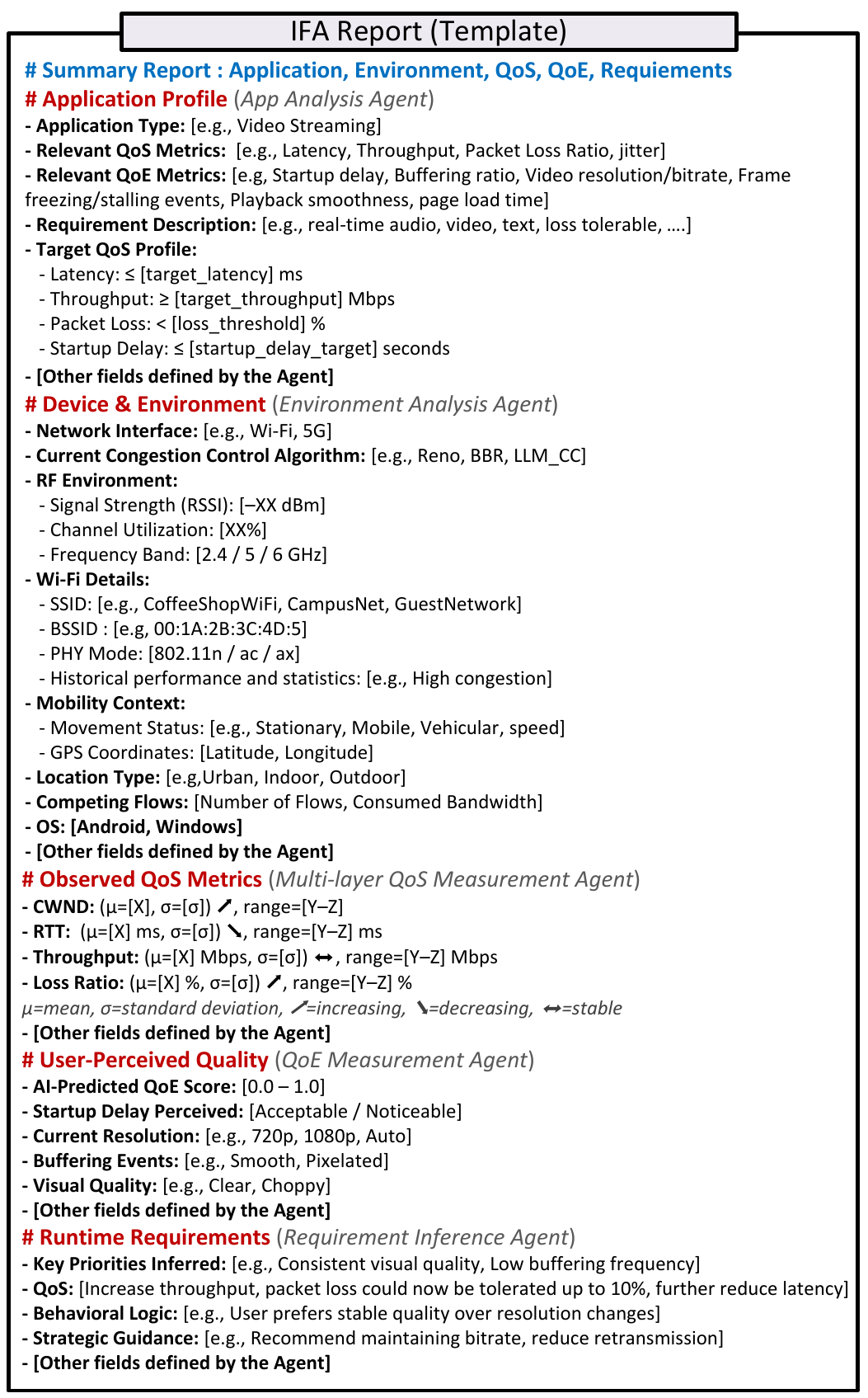}
    \caption{Template of the IFA report generated by~the~Information Fusion Agent.}
    \label{fig:IFA-Report}
\end{figure}
%==============================================================================
\section{FlexNGIA 2.0 AI Agents: Design, Proof of Concept, and Preliminary Experimental Results}\label{sec:AgenticAIinAction}
%=============================================================================
This Section presents the core design principles of~the~FlexNGIA 2.0 AI agents, each supported by initial proof-of-concept experiments. These experiments provide preliminary insights into the efficiency of the AI Agents, laying the groundwork for further optimization and validation.
%“Preliminary” signals that the results are early-stage. Hence, there should be further experiments, optimizations, and validations in the future.
%A Proof of Concept (POC) implementation involves creating a simplified, preliminary version of a product or idea to demonstrate its feasibility and potential value.A POC is used to test the core idea, demonstrate its viability to stakeholders, and identify potential challenges early on. 

%------------------------------------------------------
\subsection{Information Fusion Agent}
%------------------------------------------------------
The Information Fusion Agent acts as~a~centralized intelligence synthesizer. It is powered by~an~LLM that~aggregates heterogeneous telemetry data collected from other agents (Fig.~\ref{fig:AI-CC-Framework}) to~generate a~single, unified report (called \textit{IFA~Report}). This report provides a coherent snapshot  of the system that serves as input for downstream agents.

To ensure a~consistent structure for the~IFA report, we~designed a report template (Fig.~\ref{fig:IFA-Report}) that provides to~the~IFA's LLM the~key~sections and metrics to be included in the report. Of course, the LLM  may identify and add other metrics that it deems relevant. As~shown in the figure, the~template is organized into sections. The Application Profile section describes the~application's type (e.g.,~eXtended Reality, file transfer), QoS and QoE metrics, performance requirements, and the~target QoS. The~Device and Environment section captures details about the environment where the application is running, such as network technology, radio frequency conditions, mobility status, device type, operating system, and competing flows. The~Observed QoS Metrics section include key QoS measurements. The User-Perceived Quality section reflects inferred or measured QoE. Finally, the~Runtime Requirements section outlines current requirements, priorities, and strategic guidance for~runtime adaptation.

This structured report serves as the primary input for~other agents, providing essential context on the state of~the~performance, network, application, and~device.

%As illustrated in the template (Fig.~\ref{fig:IFA-Report}), its consistent format and semantic labels enable agents to interpret both metrics and context effectively, supporting real-time, adaptive congestion control decisions within FlexNGIA.

%The Information Fusion Agent (IFA) plays a central role in orchestrating decisions in FlexNGIA by aggregating inputs from various sources, including the Multi-layer QoS Measurement Agent. Among its most critical inputs to the Congestion Control Agent are live measurements such as congestion window (CWND), round-trip time (RTT), throughput, and packet loss.

%It provides a synthesized view of the system state across network, application, and device dimensions.

%\vspace{0.5em}

%------------------------------------------------------
\subsection{AI-Driven SFC \& Protocol Agent}\label{ssec:AI-DrivenSFCAgent}
%------------------------------------------------------
\begin{figure}[!b]
    \centering
    \includegraphics[width=0.9\columnwidth]{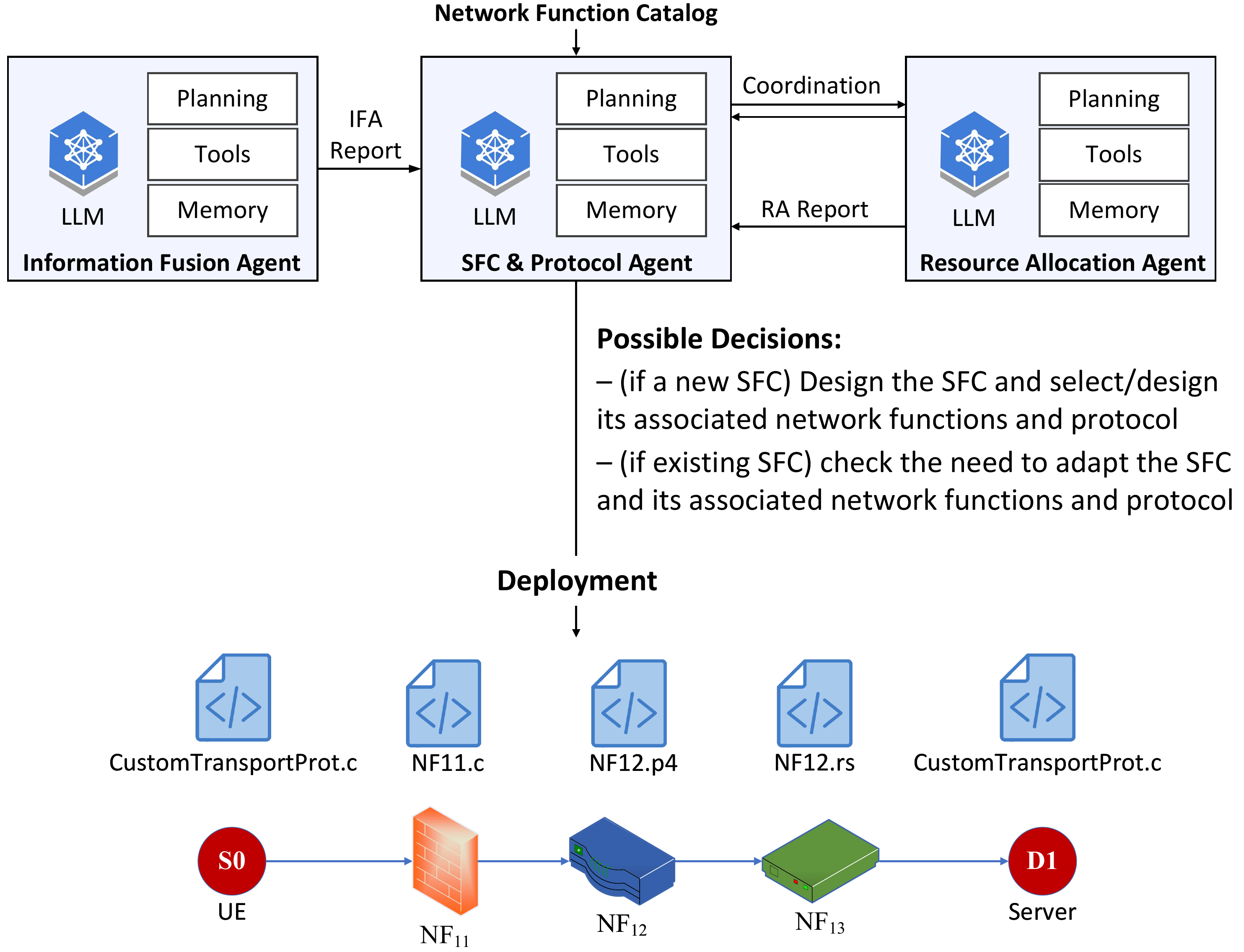}
    \caption{Operation of the SFC \& Protocol Agent.}
    \label{fig:SFCAgent-Architecture}
\end{figure}

The~SFC~\&~Protocol Agent is in~charge of~the~design, deployment, and runtime adjustment of~an~end-to-end~SFC to~support the~application including its composing NFs and communication protocol. As~shown~in~Fig.~\ref{fig:SFCAgent-Architecture}, this agent can dynamically design the~SFC associated with a~given application. It~determines which network functions should be included in the chain, either by~selecting from the Network Function Catalog, a~repository of pre-implemented and ready-to-use functions, or~by~designing and implementing new functions as needed. In parallel, the agent selects the most suitable communication protocol from legacy protocols (e.g.,~UDP, TCP, QUIC) or~designs a custom one, then fine-tunes its~parameters according to current network conditions and application requirements. 

In the following, we present the configuration and system prompt of~the~LLM associated with~this~agent, along with experimental results that demonstrate its operation.

%---------------------------------------------------------
\vspace{0.5em}\noindent\textbf{$\bullet$ Prompt Structure for the the~SFC~\&~Protocol~Agent:}
The~SFC~\&~Protocol~Agent is guided by a structured system prompt that instructs it to design the~optimal SFC and communication protocol for a given application, using inputs such as the IFA report, the~NF~catalog, and the RA Report containing potential paths to host the SFC's network functions. 

\begin{figure}[!b]
    \centering
    \includegraphics[width=0.80\columnwidth]{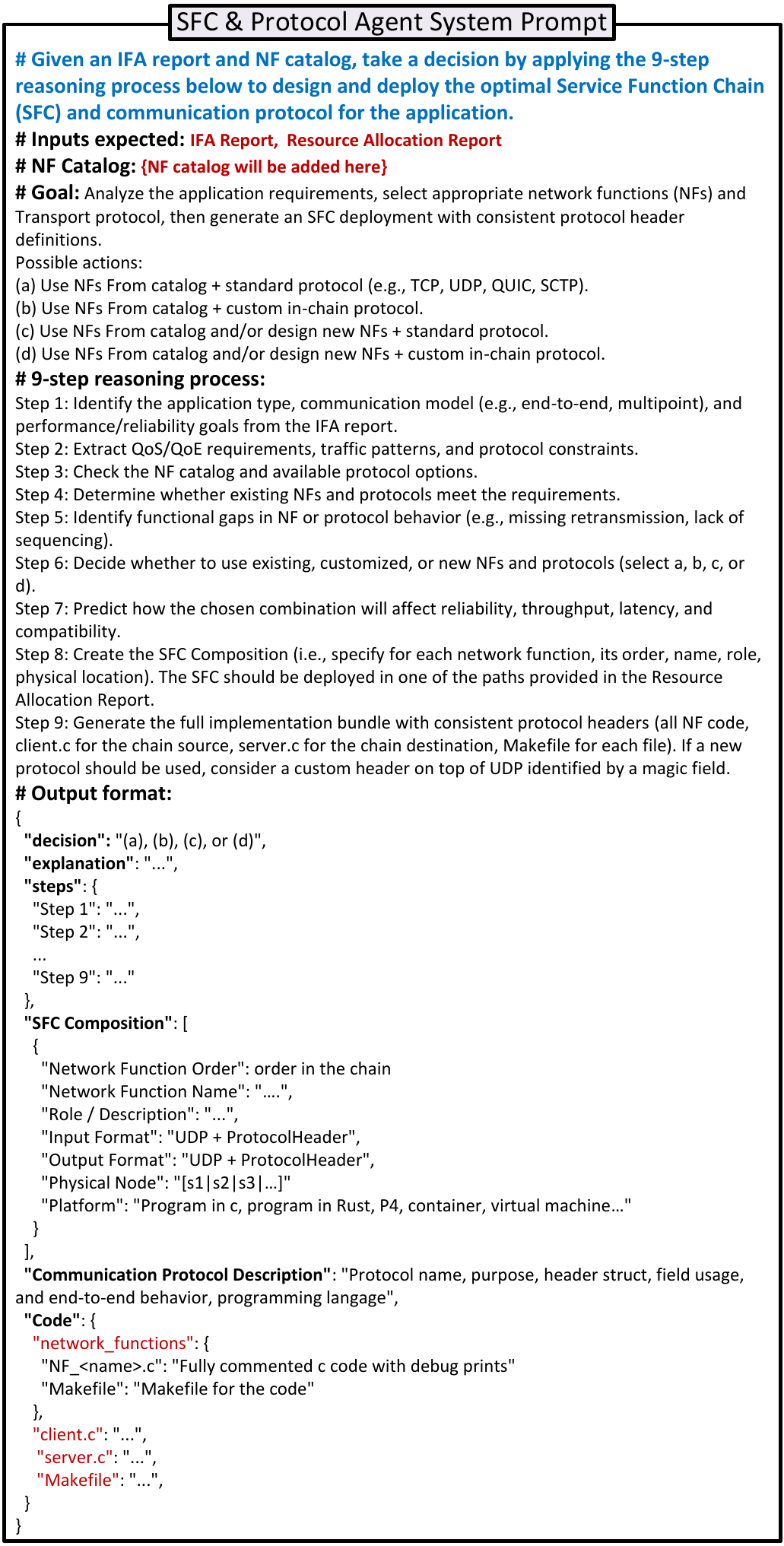}
    \caption{System prompt of the~SFC~\&~Protocol~Agent.}
    \label{fig:SFC-prompt-template}
\end{figure}

% The prompt defines the agent’s goal, to analyze application requirements, select appropriate network functions and protocols, and generate an SFC deployment with consistent protocol headers, and enumerates possible actions, including using existing NFs with standard or custom protocols, or creating new NFs with standard or custom protocols. To achieve this, the agent applies a 9-step reasoning process, starting with identifying the application type, communication model, and performance goals, extracting QoS/QoE requirements, and checking available NFs and protocols. 

As shown~in~Fig.~\ref{fig:SFC-prompt-template}, the prompt defines the agent’s goal: to analyze application requirements, select or~design and~implement appropriate NFs and transport protocols, and a~deployment-ready SFC. It~also lists actions, from reusing catalog NFs with standard protocols to combining catalog or~new NFs with standard or~custom protocols. 

The agent follows a 9-step reasoning process~(Fig.~\ref{fig:SFC-prompt-template}) that starts with analyzing the application and its requirements, then checks whether available NFs and protocols fit to these requirements. It then decides to use them or design and~implement new ones, and predicts the impact of~the~chosen solution on performance metrics. Finally, it~creates the~SFC and generates a complete implementation code bundle including the NFs, client/server programs, and Makefiles.

\begin{figure}[!t]
    \centering
    \includegraphics[width=0.8\columnwidth]{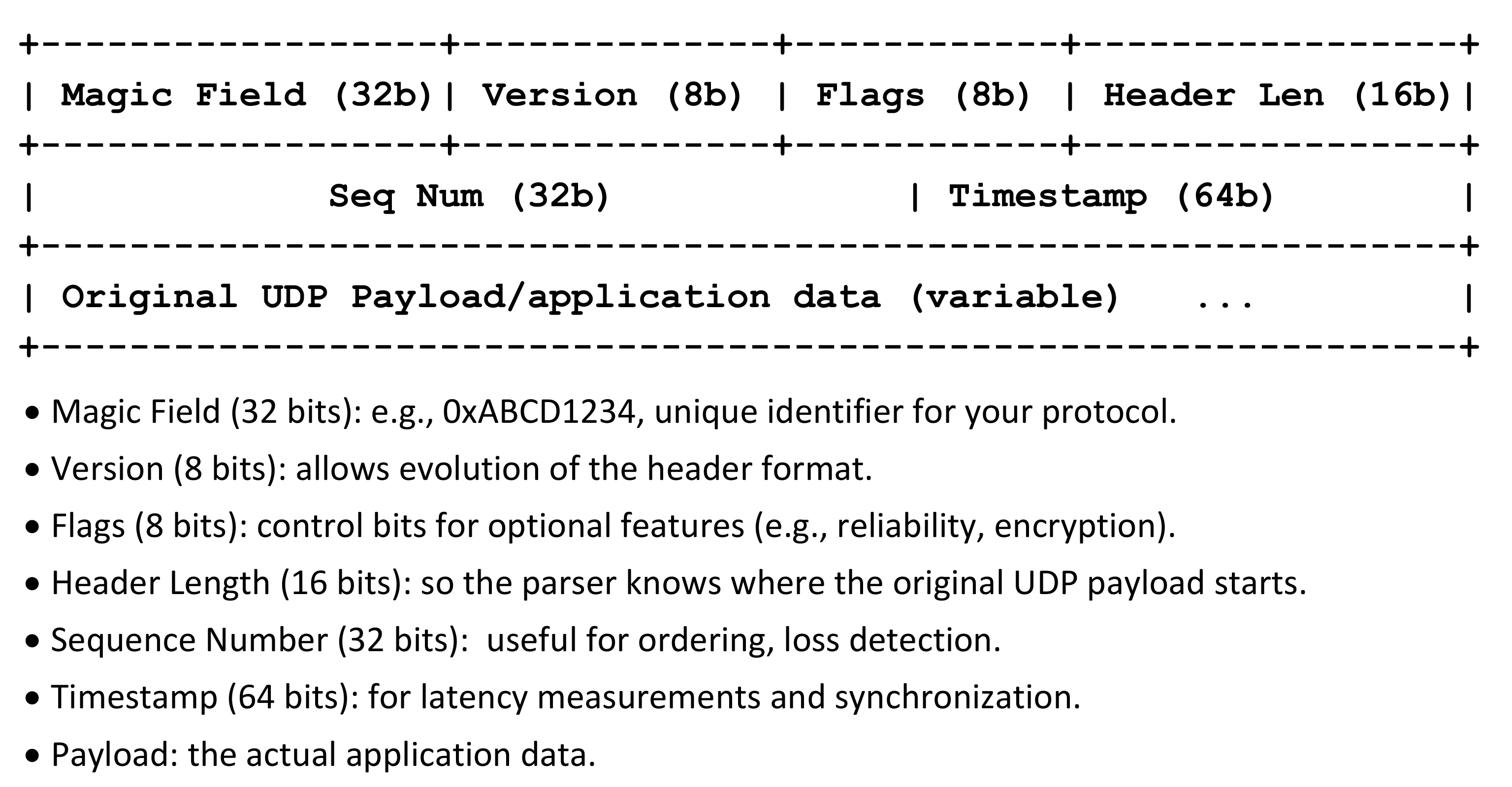}
    \caption{Example of a custom header generated by~the~SFC~\&~Protocol~Agent }
    \label{fig:SFC-CustomHeader}
\end{figure}

\noindent\textbf{$\bullet$ Protocol Generation:} When the~SFC~\&~Protocol~Agent decides to generate a new custom transport protocol rather than using legacy protocols (e.g.,~UDP, TCP, QUIC, SCTP),  we opted to create~it on~top of~UDP to~ensure compatibility with today's networks. The custom protocol is realized by embedding a custom header within the UDP payload. This is performed by inserting a magic field, i.e.,~a fixed-value sequence (e.g.,~0xABCD1234), at~the~start of the UDP payload as~proposed in~\cite{draft-herbert-udp-magic-numbers-00}. By~checking this~field, intermediate NFs can recognize and parse the custom header (see~Fig.~\ref{fig:SFC-CustomHeader}). If~the~magic field is missing, then the packet is treated as~a~traditional UDP packet with no special header. 
From an implementation standpoint, the new header can be inserted into the UDP payload either by a kernel-level module on~the~UE or~directly by~the~application itself.%, provided the application is aware of the new protocol header.

\begin{figure}[!b]
\centering
\includegraphics[width=0.9\columnwidth]{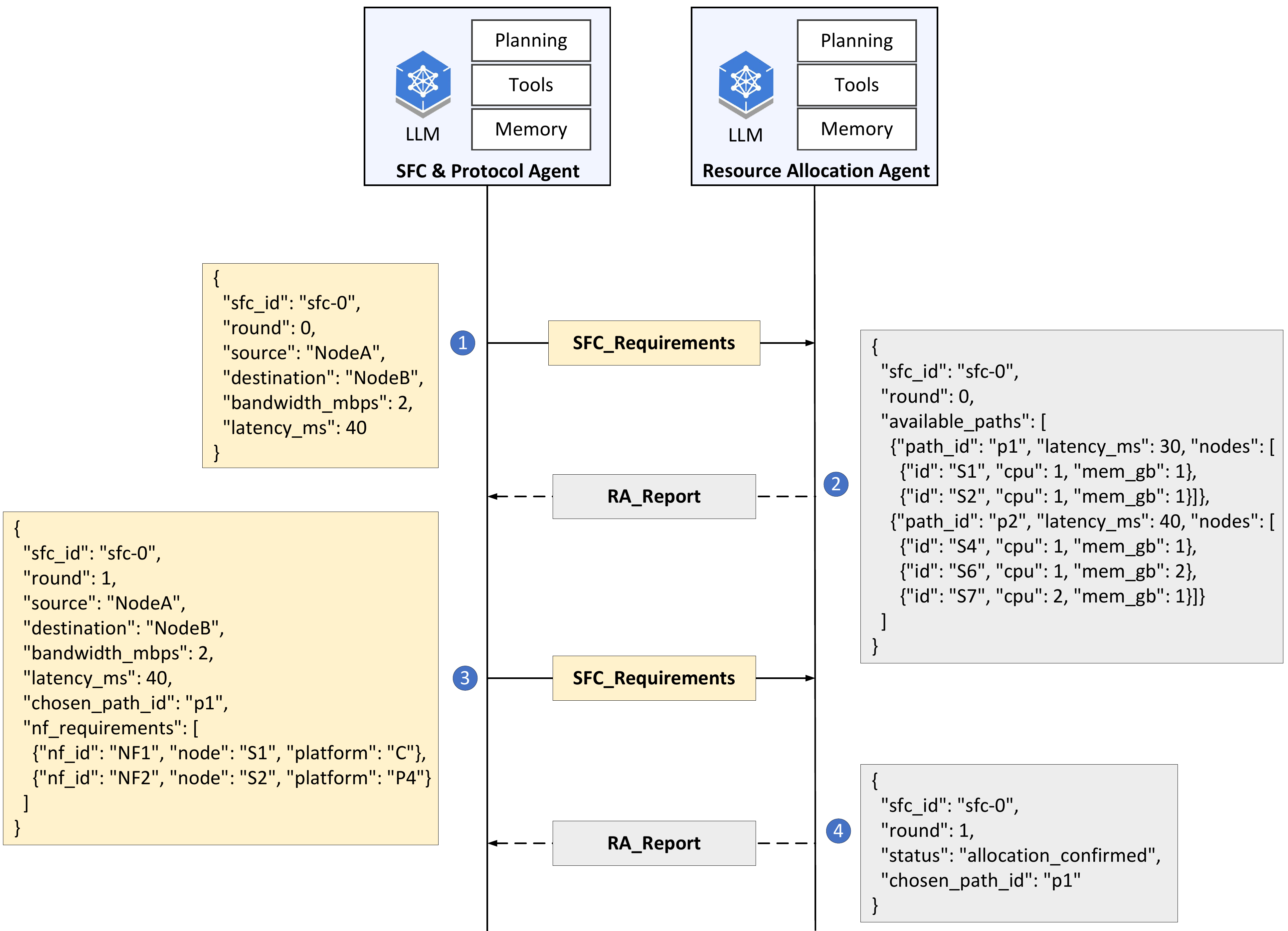}
\caption{Coordination Messages Between the~SFC~\&~Protocol Agent and the~Resource~Allocation Agent.}
\label{fig:SFC-RA-Coordination}
\end{figure}

\begin{figure*}[!t]
    \centering
    % First subfigure
    \begin{subfigure}[b]{0.6\columnwidth}
        \centering
        \includegraphics[width=\columnwidth]{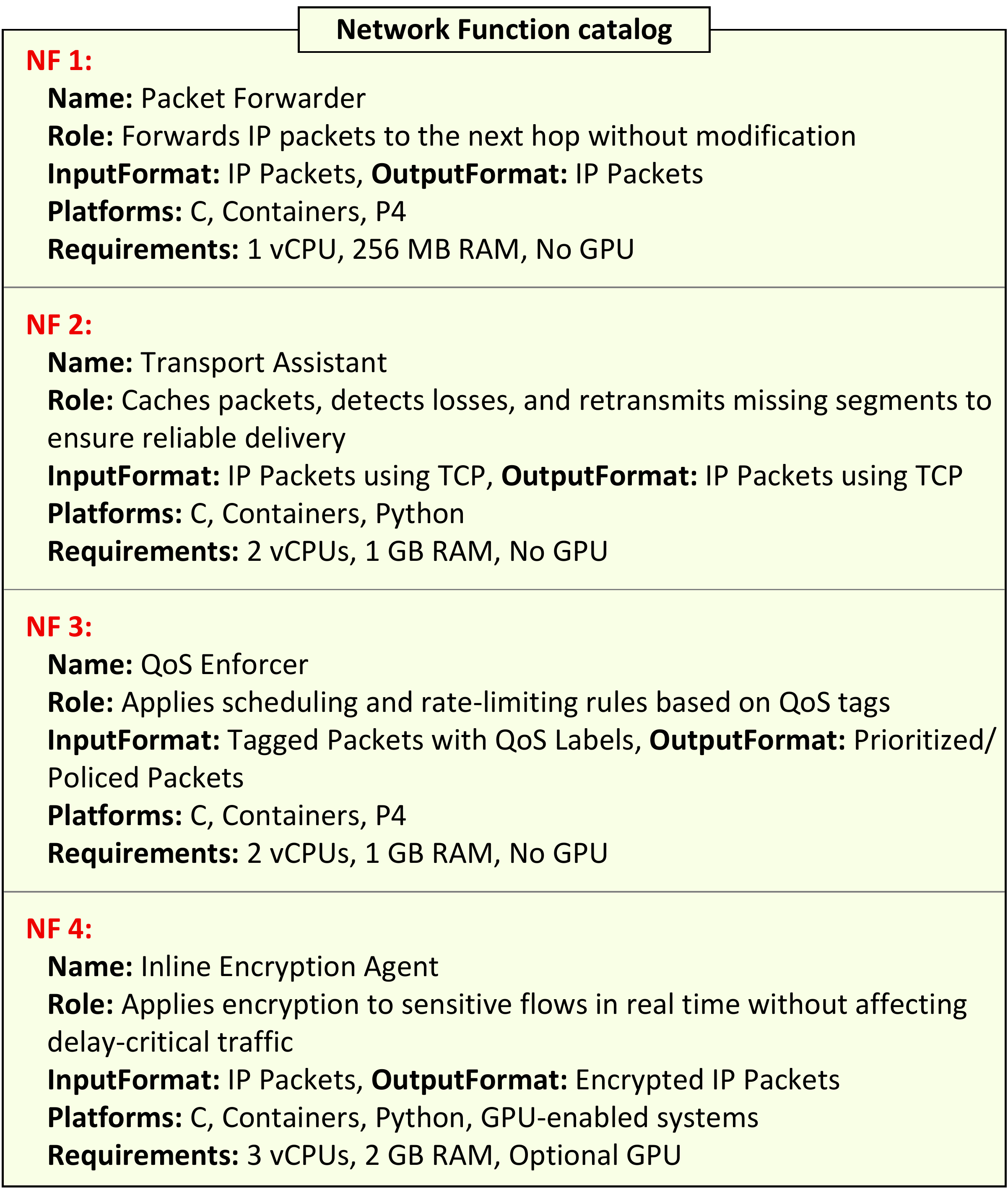}
        \caption{Network Function Catalog.}
        \label{fig:SFC-NFCatalog}
    \end{subfigure}
    \hfill
    % Second subfigure
    \begin{subfigure}[b]{1.4\columnwidth}
        \centering
        \includegraphics[width=\columnwidth]{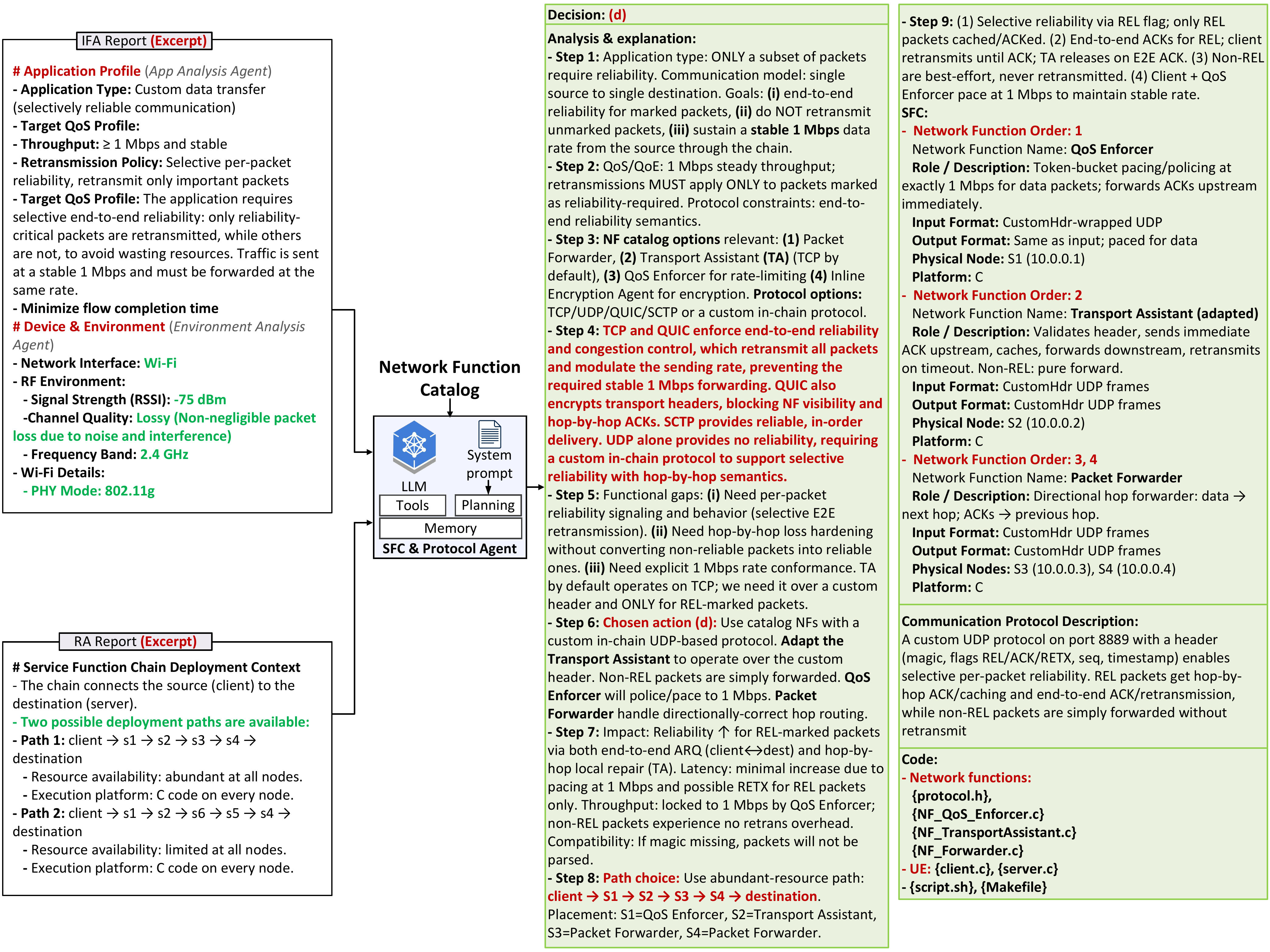}
        \caption{Input and output of the SFC \& Protocol Agent.}
        \label{fig:SFC-Output}
    \end{subfigure}
    \caption{Inputs and Outputs of the SFC \& Protocol Agent in our experiments.}
    \label{fig:SFC-Agent-Action}
\end{figure*}

\noindent\textbf{$\bullet$ Coordination with~the~Resource Allocation Agent:} The~SFC~\&~Protocol Agent works in close coordination with the Resource Allocation Agent to ensure that the SFC composition, network functions, and protocol configuration are aligned with available computational, bandwidth, and energy resources. This tight integration enables joint optimization, where protocol selection, SFC design, and resource allocation reinforce each other, maximizing QoS guarantees.

Fig.~\ref{fig:SFC-RA-Coordination} illustrates an example of a~basic coordination protocol between the two agents. In this example, the~SFC~\&~Protocol agent provides the basic requirements of~the~SFC (e.g.,~source, destination, required latency and bandwidth) to the RA agent, which in turn supplies it with the paths and resources available in~the~physical infrastructure between the~SFC's source and destination. This information is~taken into account by~the the~SFC~\&~Protocol agent to carefully craft an~adapted SFC.
%depends on factors such as path availability, resource capacity, and the platforms and technologies available for hosting the network functions.

%the  that enable joint SFC design, protocol selection, and resource allocation.

%---------------------------------------------------------------------------------------------
\vspace{0.5em}
\noindent\textbf{$\bullet$ SFC \& Protocol Agent in Action -- Experimental Proof of Concept:}  
%---------------------------------------------------------------------------------------------
To demonstrate the operation of~the~SFC~\&~Protocol~Agent, we implemented a~prototype powered by the LLM \texttt{GPT-5 Thinking} to leverage its advanced reasoning capabilities~\cite{openai_gpt5_2025,openai_introducing_gpt5_2025}. 
%
%(e.g., packet re packet
%. The figure provides the catalog used with the four NFs:  
%\textbf{(1) Packet Forwarder} -- Forwards IP packets to the next hop without modification; parses IP packets. \textbf{(2) Transport Assistant} -- Caches packets, detects losses, and retransmits missing segments to ensure reliable delivery; parses IP packets with TCP. \textbf{(3) QoS Enforcer} -- Applies scheduling and rate-limiting rules based on QoS tags. \textbf{(4) Inline Encryption Agent} -- Performs real-time encryption of sensitive flows while avoiding additional delays for latency-critical traffic.
%
%\vspace{0.5em}
%\noindent\textbf{$\bullet$ Experimental Results:}  
Fig.\ref{fig:SFC-Agent-Action} illustrates the inputs to~the~SFC~\&~Protocol~Agent (i.e.,~the~IFA and~RA reports, NF catalog) used in our experiment and the corresponding output. Fig.~\ref{fig:SFC-NFCatalog} shows the~NF catalog, which includes four NFs. Each function is defined by its name, role, input/output packet formats, platform, and resource requirements. The catalog spans from basic to advanced functions, addressing reliability, QoS, and security.

Fig.~\ref{fig:SFC-Output} shows the IFA report that describes the~application's requirements: it~requires selective reliability, i.e., important packets must be retransmitted upon loss, while non-critical packets are delivered on a best-effort basis. In addition, the application demands a stable throughput of at least 1~Mbps and flow completion time should be~minimized. The figure also shows the RA report, provided by the Resource Allocation Agent, which contains information about the paths that could potentially host the application’s SFC. As shown, the report indicates two possible paths, detailing the characteristics of each path, its constituent nodes, and the available resources.

%The LLM produced the output in~under 10~s. 
As~shown in~the~output (Fig.~\ref{fig:SFC-Output}), the LLM agent performed multi-step reasoning, it identified the application's requirements (step~1 and~2), determined the required network functions (step~3), and~decided, with justification for not using legacy protocols (step~4), identified functional gaps (step~5), recognized the need for custom transport protocol and functions (step~6), predicted the performance (step~7), selected the path to embed the SFC, crafted the SFC with~its~composing NFs, and~generated a~complete C code bundle for~all components (steps~8~and~9).

\begin{figure}[h]
\centering
\includegraphics[width=0.9\columnwidth]{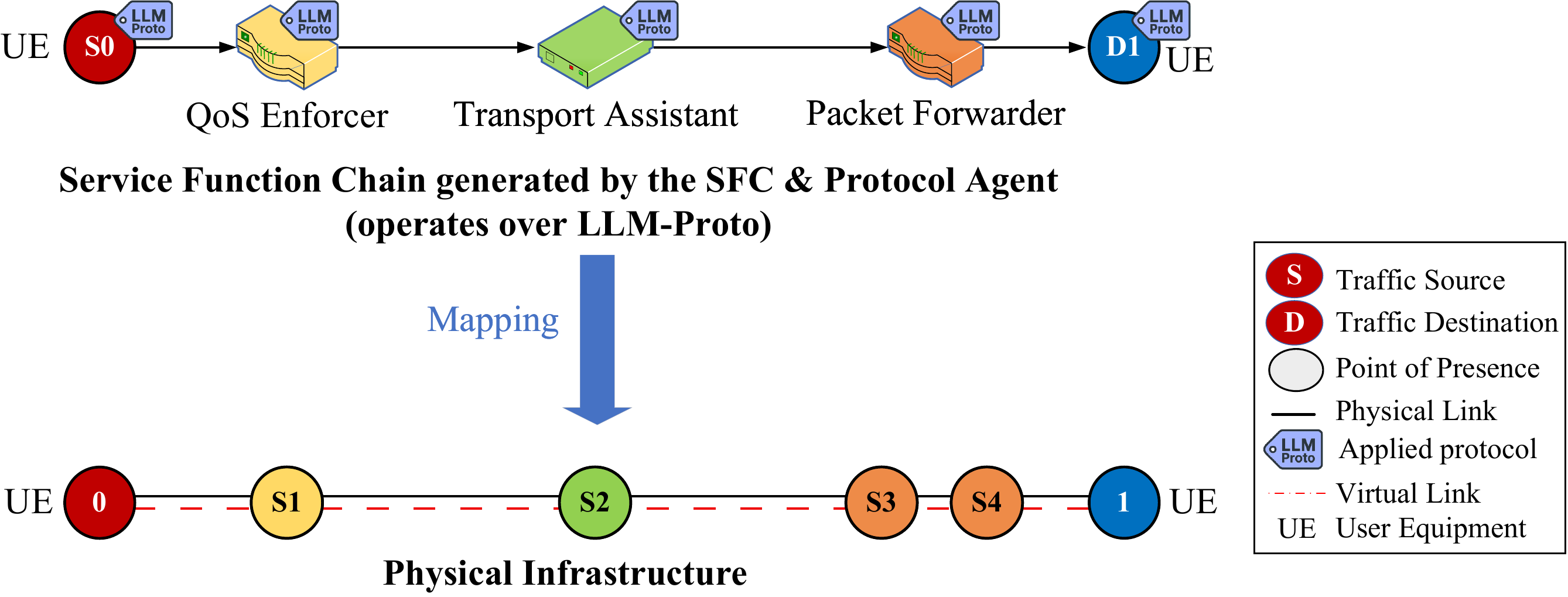}
\caption{Description of the generated SFC and custom protocol LLM-Proto.}
\label{fig:SFC-LLM-PROTO}
\end{figure}

As~it~can be seen in the~output, the agent selected action \textbf{(d)}: \emph{use of NFs from the catalog and/or~design new NFs combined with a custom in-chain protocol}. The~output traces the agent’s step-by-step reasoning, which~led it~to~(1) design LLM-Proto, a~lightweight UDP-based protocol with a custom header carrying reliability flags, sequence numbers, and timestamps, and to (2)~incorporate two existing NFs into the chain (\emph{QoS Enforcer}, \emph{Packet Forwarder}); and~(3)~adapt an existing NF, the \emph{Transport Assistant} to~implement selective reliability.  
The~output provides a detailed description of~the~SFC and~the~communication protocol designed by~the~agent, and~includes the~generated, ready-to-deploy bundle of~C~code for~the~new NFs, the~client and~the~server, a~Makefile for compilation, and~a~shell script for executing the complete SFC in Mininet. Fig.~\ref{fig:SFC-LLM-PROTO} shows the generated SFC and~its~mapping onto the physical infrastructure, all~operating using the~generated protocol,~LLM-Proto.

\begin{figure}[b]
\centering
\includegraphics[width=0.7\columnwidth]{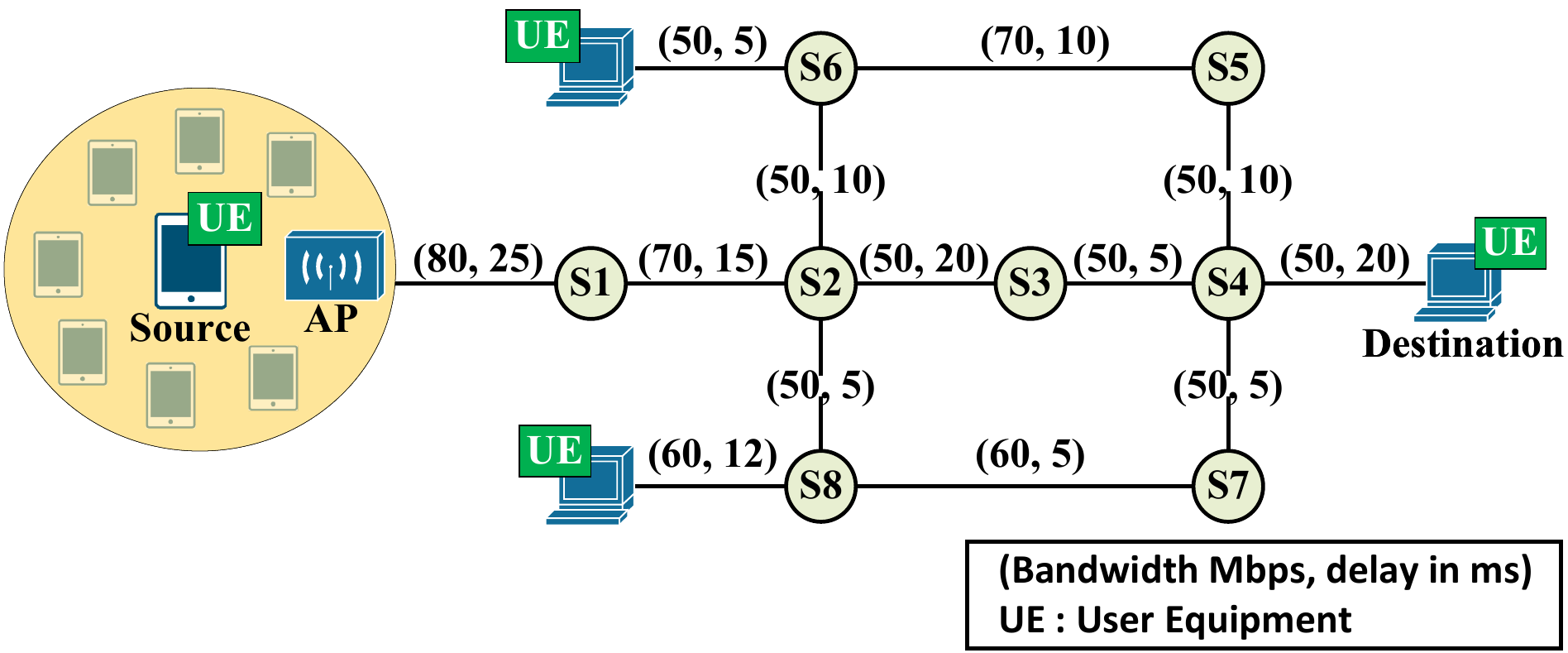}
\caption{Experimental network topology.}
\label{fig:TOPO}
\end{figure}

Now that the LLM agent has designed the~SFC and provided the implementation in~C of the~NFs, and~the~associated custom protocol, we need to evaluate their validity, and effectiveness, and~to~assess their advantages compared to existing protocols (e.g.,~TCP and UDP). To this end, we ran experiments with the \texttt{mininet-wifi} emulator~\cite{mininet-wifi}, emulating the network shown in Fig.~\ref{fig:TOPO}. The application runs on the source UE, sending data to the destination, while other hosts generate background traffic. We assume the application generates 10,000 messages of 1400~bytes each at~the~rate of~1~Mbps with 50\% of~these messages requiring reliability. To~mimic a~realistic lossy environment, the the source UE is connected through a~Wi-Fi interface operating in IEEE 802.11g~mode on channel 6, under interference conditions similar to those reported in~\cite{Korbi2024-CC-IEEEAccess}.
%configured to introduce wireless interference and dynamic background traffic, thereby inducing non-negligible packet loss.

Three scenarios were compared: (1)~no~SFC deployed between the source and destination, with TCP as~the~transport protocol; (2)~no~SFC with UDP as~the~transport protocol; and~(3)~using the~SFC (including its~NFs) and~the~custom communication protocol (LLM-Proto) generated by~the~SFC~\&~Protocol Agent.

\begin{figure}[!t]
    \centering
    \begin{subfigure}[b]{0.4\columnwidth}
        \centering
        \includegraphics[width=\textwidth]{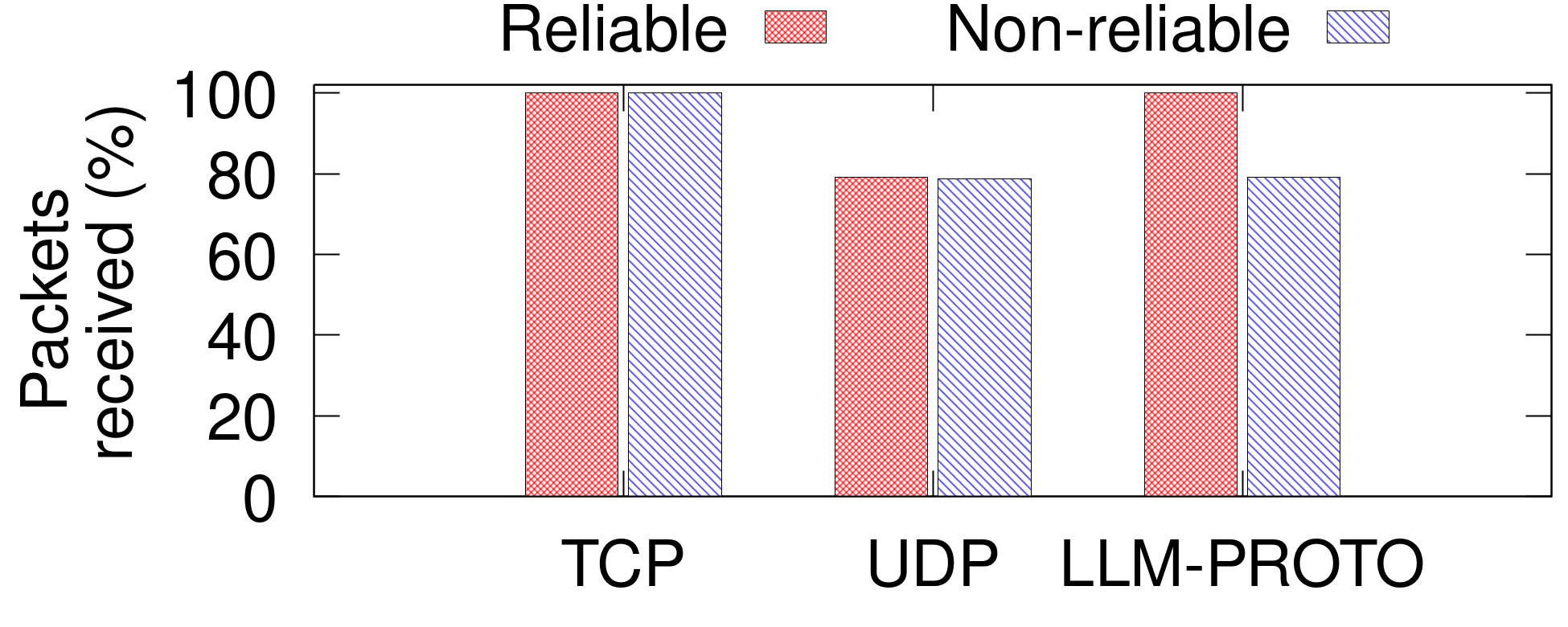}
        \caption{Packets received}
        \label{fig:SFC-rel-nonrel-stats}
    \end{subfigure}
    %\hspace{0.1cm}
    \begin{subfigure}[b]{0.4\columnwidth}
        \centering
        \includegraphics[width=\textwidth]{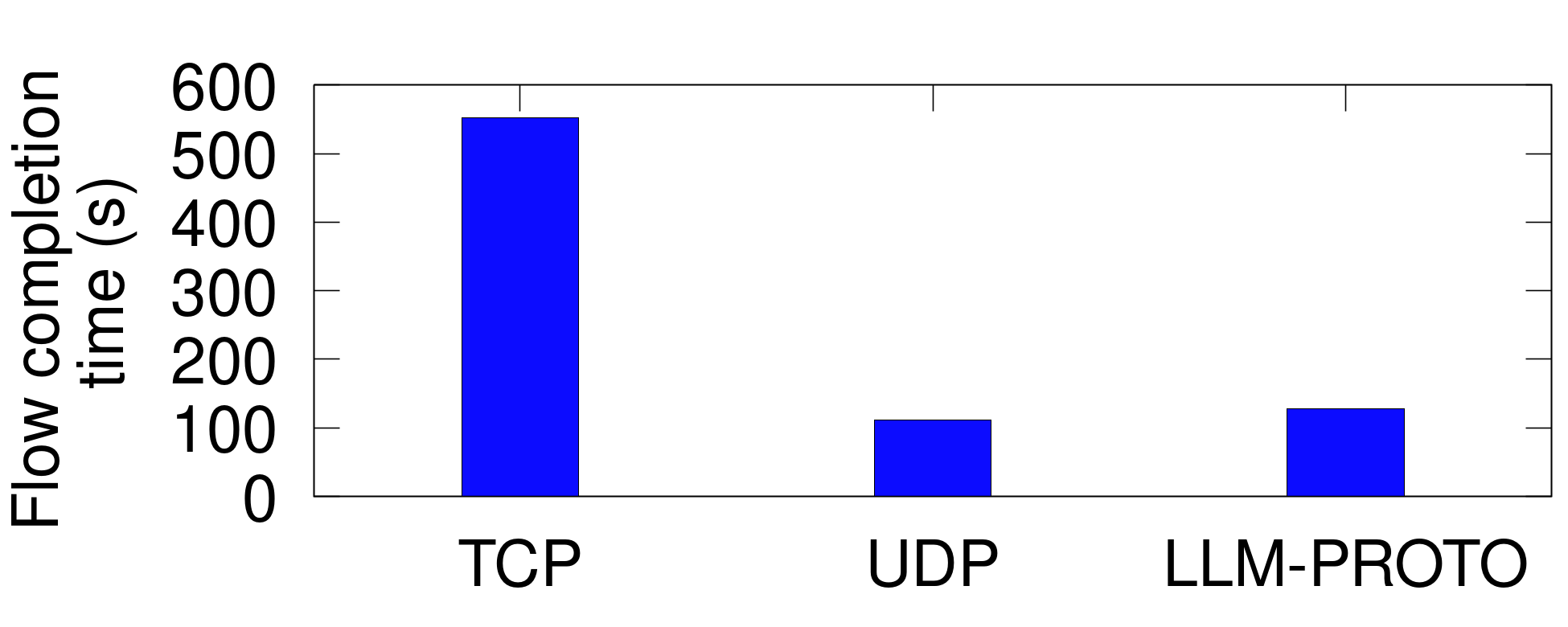}
        \caption{Flow completion time}
        \label{fig:SFC-fct-stats}
    \end{subfigure}
% \hfill
        \begin{subfigure}[b]{0.5\columnwidth}
        \centering
        \includegraphics[width=\textwidth]{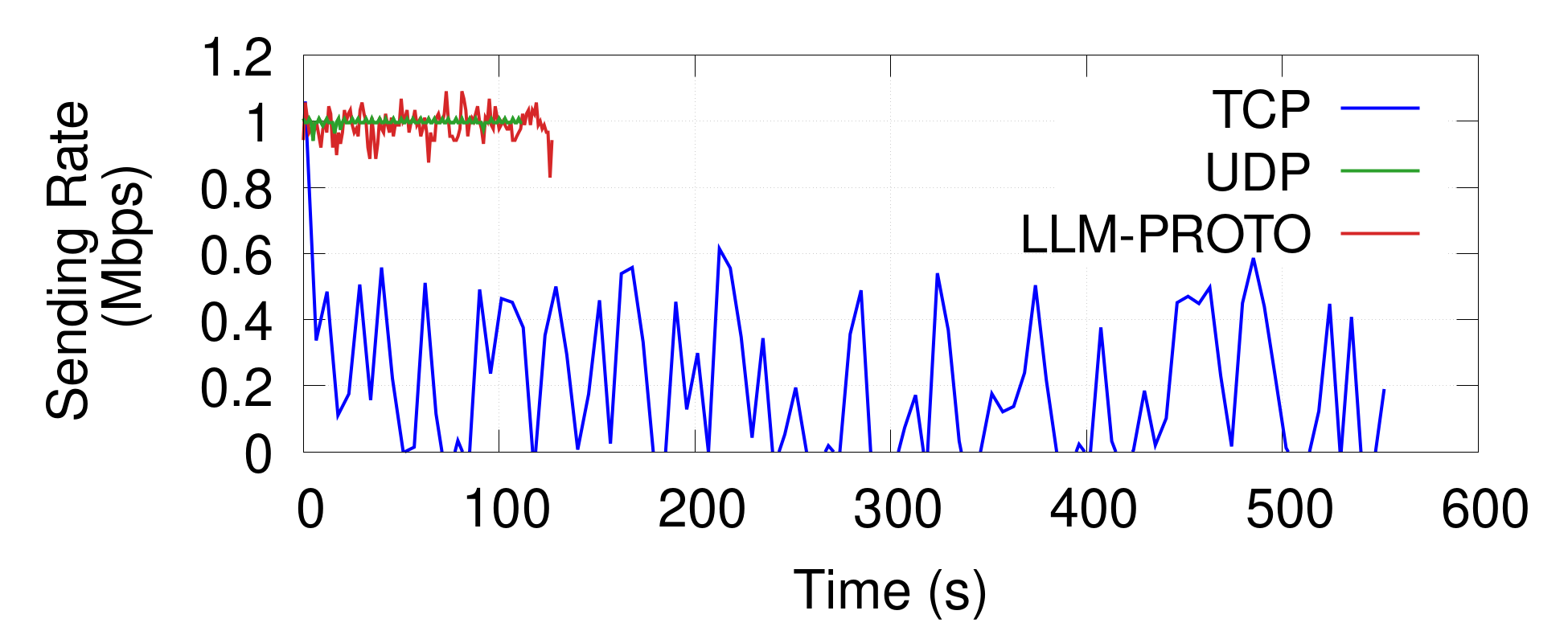}
        \includegraphics[width=\textwidth]{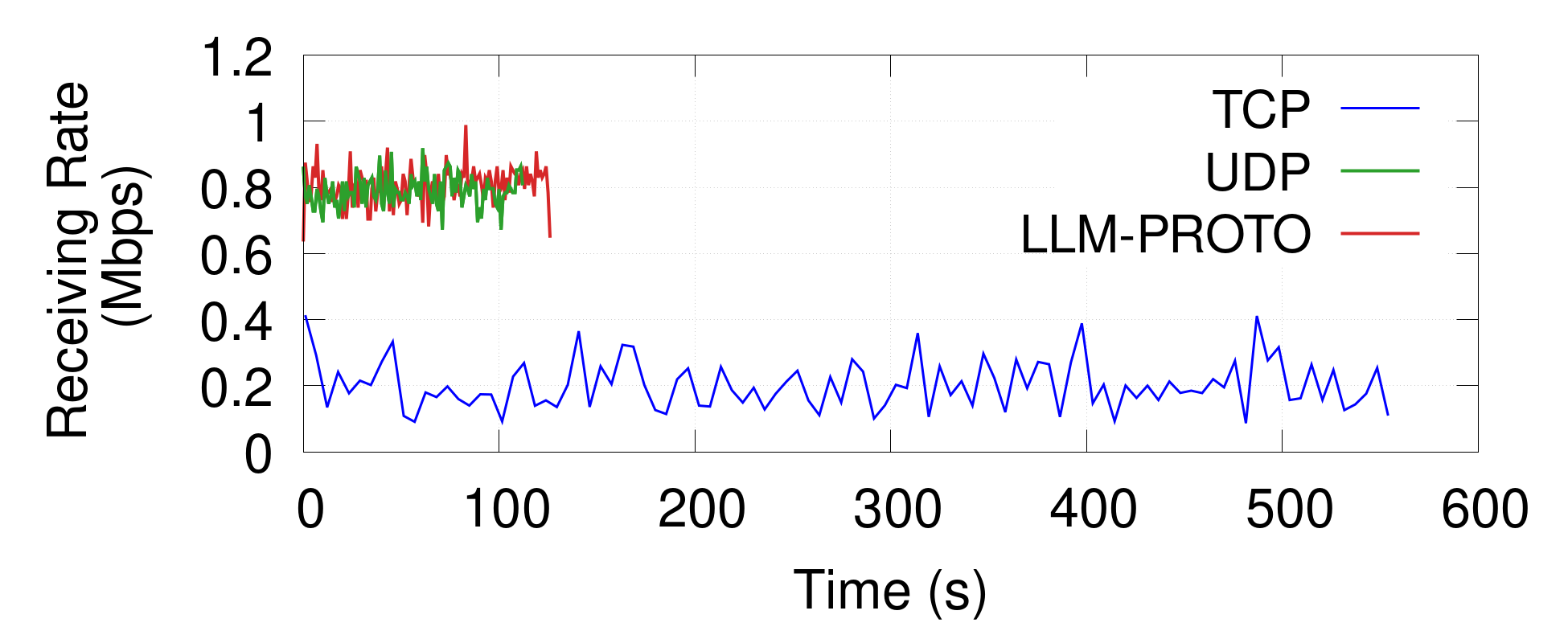}
        \caption{Sending/Receiving rate (Mbps)}
        \label{fig:SFC-sending-rate}
    \end{subfigure}

    % \begin{subfigure}[b]{0.7\columnwidth}
    %     \centering
    %     \includegraphics[width=\textwidth]{Figures/SFC/experiments/receiving_rate.png}
    %     \caption{Receiving rate(Mbps)}
    %     \label{fig:SFC-receiving-rate}
    % \end{subfigure}  
    \caption{Performance of~the~SFC and Protocol generated by~the~SFC~\&~Protocol~Agent~(LLM-Proto) compared to~legacy protocols UDP and TCP.}
    \label{fig:SFC-results}
\end{figure}

Our first observation is that the C code generated by \texttt{GPT-5 Thinking} ran successfully without any errors, and~each NF correctly implemented its~functional requirements as~defined by~the~agent. We~ran experiments to evaluate performance metrics across the studied scenarios as~shown in~Fig.~\ref{fig:SFC-results}: percentage of reliable/non-reliable packets received  by~the~destination, flow completion time (i.e.,~time to~transfer the data), and sending/receiving rates at~the~source/destination.
%TCP, UDP, and the protocol generated by~the~SFC~\& Protocol Agent (LLM-Proto).

Fig.\ref{fig:SFC-rel-nonrel-stats} shows that, as expected, TCP delivers both reliable and non-reliable packets with 100\% success, but at the cost of a~high flow completion time (Fig.~\ref{fig:SFC-fct-stats}) and unstable sending/receiving rates far below 1Mbps (Fig.~\ref{fig:SFC-sending-rate}), violating the~throughput and latency requirements. This~behavior is due to TCP’s CC~scheme, Reno, which leads to highly variable throughput and~to~the~mandatory retransmission of~all packets, increasing the flow completion time. In~contrast, UDP delivers packets unreliably ($\approx$80\%), but achieves low completion time and stable rates near 1Mbps, meeting timeliness requirements but not reliability. Finally, using the SFC and LLM-Proto generated by~the~SFC~\&~Protocol Agent, all objectives are met: reliable packets received with 100\% success, best-effort for non-reliable packets ($\approx$79\%), ~lower flow completion time, and~stable sending/receiving rates close to 1Mbps (Fig.~\ref{fig:SFC-fct-stats} and~Fig.~\ref{fig:SFC-sending-rate}), thereby providing selective reliability while meeting throughput and~minimizing flow completion time.

In summary, the~SFC~\&~Protocol Agent successfully analyzed the application, decided not to use legacy protocols, and generated a fully functional SFC, with custom NFs and transport protocol, tailored to the application’s specific performance, reliability, and behavioral requirements, and outperforming legacy protocols in~terms of all requirements.

%------------------------------------------------------
\subsection{AI-Driven Congestion Control Agent}\label{ssec:AI-DrivenCC}
%------------------------------------------------------
Congestion control, the process of adjusting the sending rate at the source, has been a persistent challenge since the~inception of the Internet. Numerous CC algorithms, such as TCP Reno~\cite{jacobson1988}, CUBIC~\cite{ha2008}, BBR~\cite{cardwell2017bbr}, and Westwood+~\cite{casetti2002westwood}, have been proposed to adapt the congestion control behavior to different factors like network conditions (e.g., wireless vs. wired, WAN vs. data centers). Recent studies~\cite{Korbi2024-CC-IEEEAccess} have also confirmed that the performance of a CC scheme is highly sensitive to the environment in which it operates. 
%At the same time, key performance metrics are significantly influenced by the choice of congestion control scheme.

As a result, no single CC algorithm performs optimally across all environments, and there is currently no~straightforward mechanism to dynamically select or~switch between CC schemes based on real-time conditions. In~practice, most operating systems adopt a fixed default CC scheme regardless of the environment. For instance, Linux has used CUBIC as the default CC scheme~\cite{linux-default-cc-2025}, while Windows uses CTCP and DCTCP \cite{microsoft_set_nettcpsetting_2025} or CUBIC depending on the version~\cite{microsoft_cubic, ns3tcp2022}. Newer versions of macOS and iOS default to BBR or CUBIC~\cite{ns3tcp2022}. 
These observations highlights two key limitations: (1)~the underutilization of existing CC~algorithms designed for diverse conditions, and~(2)~the~lack of dynamic mechanisms to adapt CC behavior to real-time network and performance requirements.

%These dynamic conditions render traditional congestion control mechanisms inadequate, as they are static, manually tuned, and often overly conservative or misaligned with real-time application needs.

To address these limitations, FlexNGIA 2.0 introduces an~AI-driven Congestion Control Agent capable of analyzing real-time conditions (e.g., execution environment, network status) and dynamically adjusting the parameters of the current CC scheme or switching between different schemes at runtime. Beyond that, it can also design and deploy custom CC scheme on the fly, tailored to~the~prevailing network conditions.

In the following, we present more details about the operation of the CC agent, the prompt for the associated LLM, and experimental results showcasing the operation of~the~CC~agent.

\begin{figure}[b]
    \centering
    \includegraphics[width=1.0\columnwidth]{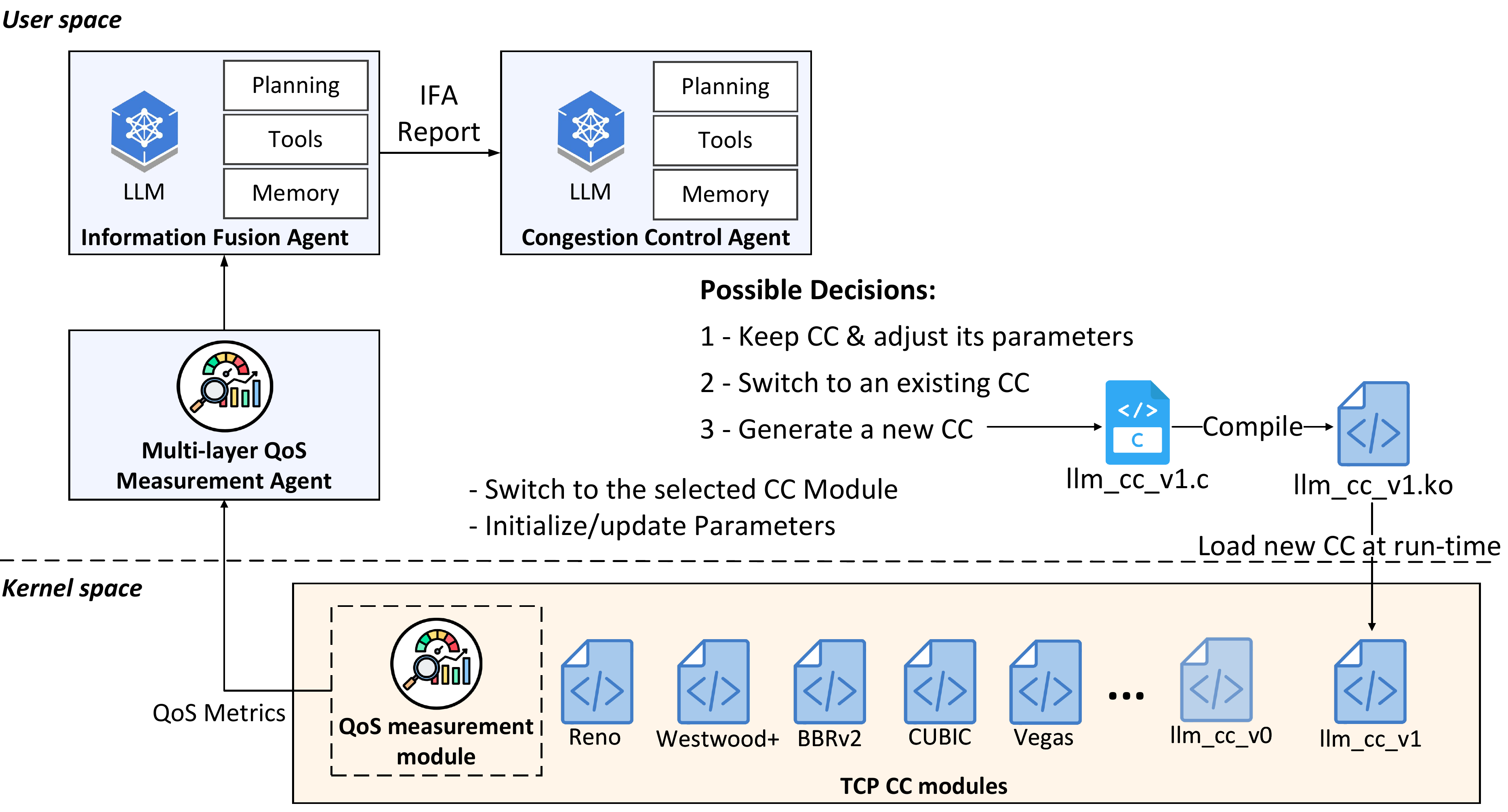}
    \caption{Operation of the Congestion Control Agent integrated into the User Equipment.}
    \label{fig:CCA-Architecture}
\end{figure}

\vspace{0.5em}\noindent\textbf{$\bullet$ Operation of the CC Agent:} Fig.~\ref{fig:CCA-Architecture} illustrates the operation of the CC~Agent and its interactions with other agents within the FlexNGIA 2.0 framework. As shown, the Information Fusion Agent provides its detailed IFA~report to~the~CC~Agent, which analyzes it to determine the~most appropriate decision: (1)~maintain the current CC~scheme without any~changes, (2)~retain it and adjust its~parameters, and~(3)~design, implement and deploy a new CC~mechanism adapted to the prevailing network conditions and~environment.

Fig.~\ref{fig:CCA-Architecture} also shows the various CC~modules already deployed in the kernel of the UE's operating system (e.g.,~Reno, Westwood+, BBRv2, CUBIC, and~Vegas). These modules are commonly available by default in most Linux distributions and the~CC~Agent can decide to~use one them as needed. Additionally, the figure shows custom CC~modules (llm\_cc\_v(\textit{i})) that were dynamically designed and~loaded into the~kernel at runtime by~the~CC~Agent. 
The~figure also depicts a QoS measurement module integrated into the~OS, which collects metrics relevant to congestion control (e.g., window size, throughput, RTT, and retransmission rate) and reports them to~the~IFA~Agent to be included in~the~IFA~agent.
%then combines them with additional data received from the other FlexNGIA~2.0 agents to~generate the IFA~report (see Fig.~\ref{fig:AI-CC-Framework}).

%Note that the IFS report is a comprehensive summary that aggregates insights from other agents in the FlexNGIA 2.0 Agentic AI-based Management Framework (see Fig.~\ref{fig:AI-CC-Framework}), such as the Multi-layer QoS Measurement Agent and others.

\begin{figure}[hb]
    \centering
    \includegraphics[width=0.80\columnwidth]{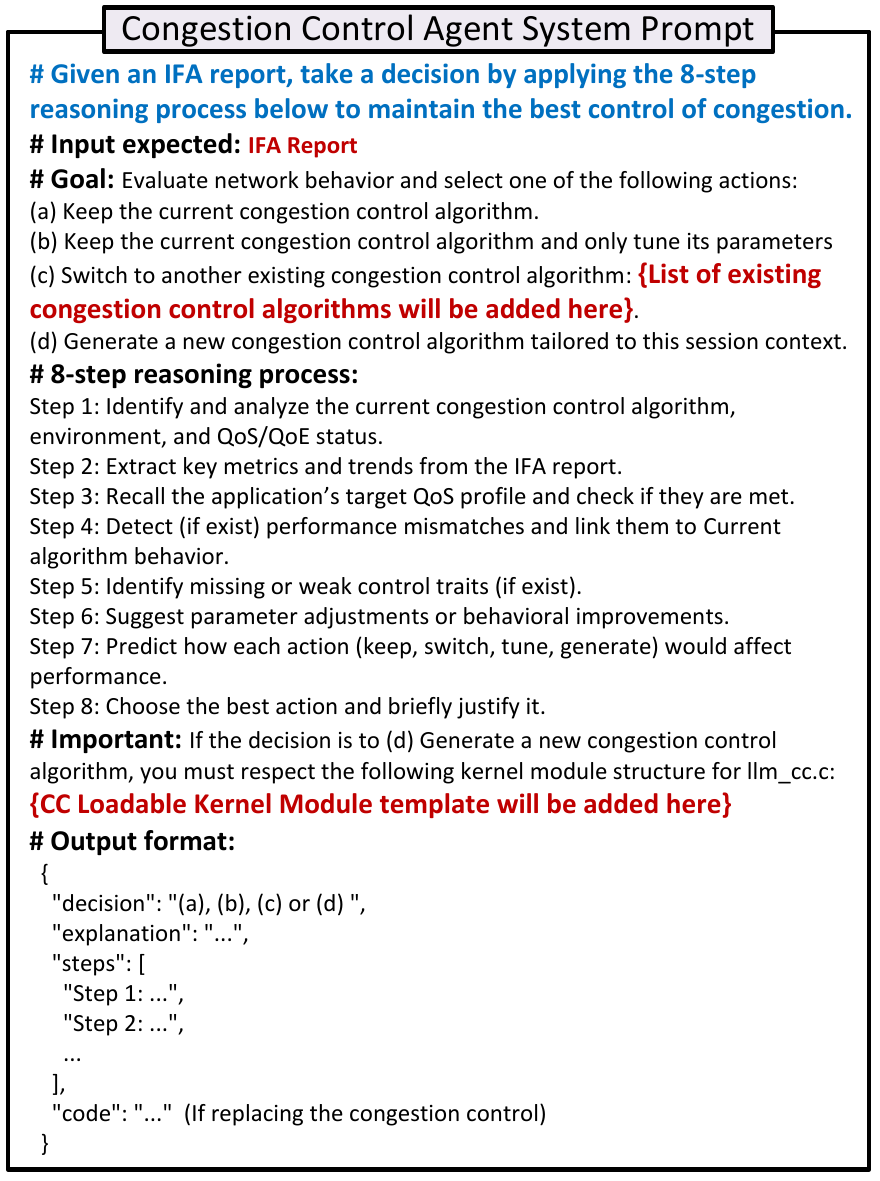}
    \caption{System prompt of the Congestion Control Agent.}
    \label{fig:CCA-prompt}
\end{figure}

%---------------------------------------------------------
\vspace{0.5em}\noindent\textbf{$\bullet$ Prompt Structure for the CC Agent's LLM:}
The~CC~Agent is prompted with a structured system prompt as~shown~in~Fig.~\ref{fig:CCA-prompt}. The~core of this~prompt embeds the~IFA~report, followed by an 8-step reasoning process that ensures consistent behavior across executions. These~steps include evaluating the current CC~scheme, identifying potential performance bottlenecks, forecasting the~outcomes of alternative schemes, and ultimately selecting and justifying the most appropriate action among four options: (a)~keep the~current scheme and~its~parameters, (b) keep the~current scheme but tune its~parameters, (c)~switch to a different CC~scheme from the~existing ones, or~(d)~generate and deploy a~new CC~algorithm tailored to the current conditions.

After completing all eight reasoning steps, the LLM generates a~\textit{CC Decision Report} using a structured format defined in the prompt (see Section \textbf{\#Output Format} in~Fig.~\ref{fig:CCA-prompt}). This report includes the selected decision, a~detailed explanation justifying that decision, and~a~summary of the reasoning process for the eight steps. If~the~decision is to design and deploy a~new~CC~scheme (decision d), the LLM~generates the~implementation code for the new scheme.

\begin{figure}[b]
    \centering
    \includegraphics[width=0.75\columnwidth]{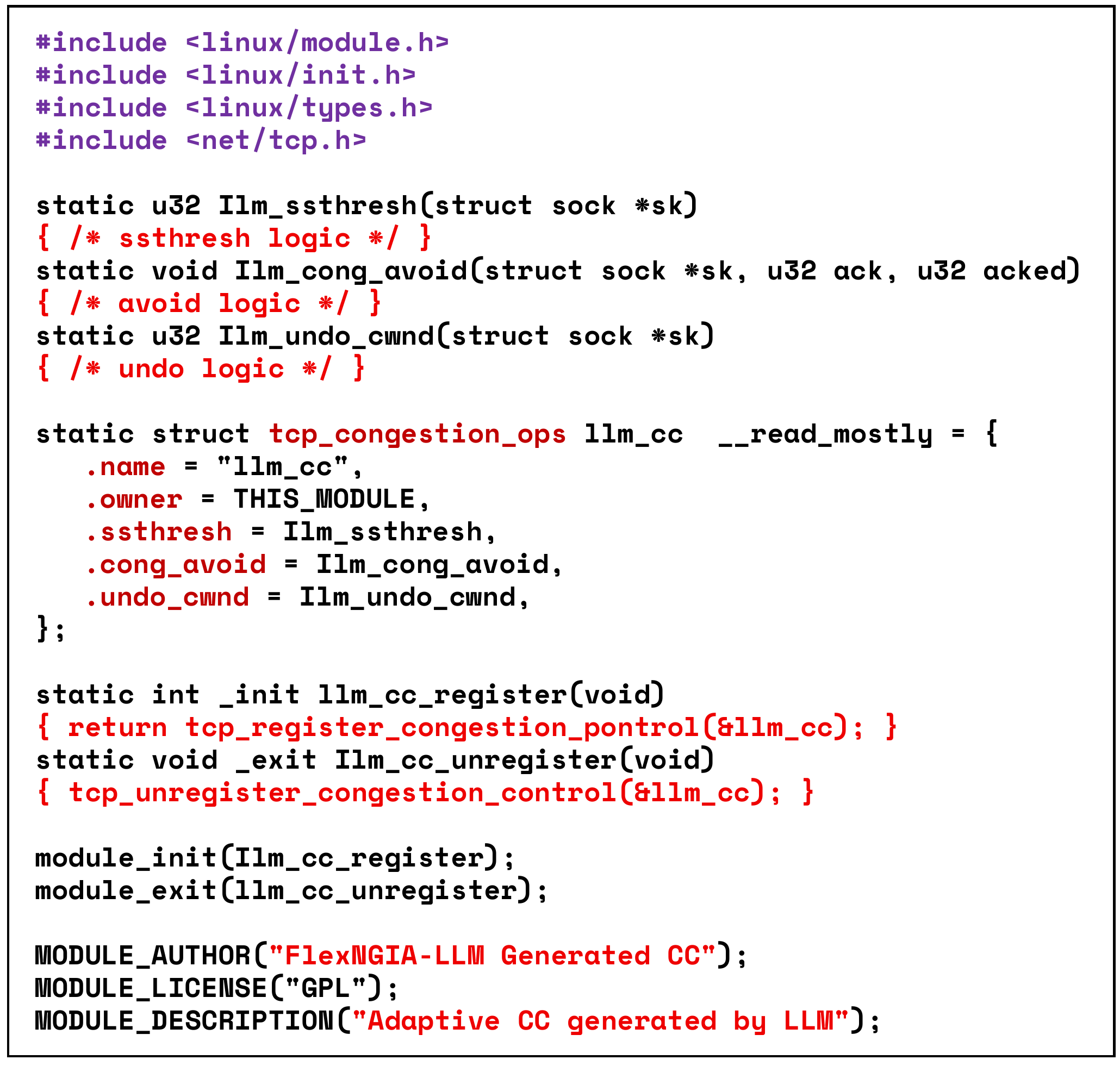}
    \caption{Template for a pluggable Linux CC module (in~C).}
    \label{fig:cc-template}
\end{figure}

To make sure the generated code is compatible with~the~Linux TCP CC~subsystem and properly exposes the required hooks via the \texttt{tcp\_congestion\_ops} interface~\cite{linux_tcp_congestion_ops_2025}, the~prompt includes a~code template used as~a~base to generate the~CC scheme's code (see~Section~\textbf{\#Important} in~Fig.~\ref{fig:CCA-prompt}). 
Indeed, we examined TCP CC implementations in~Linux Kernel~5.13.12 and created a~minimal yet valid template for~a~pluggable CC~module. Fig.~\ref{fig:cc-template}~presents this template which includes the three essential methods required by the kernel: \texttt{ssthresh}, which defines the behavior of the slow start threshold; \texttt{cong\_avoid}, which governs the growth of the congestion window during the congestion avoidance phase; and~\texttt{undo\_cwnd}, which resets the congestion window during a~rollback in the loss recovery process.

Once generated, the code is automatically loaded into the~kernel at runtime by the CC Agent to seamlessly replace the running scheme, without causing any interruption to~the~TCP~connection or requiring~it to~be~reset.

%Depending on the Linux kernel version, additional methods (e.g., \texttt{in\_ack\_event}, \texttt{pkts\_acked}) may be optional or required. For version 5.13.12, the three listed functions are the minimum required to register a new congestion control algorithm.

\vspace{0.5em}\noindent\textbf{$\bullet$ CC Agent in Action -  Experimental Results:} To~demonstrate the~operation and effectiveness of~the~CC~Agent, we conducted experiments where the~agent was tasked with managing CC~to~meet the application’s target QoS profile. 
We used \texttt{mininet-wifi} to~emulate the~network shown in~Fig.~\ref{fig:TOPO} with the~same setting described in~subsection~\ref{ssec:AI-DrivenSFCAgent}. The~CC~Agent runs on the source UE to manage the congestion control of a TCP flow toward the destination UE. The Agent was powered by~the~LLM \texttt{DeepSeek-R1-Distill-Llama-70B}~\cite{deepseek2025}, a~70-billion-parameter distilled from DeepSeek-R1. This~model was selected for~its~performance on structured, multi-step problem solving, and ability to generate explainable, chain-of-thought reasoning~\cite{deepseek2025}.

The studied TCP flow carries the traffic from a~video streaming application with strict QoS requirements. These requirements are initially identified by the App Analysis Agent and continuously monitored by the Multi-layer QoS Measurement Agent. The Information Fusion Agent generates an updated IFA~report at each evaluation interval (set~to 60~s), and~sends~it to~the~CC~Agent, which uses~it to~evaluate the~current state and take the appropriate decision.

\begin{figure}[!t]
    \centering
    \begin{subfigure}{0.38\textwidth}
        \includegraphics[width=\linewidth]{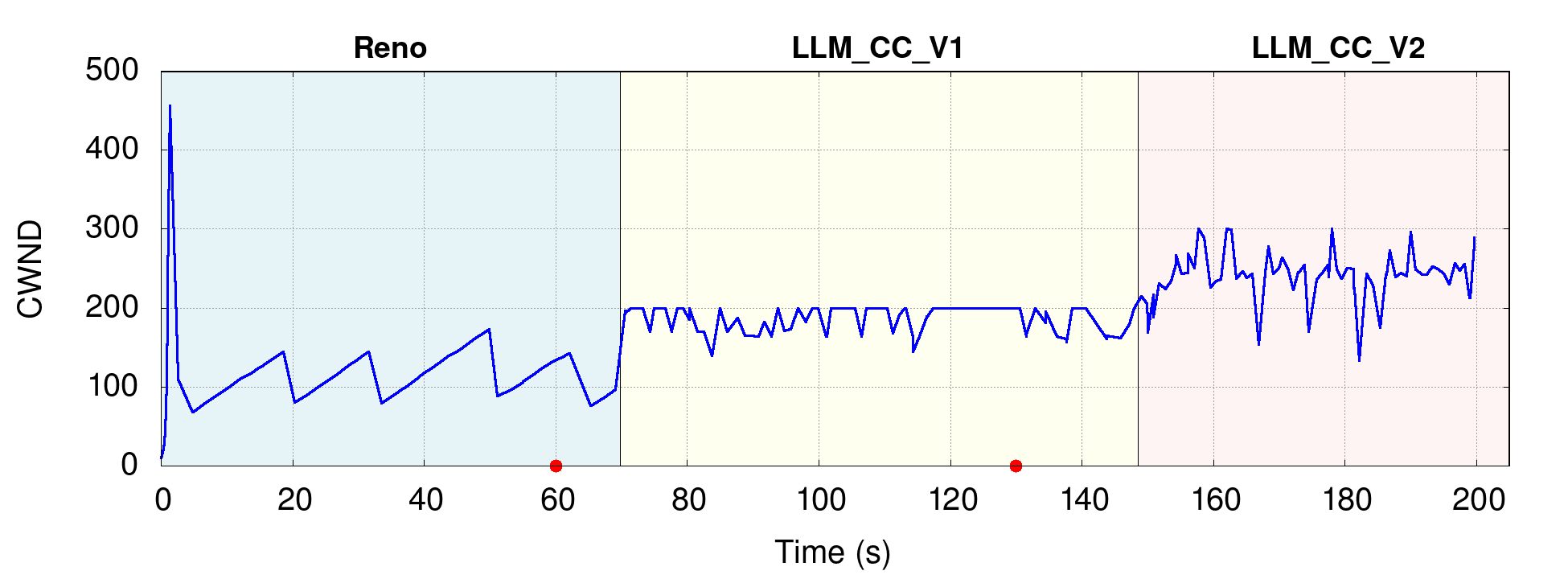}
        \caption{Congestion window}
    \end{subfigure}
    \begin{subfigure}{0.38\textwidth}
        \includegraphics[width=\linewidth]{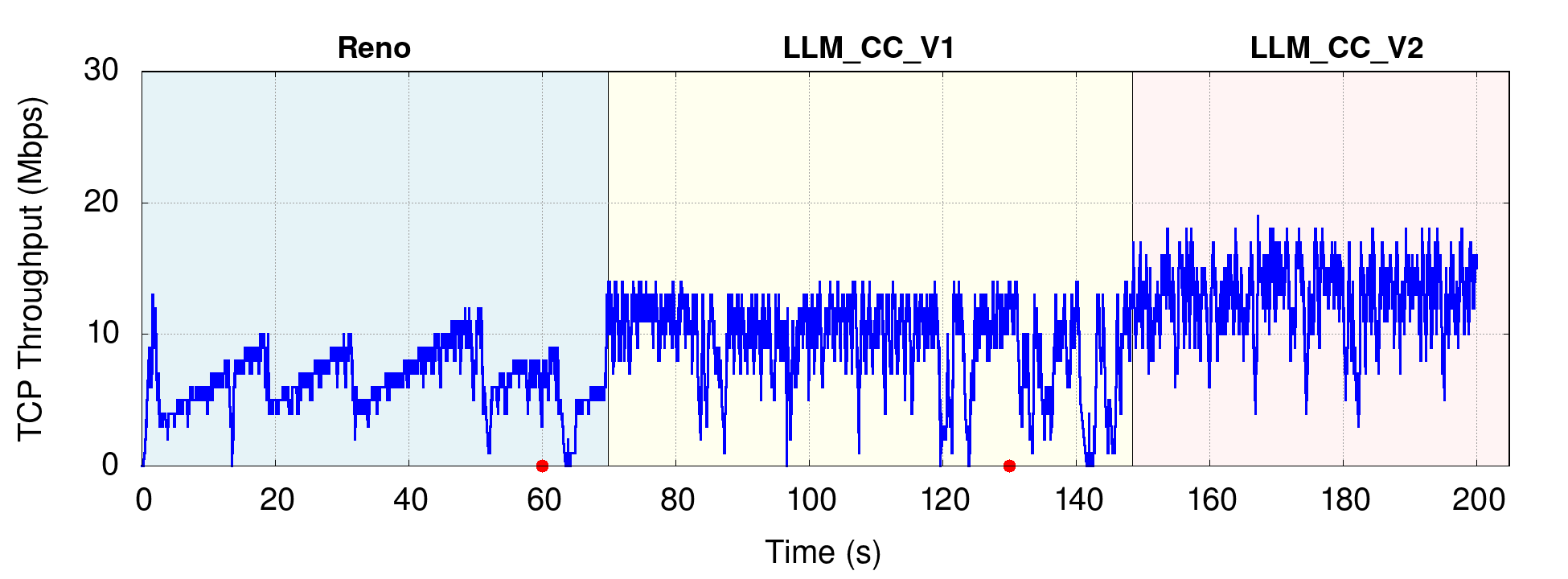}
        \caption{Throughput}
    \end{subfigure}
    \begin{subfigure}{0.38\textwidth}
        \includegraphics[width=\linewidth]{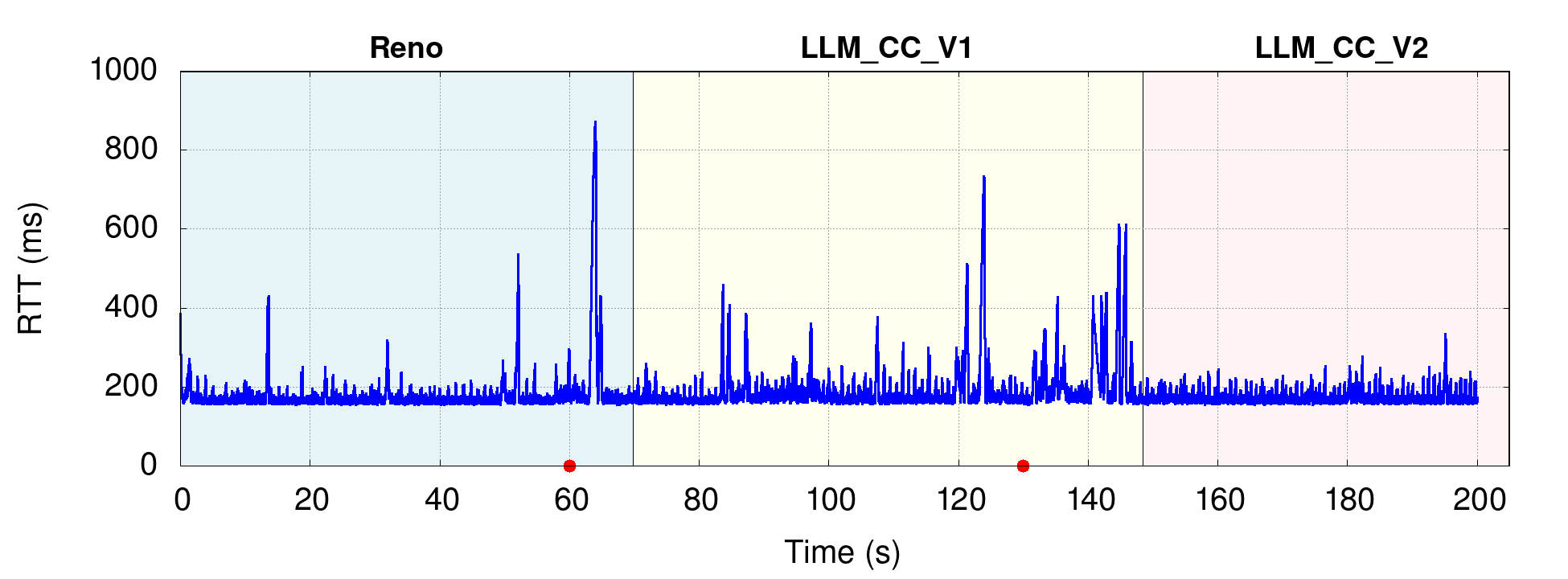}
        \caption{RTT}
    \end{subfigure}
    \begin{subfigure}{0.38\textwidth}
        \includegraphics[width=\linewidth]{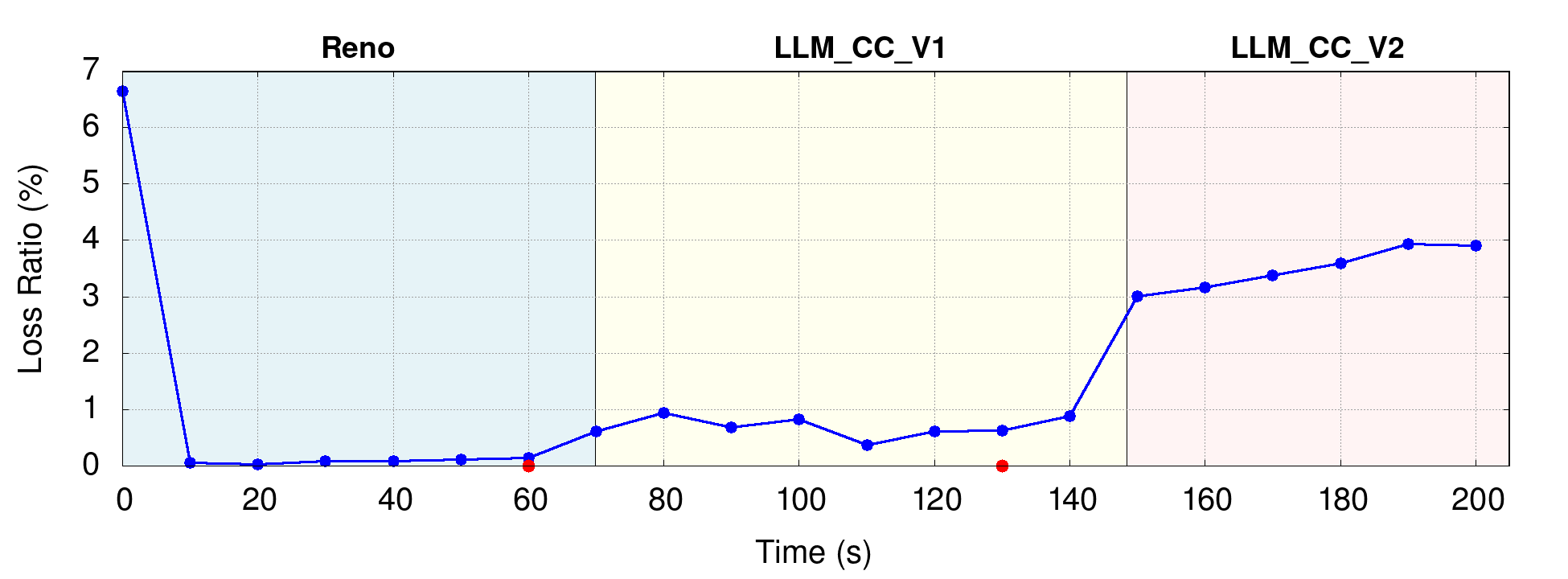}
        \caption{Loss ratio}
    \end{subfigure}
    \caption{Performance metrics over time showing the~CC~agent’s decisions to switch between an existing CC~scheme (\texttt{TCP~Reno}) and the generated schemes \texttt{LLM\_CC\_V1} and \texttt{LLM\_CC\_V2}, all at runtime. The red bullets indicate the times at~which the CC agent performed an evaluation of the~CC~scheme (evaluation \#1 and \#2).}
    \label{fig:cca_switching_results}
\end{figure}
\begin{figure*}[!htb]
    \centering
    \begin{subfigure}[b]{0.49\textwidth}
        \centering
        \includegraphics[width=\textwidth]{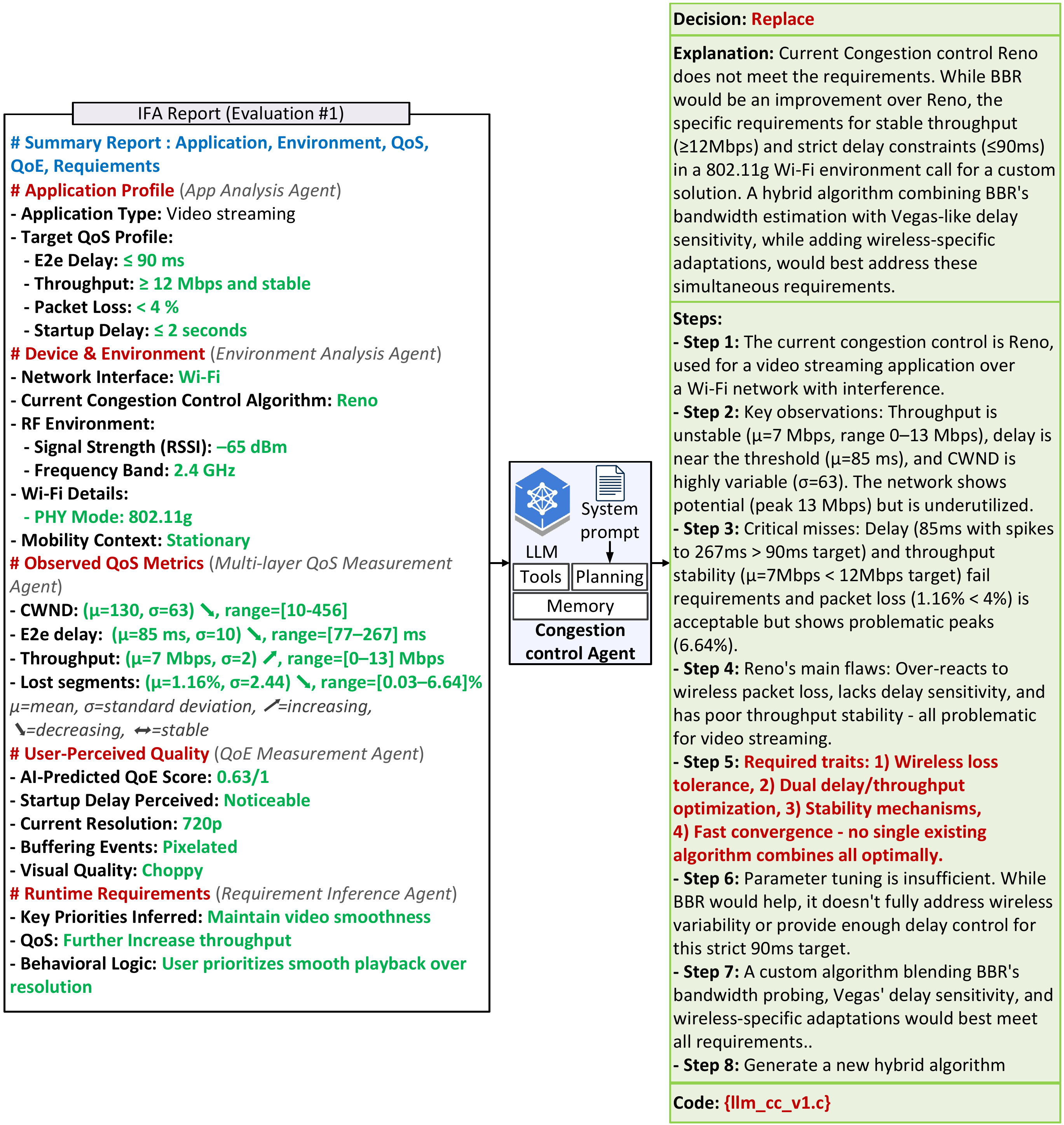}
        \caption{IFA and decision reports at the first CC evaluation}
        \label{fig:CCA-Action1}
    \end{subfigure}
    \hfill
    \begin{subfigure}[b]{0.49\textwidth}
        \centering
        \includegraphics[width=\textwidth]{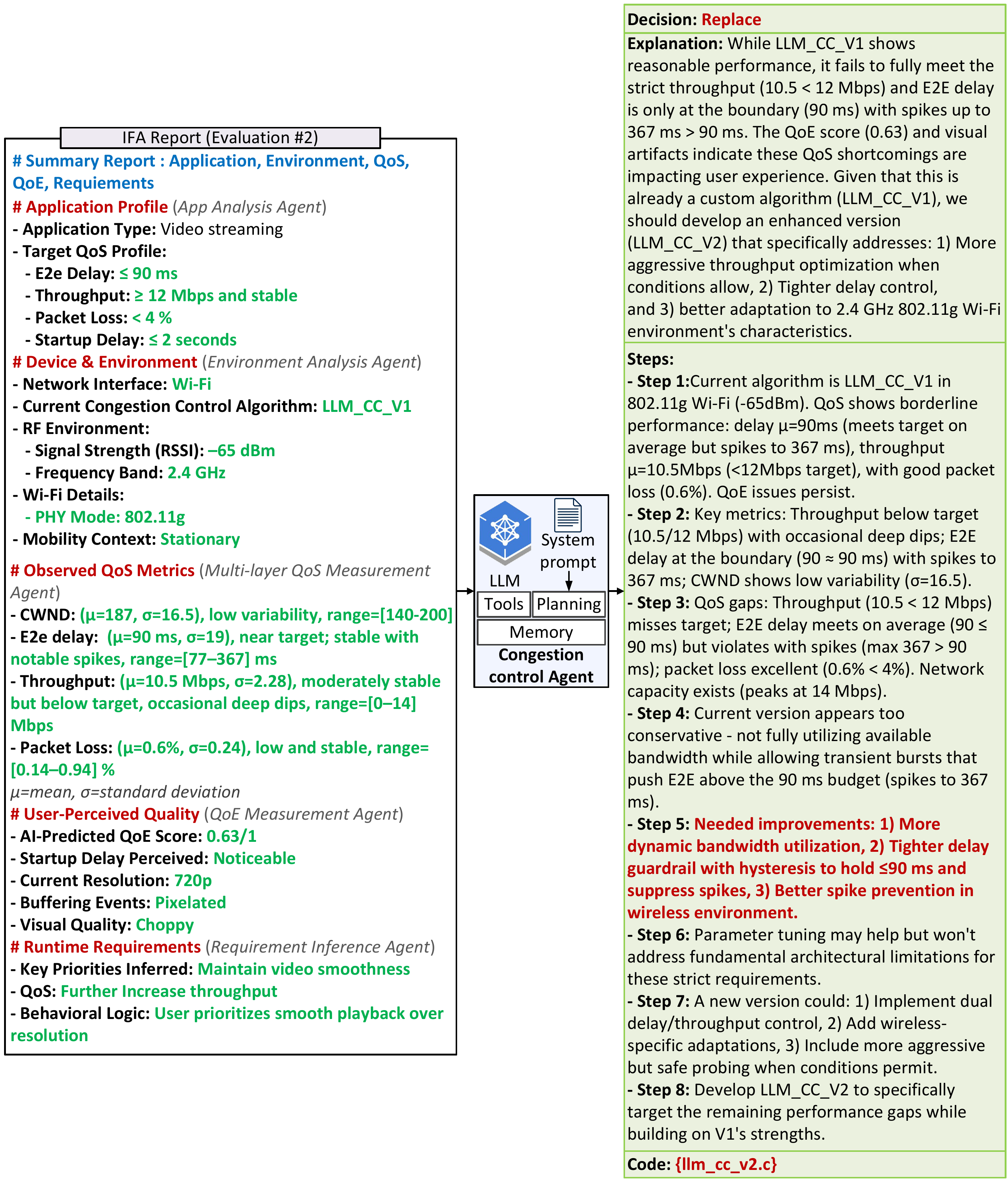}
        \caption{IFA and decision reports at the second CC evaluation}
        \label{fig:CCA-Action2}
    \end{subfigure}
    \caption{IFA Reports and their corresponding CC Agent's decision reports.}
    \label{fig:CCA-Actions}
\end{figure*}

%\vspace{0.5em}
%\noindent\textbf{$\bullet$ Experimental Results:}
Fig.~\ref{fig:cca_switching_results} presents the evolution of~key performance metrics (i.e.,~congestion window, throughput, RTT, and~packet loss ratio) throughout the experiment, which lasted 200 seconds. 
%The evolution of these metrics reflects the behavior of three CC schemes: the initial \texttt{TCP Reno}, the~first CC~agent-generated algorithm \texttt{LLM\_CC\_V1}, and its refined successor \texttt{LLM\_CC\_V2}. 
%
As shown in~Fig.~\ref{fig:cca_switching_results}, initially, the~system runs with \texttt{TCP Reno}. The CC Agent,  based on the~IFA~reports, decides to~design and deploy a~new CC scheme \texttt{LLM\_CC\_V1} ($t=70$~s) and then another one, \texttt{LLM\_CC\_V2} ($t=148$~s).
Each transition to~a~new CC~scheme corresponds to visible changes in~the~plotted metrics and~illustrates the~agent’s ability to adapt in real time to target the delay, throughput and packet loss requirements (see Section \textit{\# Application Profile} in~the~IFA report in~Subfigure~\ref{fig:CCA-Action1}). Hence, it~successfully developed CC schemes that reduced the variability of the congestion window, increased and~stabilized the~throughput around 12Mbps while keeping the packet loss below 4\% (Fig.~\ref{fig:cca_switching_results}).

Fig.~\ref{fig:CCA-Actions} provides deeper insight into the decision-making process. Subfigure~\ref{fig:CCA-Action1} shows the first IFA~report generated during the \texttt{TCP Reno} operation. It~shows also the~\textit{CC~Decision~Report} with the structured reasoning that led the CC agent to create the new CC scheme \texttt{LLM\_CC\_V1} to~replace \texttt{TCP~Reno}. The analysis highlights Reno's main flaws (step~4), the~required features (step~5),  the inadequacy of existing schemes (step~6), and the need for a~new CC scheme that addresses the identified issues by~blending BBR's bandwidth probing, Vegas' delay sensitivity, and wireless-specific adaptation (step 7~and~8). 
Subfigure~\ref{fig:CCA-Action2} shows the second IFA~report generated during the \texttt{LLM\_CC\_V1} operation. Although \texttt{LLM\_CC\_V1} has improved throughput and reduced loss, the~RTT remains unstable. The agent’s reasoning shown in~the~new \textit{CC Decision Report} recommends a refined algorithm, \texttt{LLM\_CC\_V2}.

The results clearly demonstrate how the CC agents adapts the scheme to evolving conditions by combining real-time measurements, structured reasoning, and dynamic code synthesis and deployment.

%-----------------------------------------------------------------------
\begin{figure}[!b]
    \centering
    \includegraphics[width=0.8\columnwidth]{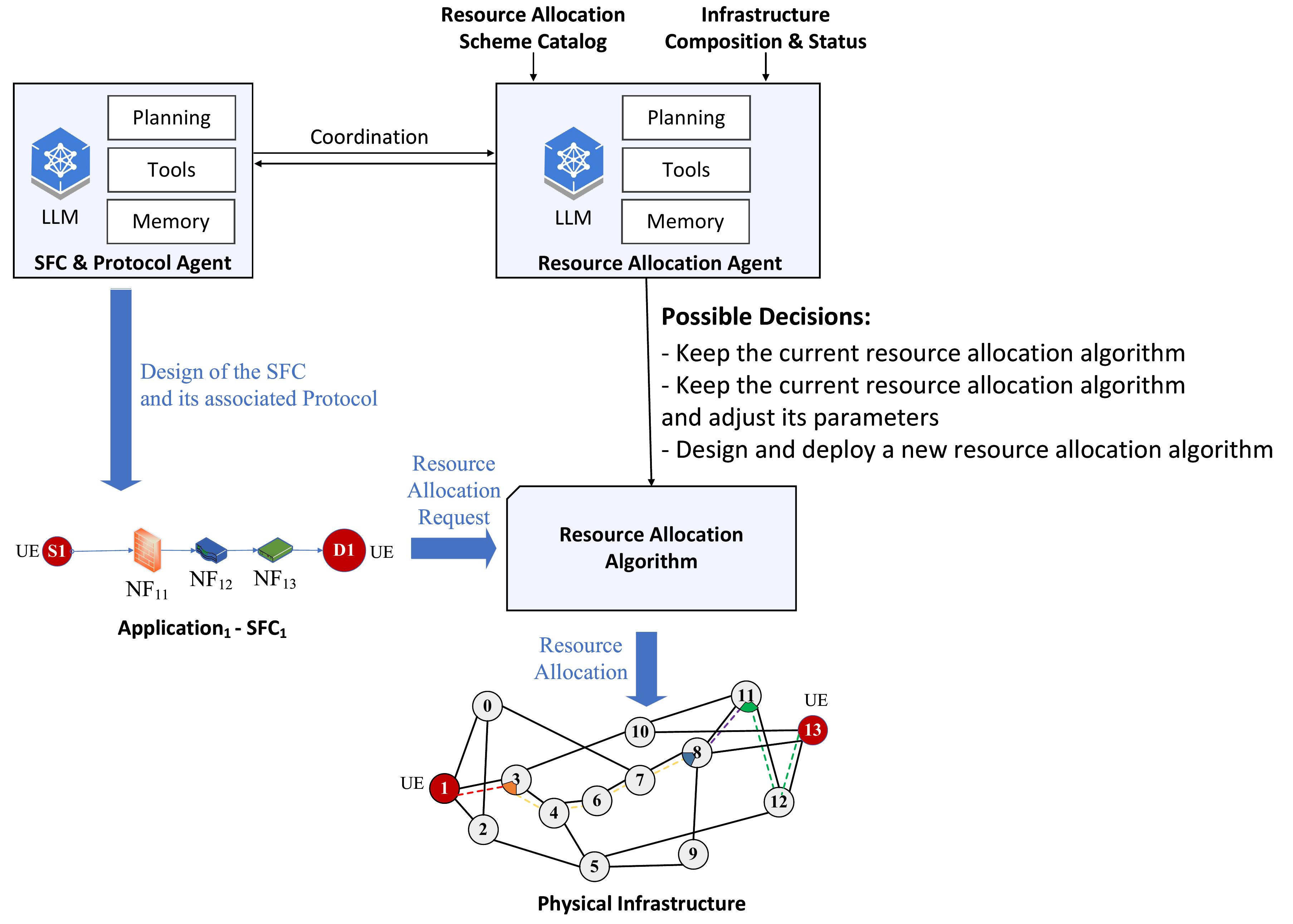}
    \caption{Operation of the Resource Allocation Agent.}
    \label{fig:RAA-Architecture}
\end{figure}
%-----------------------------------------------------------------------

%-----------------------------------------------------------------------
\subsection{AI-Driven Resource Allocation Agent}\label{ssec:AI-DrivenRA}
%-----------------------------------------------------------------------

The Resource Allocation Agent is responsible for managing and optimizing the allocation strategy of~the~ resources for SFCs while ensuring operational, sustainability, and economic objectives. As~shown in~Fig.~\ref{fig:RAA-Architecture}, the~RA~Agent is provided with detailed information about the network’s infrastructure. This includes the physical infrastructure's composition (i.e.,~defines what physical resources exist and how they are connected) such as servers, routers, switches, storage units, technologies. 
%The~catalog~includes various allocation algorithms (e.g.,~\cite{moufakirITU2022}) and their objectives. %(e.g.,~energy efficiency, maximizing the utilization of green and renewable resources, increasing the number of successfully mapped SFCs, minimizing operational costs, and maximizing revenues). 

The RA Agent relies also on~a~resource allocation scheme catalog, which is a structured set of predefined strategies, algorithms, and policies designed to efficiently assign resources to incoming SFCs across the physical infrastructure. The catalog specifies for~each algorithm the tunable parameters to~be~used to~adapt it~to different network conditions and application requirements.

The~RA~agents can dynamically select, adjust, an~RA~algorithm from the catalog or even design a~new RA~strategies to optimize performance and achieve operational, sustainability, and economic goals. Aligned with this~design, Li~et~al.~\cite{Guizani2025} presented a~solution demonstrating how LLMs can be leveraged to guide resource allocation. Specifically, they leverage LLMs to enhance the efficiency of a~genetic algorithm in~exploring the solution space for resource allocation problems. 
%By adjusting the genetic algorithm's heuristic functions, the LLM helps the algorithm explore and exploit the solution space more effectively. 
Our proposed RA Agent generalizes this~concept, enabling LLMs not~only to fine-tune heuristics but also to~select among different allocation schemes and even design new allocation strategies.

Furthermore, the RA agent receives regular statistics about the current state of the infrastructure (e.g.,~utilization, utilization fairness, operational state, performance metrics, or~faults) and other metrics pertaining to the operational and~sustainability objectives, including operational costs (e.g.,~cost of~deployment and operation of network functions), revenues, energy efficiency, use of green sources, green penalties (e.g.,~CO$_2$ emissions penalty), and profits.

\begin{figure}[!h]
    \centering
    \includegraphics[width=0.9\columnwidth]{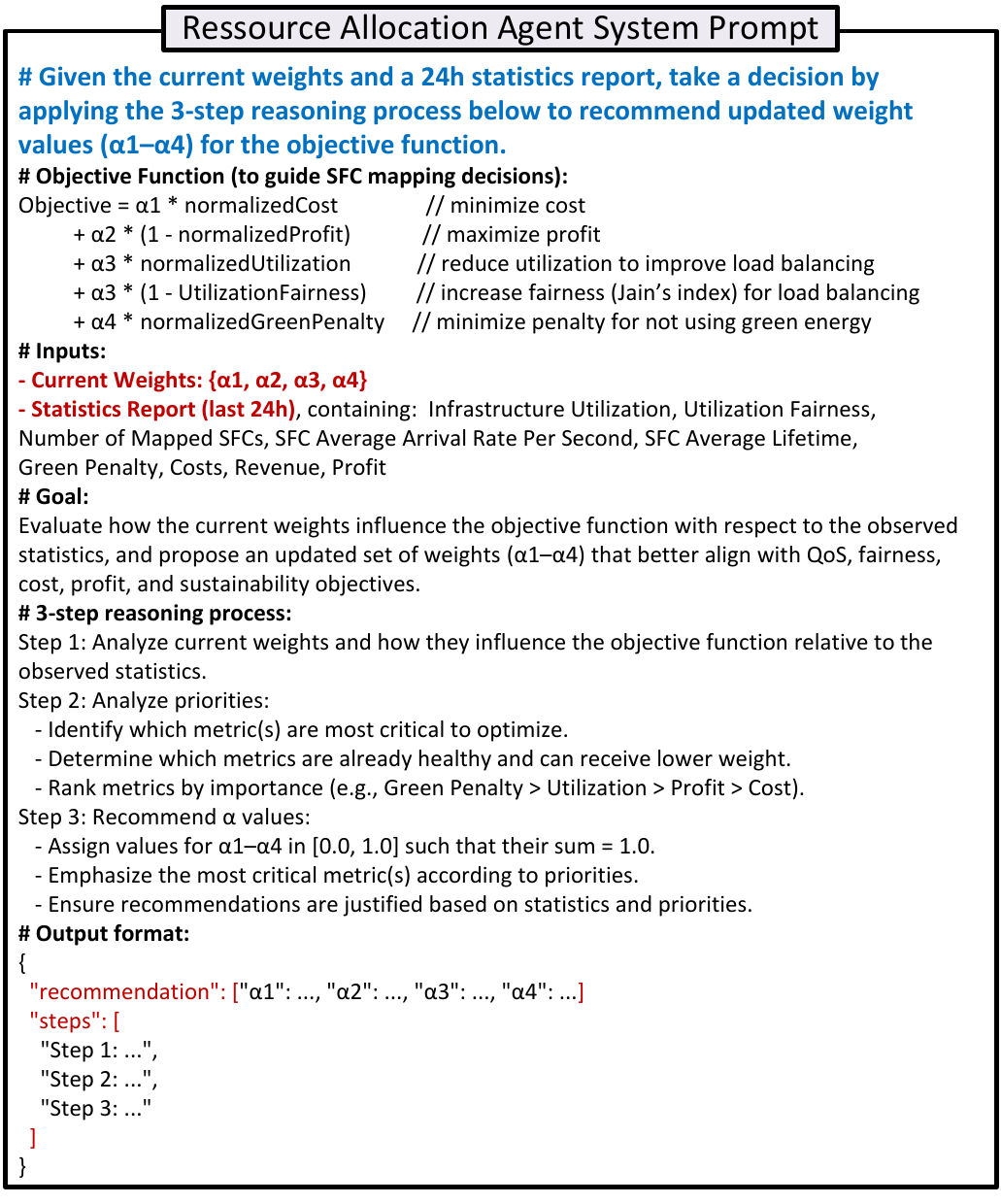}
    \caption{System prompt of the~Ressource~Allocation~Agent.}
    \label{fig:RA-prompt-template}
\end{figure}

\vspace{0.5em}\noindent\textbf{$\bullet$ Prompt Structure of the~RA Agent:}
Fig.~\ref{fig:RA-prompt-template} shows the system Prompt Structure for the RA Agent.
The prompt instructs the RA agent’s LLM to recommend updated weight values for the multi-objective function guiding SFC mapping decisions, based on the current weights and a 24-hour statistics report. The LLM follows a structured three-step reasoning process: (1) analyze how the current weights influence the objective function relative to the observed statistics, (2)~determine metric priorities and rank them according to criticality, and (3) recommend updated weights, emphasizing the most critical metrics while providing a justification based on the data. The output includes the recommended weights and a~step-by-step explanation of the reasoning process.

\vspace{0.5em}\noindent\textbf{$\bullet$ RA Agent in Action -  Experimental Results:} To~demonstrate the operation and effectiveness of the RA Agent, we conducted simulations of a~realistic scenario in~which multiple SFCs dynamically arrive over time and must be deployed across the physical infrastructure~(Fig.~\ref{fig:RAA-Architecture}). We developed a~resource allocation algorithm guided by~the~following objective function to carefully select physical nodes in the network to host the NFs of the incoming SFCs:
%, ensuring optimized cost, performance, load balancing, and~energy efficiency.
\vspace{-0.4em}  % reduce 1 line of space
\begin{align}\label{eq:objRA}
Obj & =
\boldsymbol{\alpha}_{1} \cdot Cost  
+ \boldsymbol{\alpha}_{2} \cdot \left(1 - Profit\right) \nonumber \\
& + \boldsymbol{\alpha}_{3} \cdot \left[ Utilization + \left(1 - FairnessIndex\right) \right] \nonumber \\
& + \boldsymbol{\alpha}_{4} \cdot GreenPenalty
\end{align}

%-----------------------------------------------------------------------------------
\begin{figure*}[htb]
    \centering
    \begin{subfigure}[b]{0.19\textwidth}
        \centering
        \includegraphics[width=\textwidth]{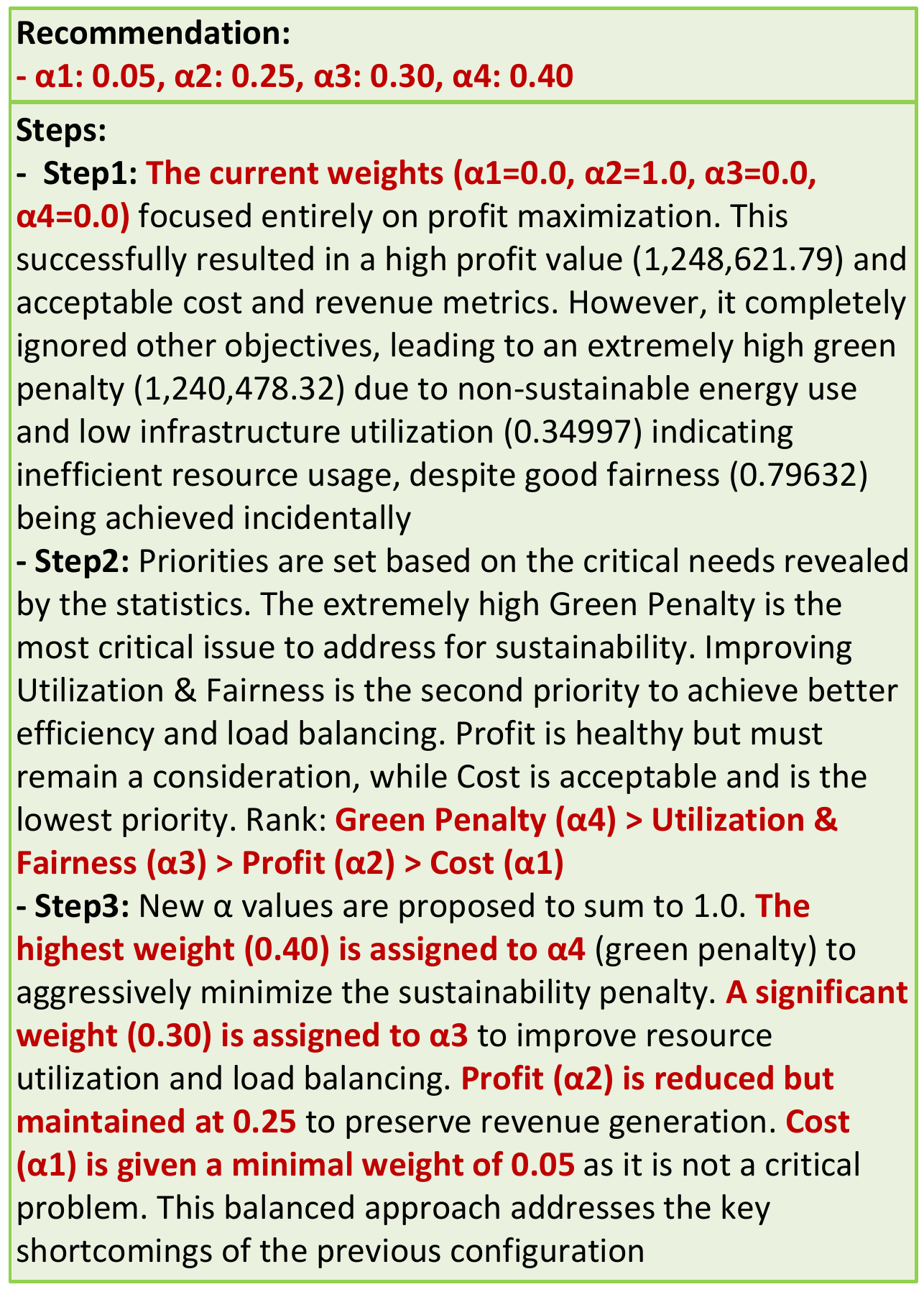}
        \caption{24h}
        \label{fig:RAA-Action1}
    \end{subfigure}\hfill
    \begin{subfigure}[b]{0.19\textwidth}
        \centering
        \includegraphics[width=\textwidth]{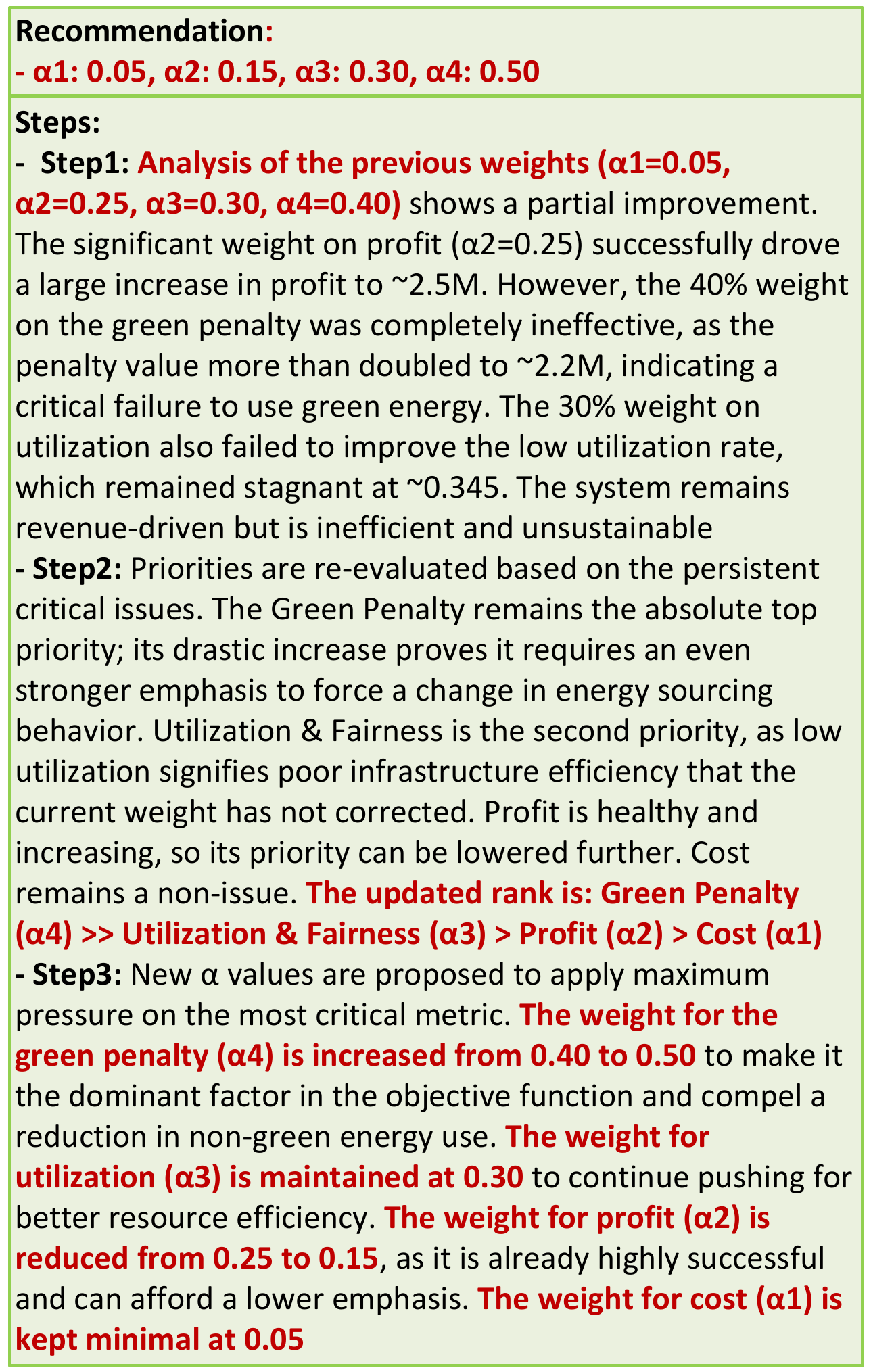}
        \caption{48h}
        \label{fig:RAA-Action2}
    \end{subfigure}\hfill
    \begin{subfigure}[b]{0.19\textwidth}
        \centering
        \includegraphics[width=\textwidth]{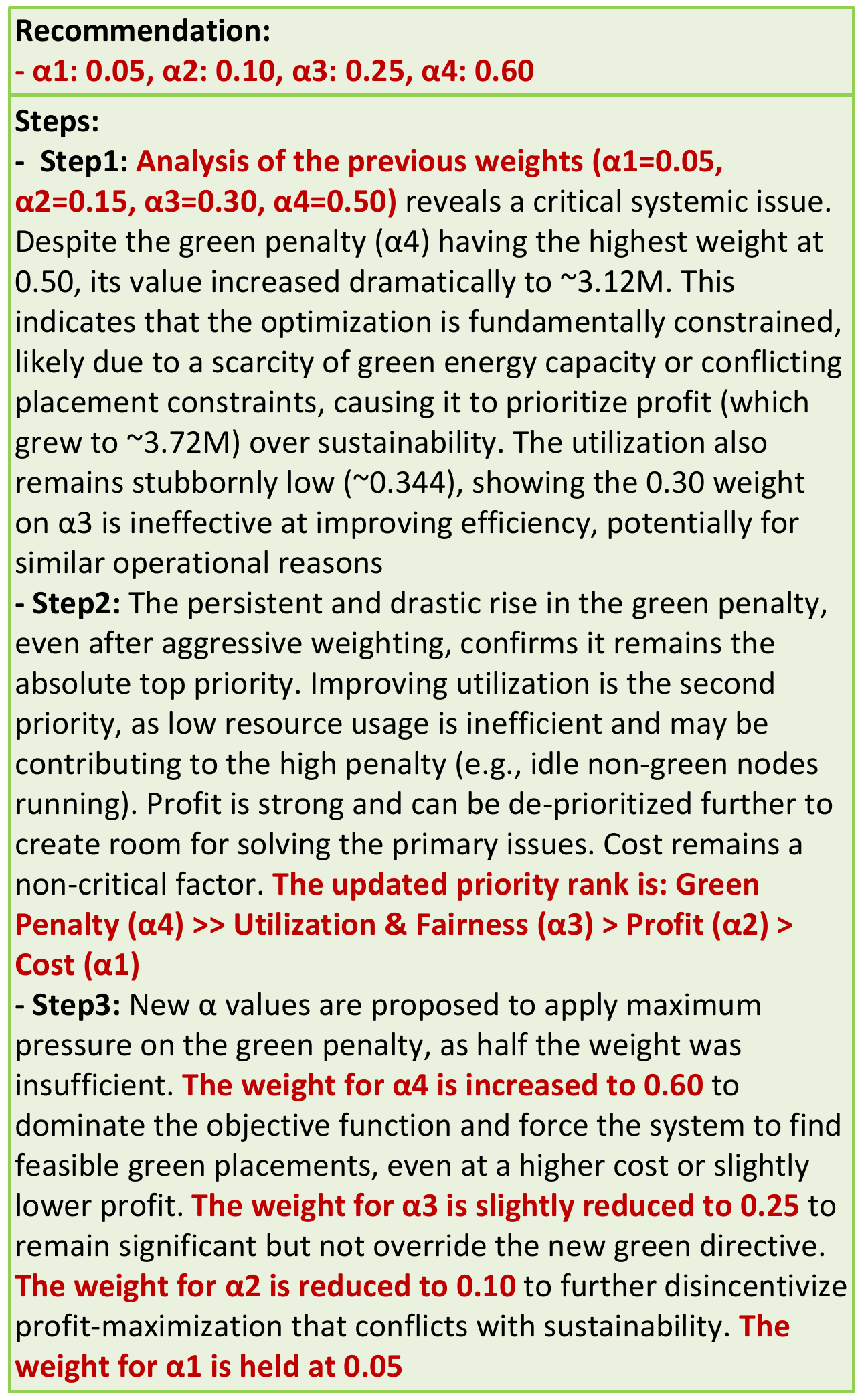}
        \caption{72h}
        \label{fig:RAA-Action3}
    \end{subfigure}\hfill
    \begin{subfigure}[b]{0.19\textwidth}
        \centering
        \includegraphics[width=\textwidth]{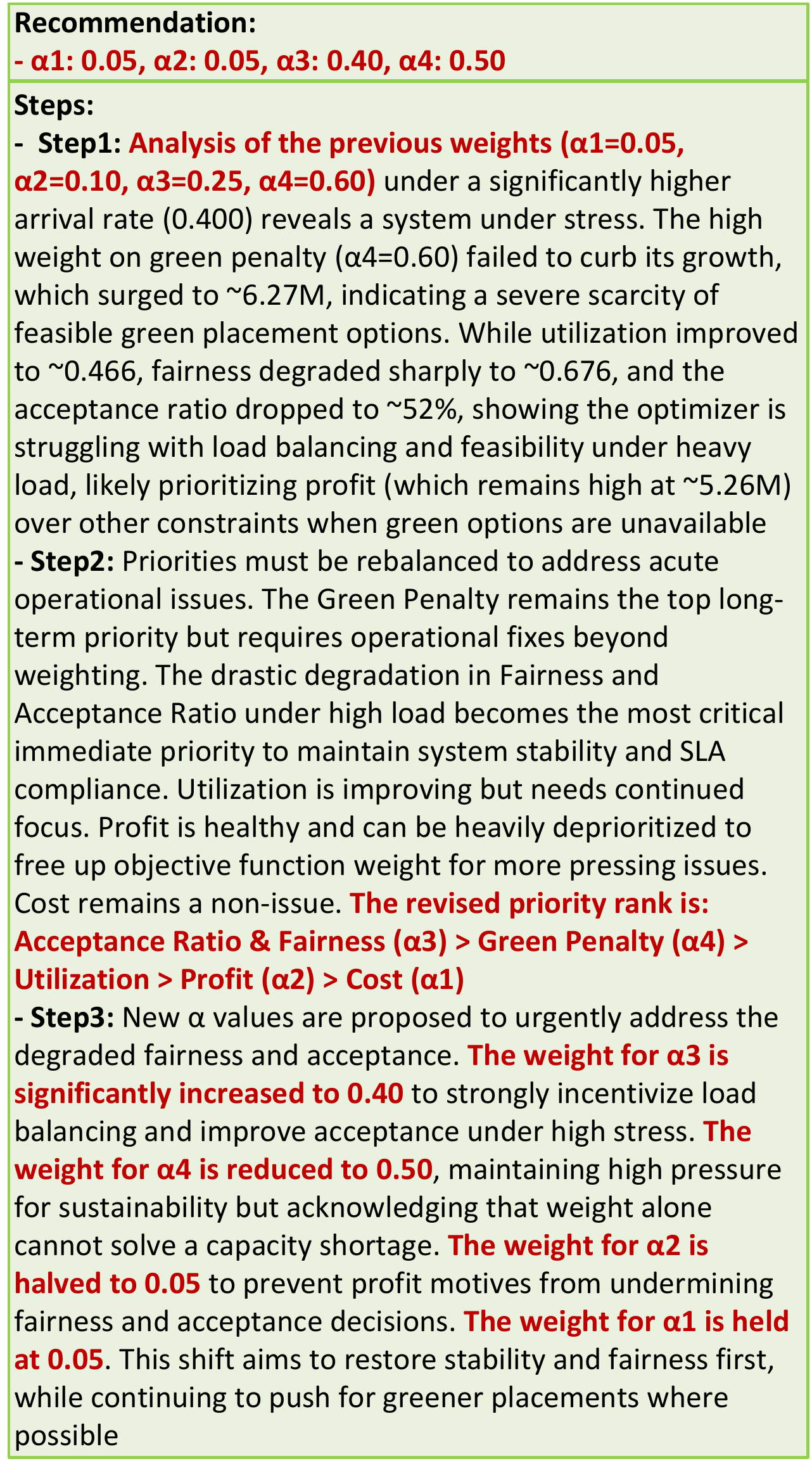}
        \caption{96h}
        \label{fig:RAA-Action4}
    \end{subfigure}\hfill
    \begin{subfigure}[b]{0.19\textwidth}
        \centering
        \includegraphics[width=\textwidth]{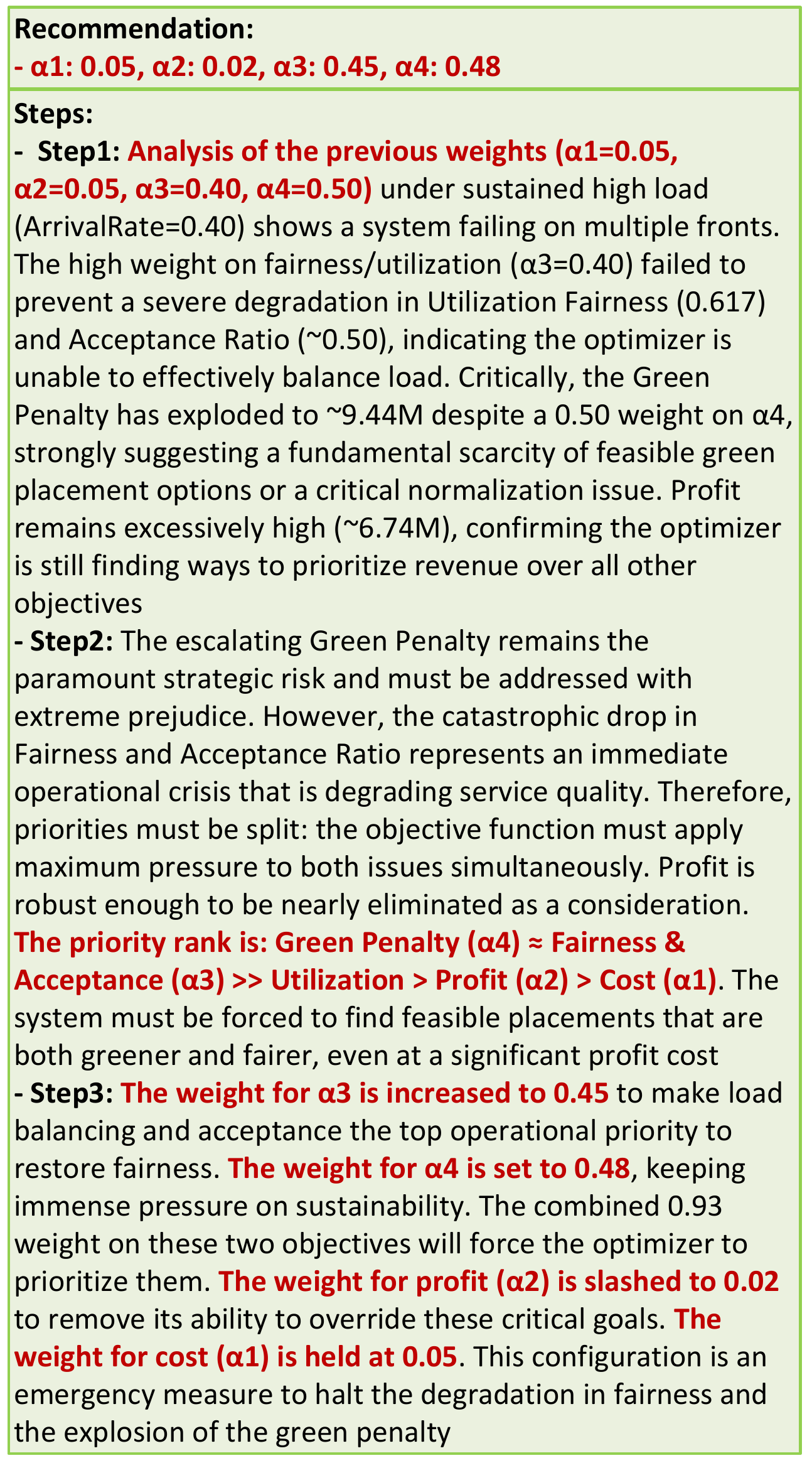}
        \caption{120h}
        \label{fig:RAA-Action5}
    \end{subfigure}
    \caption{Decision reports of the Resource Allocation Agent across five evaluation cycles.}
    \label{fig:RAA-Actions}
\end{figure*}
%
%-----------------------------------------------------------------------------------
% Define width for 3 figures per row
\newcommand{\figwidth}{0.32\textwidth} % adjust to leave space between
\begin{figure*}[!htb]
\centering

\subfloat[SFC Arrival Rate]{%
  \includegraphics[width=\figwidth]{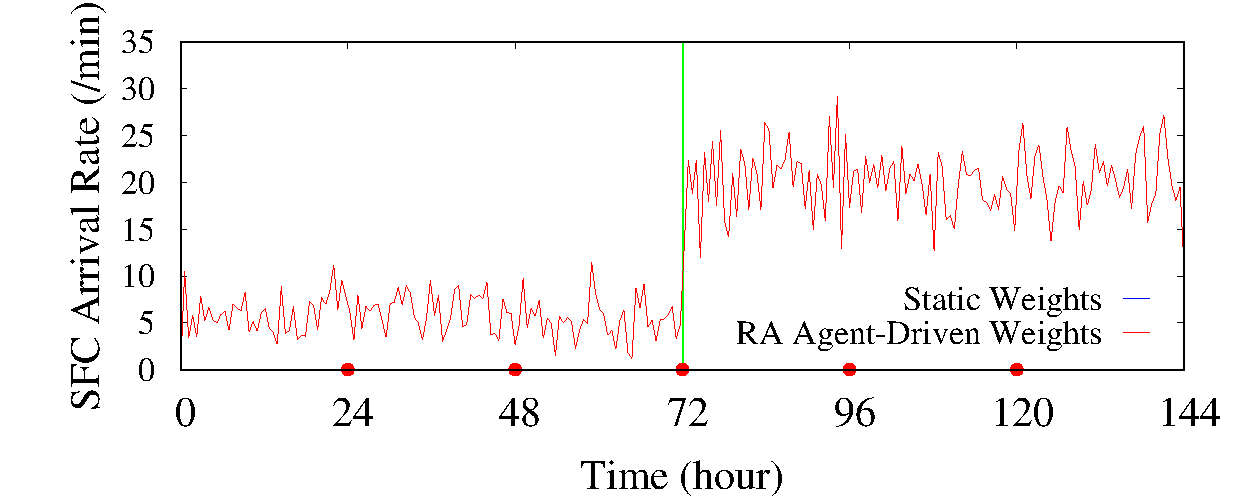}
}
\hfill
\subfloat[Cost (\$/min)]{%
  \includegraphics[width=\figwidth]{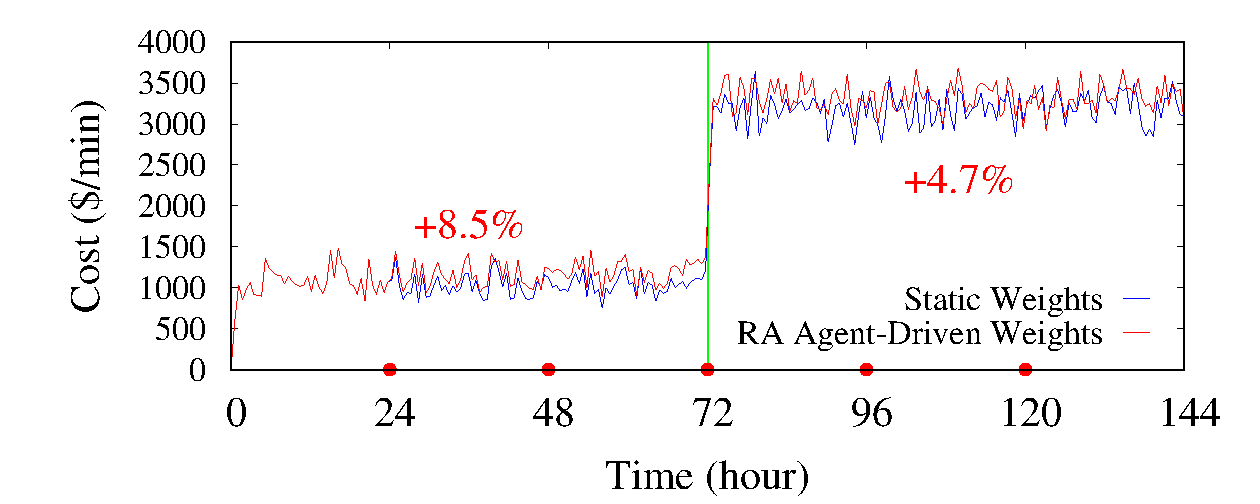}
}
\hfill
\subfloat[Revenue (\$/min)]{%
  \includegraphics[width=\figwidth]{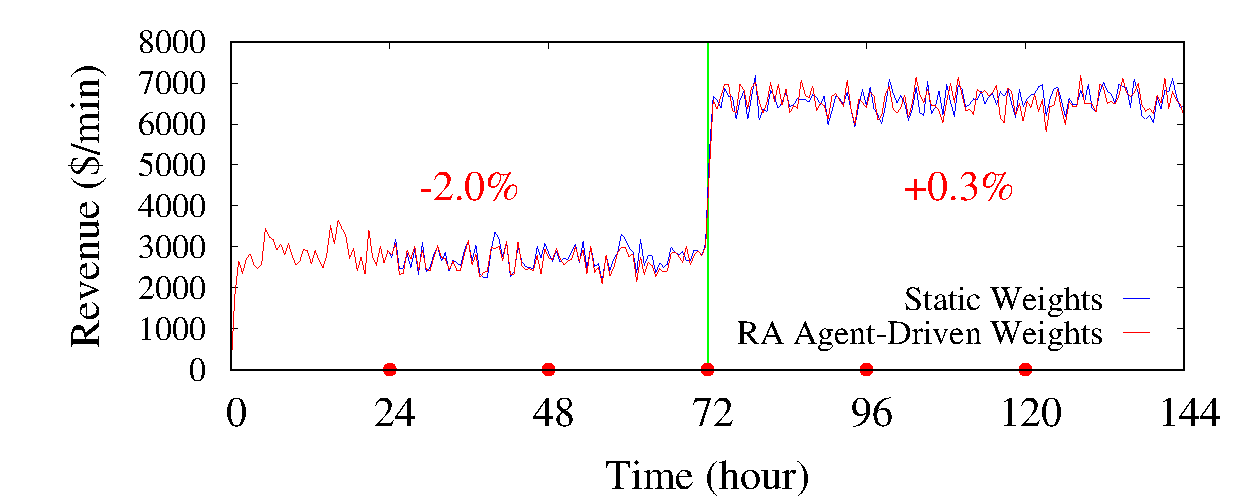}
}

\vspace{1em}

\subfloat[Green Penalty (\$/min)]{%
  \includegraphics[width=\figwidth]{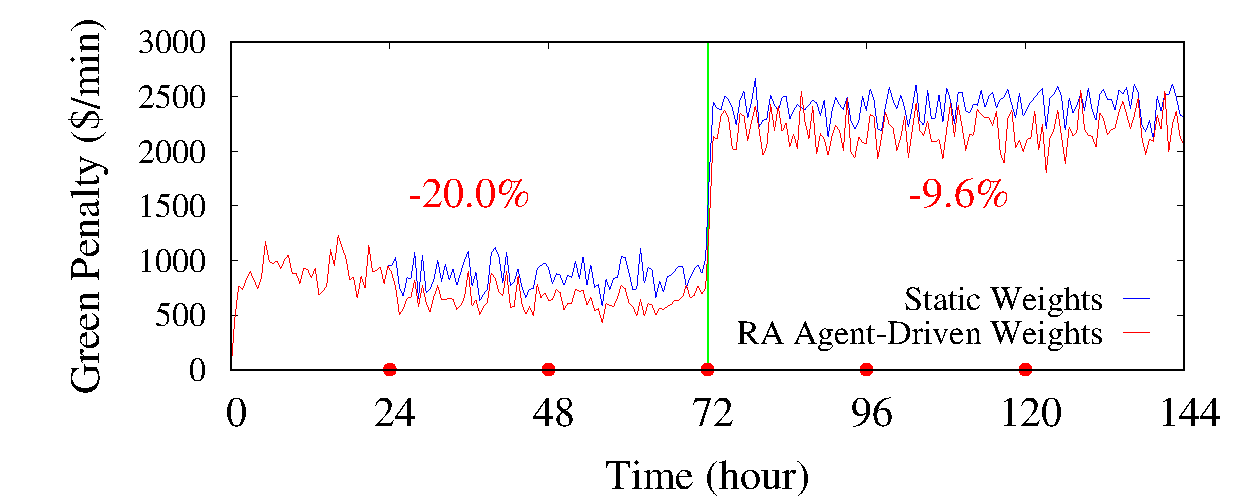}
}
\hfill
\subfloat[Profit (\$/min)]{%
  \includegraphics[width=\figwidth]{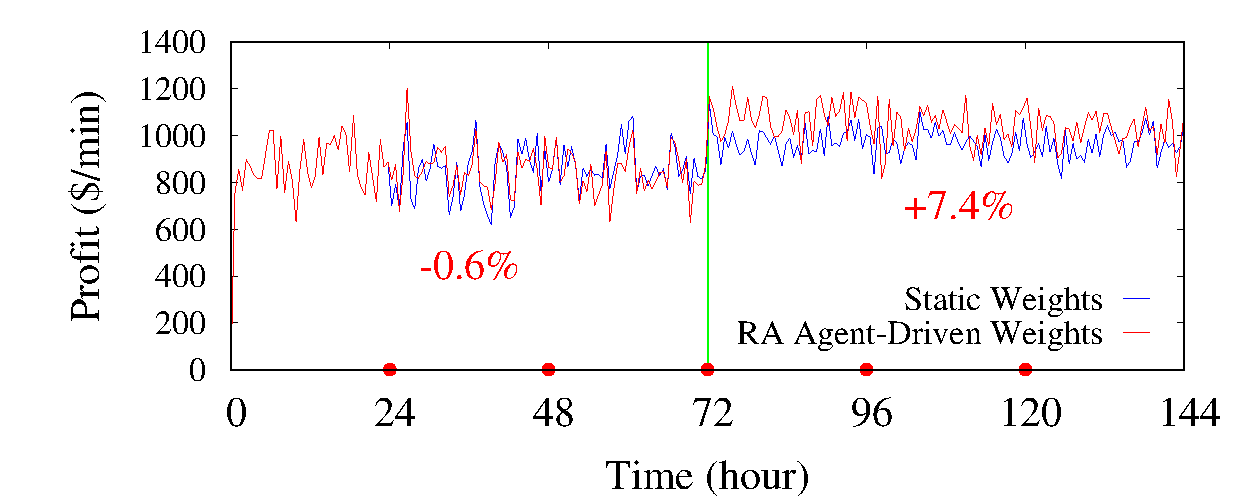}
}
\hfill
\subfloat[Infrastructure Utilization]{%
  \includegraphics[width=\figwidth]{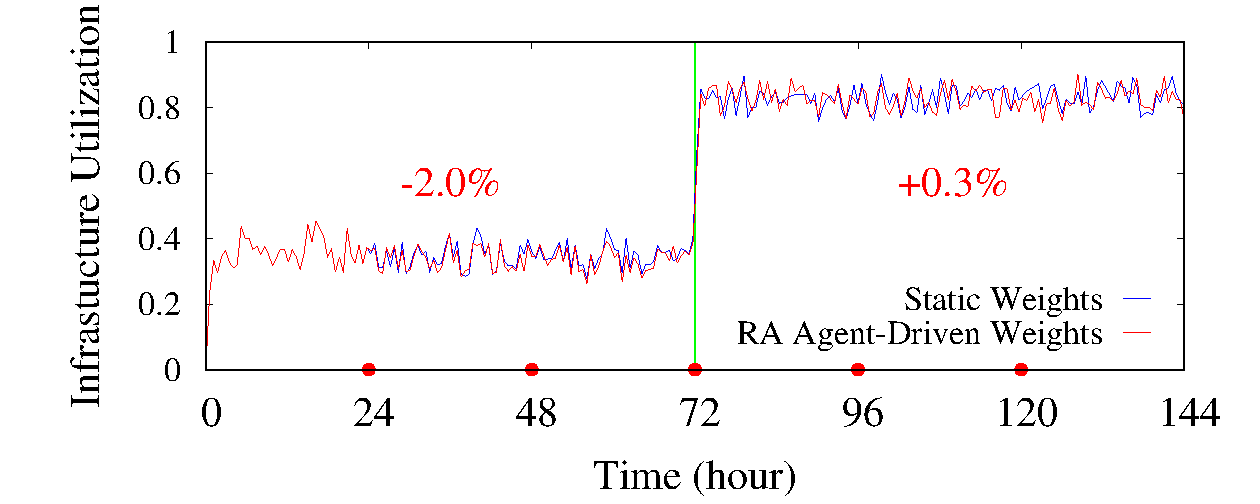}
}

\vspace{1em}

\subfloat[Utilization Fairness]{%
  \includegraphics[width=\figwidth]{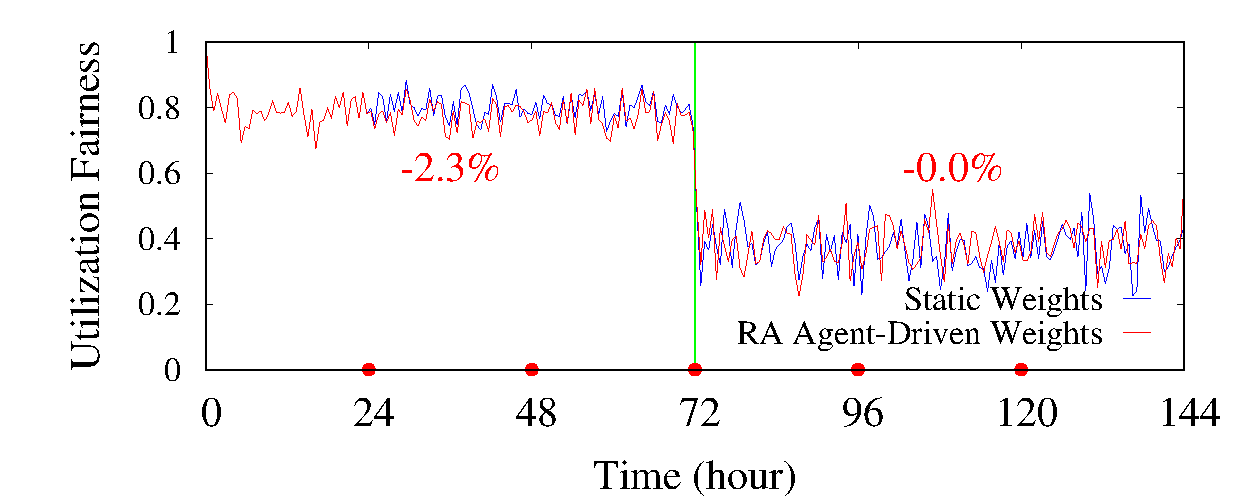}
}

\caption{Comparison of resource allocation metrics with static weights and with RA agent-managed weights. 
The red bullets indicate the times at which the RA agent updated the weights. 
The green vertical line marks the moment when the SFC arrival rate doubled. 
For each half of each subfigure, the displayed percentage represents the relative increase or decrease of the metric under RA agent-managed weights compared to the metric with static weights.}
\label{fig:RA-experiments}
\end{figure*}
%------------------------------------------------------------------------------------------------------

%
This formulation captures four complementary goals: (i)~minimizing operational and deployment cost, (ii)~maximizing profit, (iii)~avoiding resource overutilization and improving fairness across nodes by leveraging Jain’s fairness index~\cite{jain1984quantitative}, and~(iv)~enforcing sustainability by~penalizing allocations that rely on non-renewable energy sources. Each~term in the objective function is normalized using the min–max method to ensure that all components contribute comparably to the overall optimization, preventing any single metric from dominating due to scale differences. 

The RA Agent, leveraging its LLM powered by \texttt{GPT-5 Thinking}~\cite{openai_introducing_gpt5_2025}, acts as a decision engine that periodically monitors infrastructure metrics and~high-level objectives, and recommends updated weight values ($\alpha_1$ to~$\alpha_4$) for~the~multi-objective optimization function. As~such, RA Agent adjusts the~relative importance of each objective based on its analysis of runtime network conditions, costs, profits, and~green energy penalties, thereby balancing economic efficiency, performance, and~environmental sustainability. 

In our experiments, we simulate a physical infrastructure of~24~nodes connected through 57~links~\cite{moufakirITU2022}. Each node can host multiple instances (containers) and~is~characterized by its per-instance operational cost (ranging from 0.03~to~0.08~\$/s), green penalty (0.01~to~0.05~\$/s), selling price (0.07~to~0.08~\$/s), and capacity (40~to~100~instances).
SFCs are~generated dynamically over time, each connecting a single source to~a~randomly selected destination. A~chain contains between~1 and~20~NFs. The~RA~algorithm is implemented based on~the code of~\cite{moufakirITU2022} and~is~guided by the objective function~(Eq.~\ref{eq:objRA}).

Figure~\ref{fig:RA-experiments} summarizes the results of the resource allocation simulations over a period of six days (144~hours). We~compared two scenarios: one in which the objective function weights remain fixed throughout the~simulation to focus only on maximizing profit ($\alpha_{2}=1$), and another in~which the RA Agent dynamically adjusts the weights based on reports received every 24~hours. Fig.~\ref{fig:RAA-Actions}~illustrates the~reasoning performed by the RA agent’s LLM and guided by the prompt structure provided in~Fig.~\ref{fig:RA-prompt-template}. The~reasoning takes place at each evaluation cycle (set to 24 hours). For~each cycle, it~shows the analysis of the current statistics, the~decision on updated weights, and the rationale behind~it. 

Subfigure~\ref{fig:RA-experiments}(a) shows the SFC arrival rate over time (i.e.,~the average number of SFC requests per minute), highlighting the dynamic workload faced by the RA Agent. To~evaluate the agent's response to sudden demand changes, we deliberately double the~SFC arrival rate after three days of~simulation, leading to higher infrastructure utilization. Since both scenarios are executed under the same simulation setup and arrival process, the arrival rate curves are identical and~serve as~a~common reference for comparison.

We first analyze the~initial three days of~the~simulation, during which the~average arrival rate is~6~SFCs/min.
As~it~can be seen in the Subfigure~(d) and ~(e), after the first 24~hours, the RA-agent has succeeded to adjust the weights, reducing the green penalty by up to 20\% while keeping almost the same profit for~the~infrastructure provider.  This~was explicitly decided by~the~Agent's LLM in~Fig.~\ref{fig:RAA-Actions}~(a) and~(b) at~24h and 48h. The~updated weights prioritized allocations on nodes powered by greener energy sources (the ones with low green penalty), but this came at the expense of relying on resources with higher operational costs (+8\%). The revenue, infrastructure utilization, and fairness were barely impacted (-2\% as shown in Subfigure~(c), (f) and~(g)).

%-----------------------------------------
We then examine the final three days, with the SFC arrival rate doubled. As shown in Subfigures~(d) and~(e), the~RA~agent successfully adjusted the weights, increasing the~profit by~7.4\% while reducing the green penalty by 9.6\%. However, this improvement came at the expense of a moderate increase in~operational cost (+4.7\%). These adjustments were explicitly decided by the Agent's LLM in~Fig.\ref{fig:RAA-Actions}(d) and~(e) at~96h and~120h. The updated weights continued to prioritize allocations on nodes with greener energy sources (lower green penalty), however with higher weights for the utilization and~fairness compared to the first three days as the LLM increased load on the infrastructure on the infrastructure and decided to further balance the~workload and improve fairness among the physical load (see Fig.\ref{fig:RAA-Actions}(d)~and~(e)). The resulting revenue, infrastructure utilization, and fairness remained comparable for~both scenarios as observed in~Subfigures~\ref{fig:RA-experiments}~(c), (f), and~(g).

These results demonstrate that the RA agent’s LLM effectively adapts the weight allocations to dynamic workloads, by efficiently analyzing the current network status and achieving a balanced trade-off between profit, green energy usage, operational cost, and fairness.

%=============================================================================
\section{Key Research Directions}\label{sec:KeyResearchDirections}
%=============================================================================
Although promising, the design and experiments presented in this paper are still early. In the following, we present key research challenges in order to evolve FlexNGIA 2.0 toward a~full-fledged agentic AI network.

\subsection{Building Powerful and Reliable LLMs for Network Agentic~AI: Design, Customization, and Evaluation}
A key challenge in realizing FlexNGIA 2.0’s Agentic~AI vision lies in building the brain of each agent, its~LLM. This work presented proof-of-concept and preliminary experimental results where we used generic LLMs that were customized thanks to the prompts. We have also observed varying performance and outcomes when testing different LLMs (e.g.,~GPT-5, DeepSeek-R1-Distill-Llama-70B), which underscores the importance of developing standardized benchmarks to systematically evaluate and compare their~capabilities in network-related tasks. 

Beyond basic customization, it is also crucial to develop purpose-built LLMs tailored specifically for each agent’s unique role and responsibilities. TelecomGPT~\cite{zou2024telecomgpt} is a telecom-specific LLM pre-trained on domain-specific datasets derived from 3GPP and IEEE standards, as~well as arXiv publications. Such a cornerstone could support the development of powerful LLMs capable of designing protocols, algorithms, and management schemes for~communication networks from congestion control, traffic engineering to resource allocation, monitoring and fault management. Purpose-built and fine-tuned LLMs must not only understand networking concepts but also reason effectively, gather and filter relevant data from diverse sources and~existing protocols and algorithms, identify cause–effect relationships, and maintain situational awareness. The~agent’s LLM must be capable of setting both short-term and~long-term objectives, devising adaptive strategies, making informed decisions, and translating them into precise logical actions and~step-by-step plans. Furthermore, it must be able to anticipate potential collaborations with other agents, leveraging collective intelligence to achieve shared goals. 

Furthermore, ensuring the reliability and the performance of protocols, algorithms and configurations designed by these LLMs requires robust mechanisms to verify and control their logic and implementation. This includes continuous testing, validation, and performance evaluation, potentially using parallel simulation or digital twin environments at~runtime, to~safely assess their behavior and impact before full deployment. Such rigorous checks are critical to prevent unintended consequences and maintain network stability.

Designing such a cognitive core, capable of integrating domain expertise, reasoning skills, and autonomous decision-making would mark a fundamental breakthrough in~enabling FlexNGIA 2.0’s agentic AI adaptive, self-evolving network management paradigm.

\subsection{LLM Prompting for the~Design of~Intelligent Network Protocols and Management Algorithms}
Prompting plays a crucial role in effectively leveraging LLMs, as it directly shapes the quality, relevance, and~explainability of their outputs. Well-crafted prompts act as~the~interface through which an LLM understands the context, assesses the situation, and generates solutions. In~our~experiments, we observed that the structure and content of~the~prompt significantly affect the LLM’s ability to design communication protocols and~network-related algorithms.

Therefore, it is essential to develop dedicated prompt models or templates specifically tailored for the LLMs powering FlexNGIA 2.0 AI agents. Such prompt frameworks should enable the LLM to efficiently perceive network states, identify goals and strategies, define step-by-step actions and~plans, anticipate potential impact, invoke relevant measurement and analysis tools, and~apply systematic reasoning to design network components. %Moreover, the prompt should force agents to clearly justify and explain their decisions to humans or other agents.
Establishing robust prompting methodologies will be key to harnessing LLMs’ potential in creating adaptive, intelligent, and explainable networking solutions within the FlexNGIA 2.0 architecture.

\subsection{Rethinking the way Network Services, Protocols and,~Algorithms are Designed}

A key step in advancing network intelligence lies in~rethinking how services, protocols,~algorithms, and standards are designed in the AI era. AI, especially LLMs, can dramatically accelerate the entire workflow, spanning design, implementation, testing, performance evaluation, and deployment of network functions and~protocols.
Our preliminary experiments indicate that~LLMs deliver more reliable and higher-quality algorithms when guided by structured inputs and access to rich contextual information. To enable this, a comprehensive set of libraries is essential, encompassing existing services, protocols, and~algorithms, as~well~as~fundamental low-level building blocks such as packet processing primitives, measurement tools, and algorithmic components. These libraries could serve as~the~foundation for dynamically composing sophisticated network services, protocols, and algorithms, customized for~specific applications and operational scenarios.

The orchestration and synthesis of these components can be performed autonomously by LLM-powered AI~agents or~via~human-in-the-loop collaboration, combining computational efficiency with domain expertise to achieve optimized, adaptable, and robust network designs.
%---------------------------------------------------------------------------
\subsection{AI Agent Design and Coordination Challenges }
%---------------------------------------------------------------------------
Designing AI agents for FlexNGIA 2.0 introduces several fundamental challenges. Each agent must possess a robust internal architecture that supports autonomous reasoning, goal setting, adaptive planning, and learning from dynamic network conditions. Beyond intelligence, agents must be equipped with appropriate tools to perceive the environment, analyze network states, and execute actions safely. Additionally, agents must coordinate effectively with each other to ensure coherent system-wide behavior. Without such coordination, independent decisions could conflict, leading to suboptimal performance and unintended outcomes. Proper collaboration enables the overall system to align objectives, balance trade-offs, and maintain overall stability, ensuring that actions collectively enhance performance and reliability objectives.

This work presented an initial design and implementation of these AI~agents that~demonstrate their potential for~runtime adaptation and decision-making. However, substantial work remains to refine agent architectures, expand their capabilities, enhance inter-agent coordination, and~validate their performance across complex scenarios.

%---------------------------------------------------------------------------
\subsection{Designing New Breeds of Network Protocols}
%---------------------------------------------------------------------------
The ability offered by FlexNGIA to design protocols customized for each application opens the door to~a~wide array of possibilities and challenges. One compelling research direction is the development of multi-point aware protocols, which requires rethinking traditional end-to-end communication models. Such protocols must enable a single instance to handle multiple endpoints while ensuring consistency, reliability, and performance guarantees. This approach allows the network to manage application traffic more efficiently, by intelligently handling priorities, orchestrating in-path processing, and managing packet duplication across different endpoints.

Another compelling challenge is the design of network function-aware protocols, which are cognizant of the presence, capabilities, and behavior of in-path network functions. Such protocols should not only leverage these network functions effectively but may also influence or define their behavior and operational logic to optimize end-to-end communication and achieve the performance and reliability goals.

Additional challenges include ensuring scalability across heterogeneous network environments, providing robust security and privacy guarantees, supporting real-time monitoring and feedback, and enabling interoperability with legacy protocols. 
Furthermore, defining mechanisms for dynamic customization and evolution of protocol logic at runtime, potentially guided by AI agents, presents both theoretical and practical research questions in protocol design, verification, and deployment.
%==============================================================================
\section{Conclusion}\label{sec:Conclusion}
%==============================================================================

%The stringent performance requirements of immersive applications and teleoperation underscore the urgent need for a radical rethinking of~the~Internet’s architecture. 
%The rigid protocols, static control mechanisms, and best-effort delivery model that characterize today’s Internet cannot meet the stringent performance and reliability requirements of~these~emerging applications. 
%
In~this~paper, we proposed FlexNGIA 2.0, an~Agentic~AI-driven architecture for the future Internet that~embeds cognitive intelligence and reasoning capabilities into the core of the~network. 
This is achieved by introducing LLM-based~AI~agents capable of~autonomously evaluate the~network conditions and application requirements, and~to~design, implement, and~adapt service function chains, network functions, communication protocols, congestion control strategies, and resource allocation algorithms.

Our prototype and experimental evaluation demonstrated the~ability of these agents to~design and~implement end-to-end custom SFCs, to~craft tailored transport protocols, to~select or design and implement congestion control schemes, and~to~optimize resource allocation, all at runtime. These results highlight the transformative potential of integrating generative and reasoning-driven agents into the~network architecture, paving the way toward networks that are not only programmable but~also self-learning and self-adaptive.
%, transforming the Internet from a static and pre-programmed system into~a~dynamic, self-evolving ecosystem.

While this work is a major step toward agentic AI-driven networking, numerous key challenges remain. Future efforts must move beyond prompt-based customization toward purpose-built LLMs, with strong domain knowledge, reasoning, planning, and algorithm design capabilities adapted to network, services, protocols and management. %These models should design protocols, anticipate collaboration, and adapt dynamically, while their outputs require rigorous validation through testing and digital twins. Standardized benchmarks, robust prompting frameworks, and rich libraries of network components are also essential. 

%===================================================================
%\bibliography{bibtex/MyOwnRef-Journals,bibtex/MyOwnRef-Conferences,bibtex/References-FlexNGIA1, bibtex/References-FlexNGIA2, bibtex/cc_references} 
%\bibliographystyle{IEEEtran}

% Generated by IEEEtran.bst, version: 1.14 (2015/08/26)

\end{document}